\documentclass{aa}
\pdfoutput=1

\usepackage{amsmath}
\usepackage[varg]{txfonts}
\usepackage{graphicx}
\usepackage{grffile}
\usepackage{subcaption}
\usepackage{natbib}
\usepackage{xspace}
\usepackage{nicefrac}
\usepackage{xcolor}
\usepackage{listings}
\usepackage[colorlinks=true,linkcolor=blue,urlcolor=blue,citecolor=blue]{hyperref}

\authorrunning{}
\titlerunning {Mixed-mode coupling in the Red Clump I}

\newcommand{\Sred}{\ensuremath{\tilde{S}}\xspace}
\newcommand{\Nred}{\ensuremath{\tilde{N}}\xspace}
\newcommand{\Nu}{\ensuremath{\mathcal{V}}\xspace}
\newcommand{\A}{\ensuremath{\mathcal{A}}\xspace}
\newcommand{\dd}{\ensuremath{\, \mathrm{d}}}

\newcommand{\numax}{\ensuremath{\nu_{\mathrm{max}}}\xspace}
\newcommand{\msol}{\ensuremath{\mathrm{M}_\odot}\xspace}
\newcommand{\rsol}{\ensuremath{\mathrm{R}_\odot}\xspace}
\newcommand{\Lsol}{\ensuremath{\mathrm{L}_\odot}\xspace}
\newcommand{\FeH}{\text{[Fe/H]}\xspace}
\newcommand{\deltanu}{\ensuremath{\Delta\nu}\xspace}
\newcommand{\uHz}{\ensuremath{\mathrm{\mu Hz}}\xspace}
\newcommand{\dqdnu}{\ensuremath{\langle \dd q/\dd\nu_\mathrm{q}\rangle}\xspace}

\newcommand{\MESA}{\texttt{MESA}\xspace}

\newcommand{\brunt}{Brunt-V\"ais\"al\"a\xspace}
\newcommand{\Kepler}{\emph{Kepler}\xspace}
\newcommand{\CoRoT}{{CoRoT}\xspace}
\newcommand{\TESS}{{TESS}\xspace}
\newcommand{\Teff}{\ensuremath{T_\mathrm{eff}}\xspace}

\renewcommand{\thesubfigure}{\Alph{subfigure}}

\begin{document}

\title{Mixed-mode coupling in the red clump}
\subtitle{I. Standard single star models}

\author{Walter E. van Rossem \inst{1,2}
	\and Andrea Miglio \inst{1,2,3}
	\and Josefina Montalbán \inst{1,2}}

\institute{Department of Physics \& Astronomy "Augusto Righi", University of Bologna, via Gobetti 93/2, 40129 Bologna, Italy \and School of Physics and Astronomy, University of Birmingham, Edgbaston B15 2TT, UK \and INAF-Astrophysics and Space Science Observatory of Bologna, via Gobetti 93/3, 40129 Bologna, Italy}

\date{}

\abstract{The investigation of global, resonant oscillation modes in red giant stars offers valuable insights into their internal structures. In this study, we investigate in detail the information we can recover on the structural properties of core-helium burning (CHeB) stars  by examining how the coupling between gravity- and pressure-mode cavities depends on several stellar properties, including mass, chemical composition, and evolutionary state.

Using the structure of models computed with the stellar evolution code \MESA, we calculate the coupling coefficient implementing analytical expressions, which are appropriate for the strong coupling regime and the structure of the evanescent region in CHeB stars.

Our analysis reveals a notable anti-correlation between the coupling coefficient and both the mass and metallicity of stars in the regime $M \lesssim 1.8~\msol$, in agreement with \Kepler data. We attribute this correlation primarily to variations in the density contrast between the stellar envelope and core. The strongest coupling is expected thus for red-horizontal branch stars, partially stripped stars, and stars in the higher-mass range exhibiting solar-like oscillations ($M \gtrsim 1.8~\msol$).
While our investigation emphasises some limitations of current analytical expressions, it also presents promising avenues. The frequency dependence of the coupling coefficient emerges as a potential tool for reconstructing the detailed stratification of the evanescent region.}

\keywords{asteroseismology -- stars: evolution -- stars: interiors -- stars: oscillations}

\maketitle

\section{Introduction}\label{sec:Intro}
Over the past decade, space-based missions such as \CoRoT \citep[{Convection Rotation \& Planetary Transits};][]{CoRoT}, \Kepler/K2 \citep{Kepler, K2}, and now \TESS \citep[{Transiting Exoplanet Survey Satellite};][]{TESS}, have provided photometric time series of sufficient precision and duration to enable the detection and precise characterization of stellar oscillations in different classes of stars. Particularly information-rich oscillation spectra have been detected in evolved stars showing solar-like oscillations. In the case of solar-like oscillations, the observed oscillation frequencies are typically described by a Gaussian-shaped power excess centred at the frequency of maximum oscillation power, $\numax$. These oscillations can provide valuable insights into the star's internal physics (e.g.~\citealt{Chaplin2013, Montalban2013a, Mosser2014, DiMauro2016, Hekker2017} and references therein).

In main sequence (MS) sun-like stars, solar-like oscillations consist primarily of acoustic modes (p-modes). When stars leave the MS, modes that exhibit both gravity- and pressure-like characteristics start populating observed pulsation spectra and become accessible to our investigations. These so-called mixed modes carry information on the structure of both the inner, dense radiative core (probed predominantly by g-modes) and the envelope (which largely determines the frequencies of p-modes). The detection of mixed modes in red-giant branch (RGB) and red-clump (RC) stars \citep{Bedding2010,Beck2011,Mosser2011} has enabled the determination of the star's evolutionary stage \citep{Montalban2010,Bedding2011,Mosser2011,Mosser2014,Hekker2018}, the rotation rate of stellar cores \citep[e.g.][]{Deheuvels2012,Beck2012,Eggenberger2012,DiMauro2016paper,Gehan2018}, together with inferences of the structure of the core, and near-core mixing \citep[e.g.][]{Bedding2011,Montalban2013,Mosser2015,Bossini2015,Cunha2015,Deheuvels2016,Deheuvels2020,Noll2024}.

Two complementary approaches are available to explore and investigate the interior of stars. A numerical approach to simulate modes in stellar models can be taken, as well as a more analytical approach in which the properties of modes are approximated, yet they provide a more explicit connection between the modes' properties and the internal structure of the star. In this study, we predominantly adopt the latter approach, focussing specifically on exploring the information carried by the coupling between the pressure- and gravity-mode cavities.

The coupling coefficient is an essential quantity in the study of mixed modes. It is a measure of how much energy can be transferred between the acoustic- and gravity-mode cavities, which is shown in the inertias of mixed modes in the cavities. One limitation in the interpretation of this coupling is that the analytical description of the coupling developed by \citet{Shibahashi1979} is only valid in the weak-coupling regime. \citet{Takata2016} developed a prescription for the strong-coupling regime, facilitating the extension of mixed-mode coupling studies to the early RGB, and as we show in this work, also the RC. While stellar model-based investigations in the strong-coupling regime have been reported for RGB stars \citep[e.g.][]{Hekker2018, Pincon2020, Jiang2022}, a systematic exploration of the structural information available for RC models has yet to be conducted. This is the primary focus of our work.

This paper is structured as follows. In Sect.~\ref{sec:Method} we first review the analytical approximation of the coupling coefficient and then we discuss our implementation and limitations encountered when estimating the coupling in the grid of models considered. In Sect.~\ref{sec:Results} we describe how the coupling coefficient depends on stellar properties. In Sect.~\ref{sec:Observations} we compare the results of this dependence with observations. Finally, in Sect.~\ref{sec:SummaryConcl} we discuss the implications of this work and conclude.

\section{Method}\label{sec:Method}

\subsection{Analytic approximations of the mixed-mode coupling coefficient}\label{sec:Method1}
Non-radial adiabatic stellar oscillations are governed by a fourth-order system of differential equations. To obtain approximate solutions or to understand the behaviour of non-radial waves in complex environments such as stellar interiors, one usually reduces that problem to one of second order which can then be studied using standard analytical techniques.
Ignoring perturbations to the gravitational potential in the oscillation equations (Cowling approximation, \citealt{Cowling1941}), allows approximate solutions to be developed. An asymptotic analysis of the resulting system, assuming that the scale of variation of the equilibrium quantities is much longer than the wavelength of studied oscillation, leads to a wave equation with a dispersion relation:
\begin{equation}
	k^2_r\approx \frac{\omega^2}{c_\mathrm{s}^2}\left(\frac{S_\ell^2}{\omega^2}-1\right)\left(\frac{N^2}{\omega^2}-1\right)\mathrm{,}
	\label{eq:dispersionrelation}
\end{equation}
\noindent where $k_r$ is the local radial wavenumber, $c_\mathrm{s}$ is the sound speed, and $\omega$ the angular oscillation frequency. This dispersion relation involves two characteristic frequencies, the \brunt frequency ($N$), and the Lamb frequency ($S_\ell$). The zeros of this relation (also called turning points) define the internal boundaries of the oscillatory propagation regions ($k_r^2 > 0$) of the perturbation with respect to adjacent regions of evanescent propagation in evanescent zones (EZ), ($k_r^2 <0$).
\begin{figure}
	\centering
	\includegraphics[width=\linewidth]{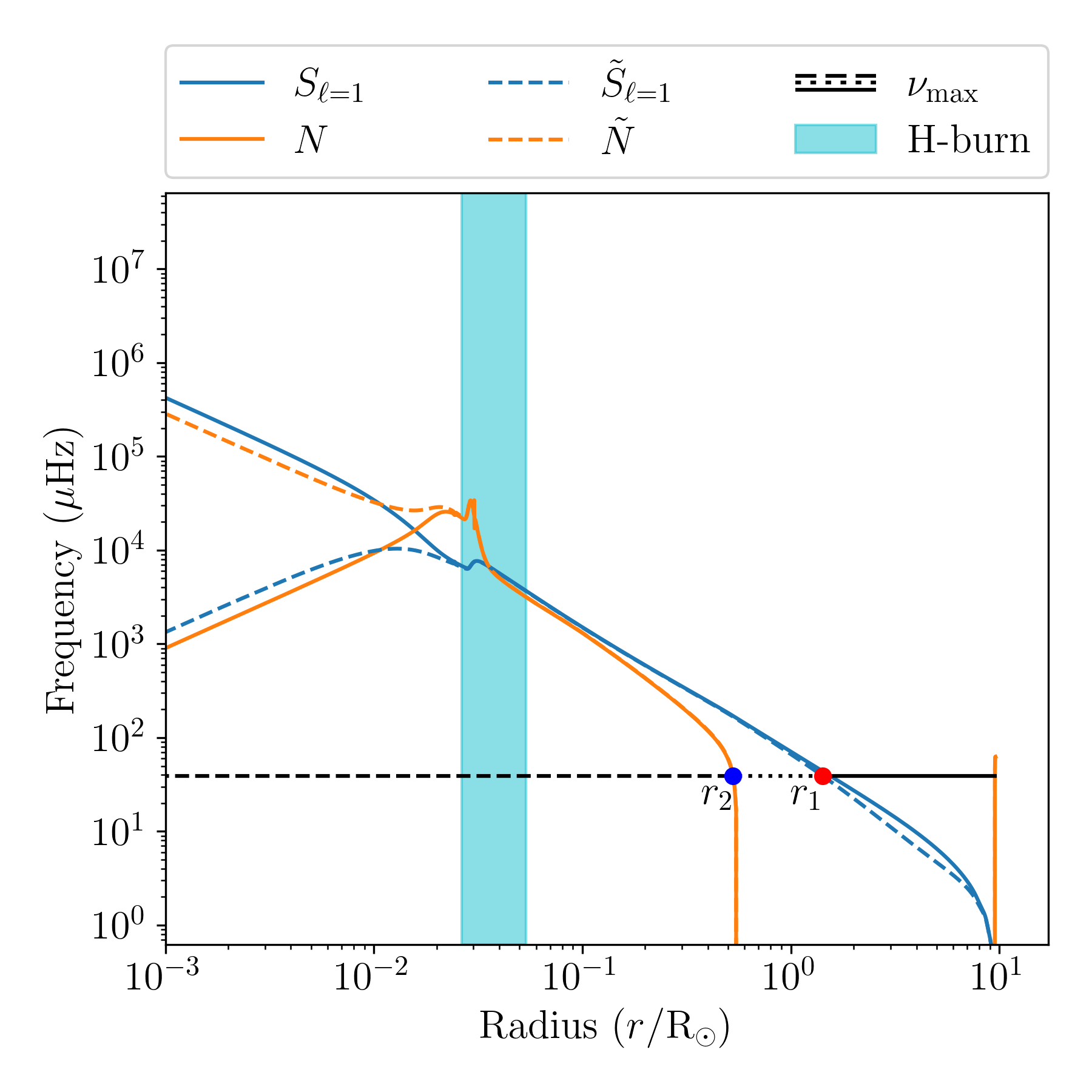}
	\caption{Propagation diagram of a 1\msol star in the RGB, just after the RGB bump. The Lamb and \brunt frequencies are shown as the solid blue and orange lines respectively. Their reduced counterparts are shown as dashed blue and orange lines respectively. A mixed mode with a frequency of \numax is shown as a dashed black line in the g-like part, as a solid black line in the p-like part, and as a dotted black line in the evanescent zone. The boundaries of the evanescent zone, $r_1$ and $r_2$ are shown as red and blue dots respectively. The hydrogen-burning shell is shown as the cyan shaded region.}
	\label{fig:propdiagramexplanation}
\end{figure}
Figure~\ref{fig:propdiagramexplanation} shows the propagation diagram of a perturbation with a frequency close to the typical frequency of stochastic oscillations in a red giant (\numax, $\omega = 2 \pi \nu$). This perturbation can, in principle, propagate as a g-character mode in the interior region (g-cavity where $\omega < N, S$), and as an acoustic (p-character) mode in the outer region (p-cavity where $\omega > N,S$). In fact, a wave in the g-cavity can tunnel through the turning point and reach the p-cavity with an amplitude high enough to have a significant amplitude in that cavity as well and propagate as an acoustic wave. \citet{Shibahashi1979} performed an asymptotic analysis of the 2$^\mathrm{nd}$ order system for these mixed modes. The JWKB solution under the assumption of a large evanescent zone (large with respect to the wavenumber) of these mixed modes, leads to the resonance condition for mixed modes, which can be written as:
\begin{equation}
	\cot \Theta_\mathrm{g} \tan \Theta_\mathrm{p} = q,
	\label{eq:resonance}
\end{equation}
where $\Theta_\mathrm{g}$ and $\Theta_\mathrm{p}$ are the integrals of $k_r$ in the p- and g-cavities respectively, together with the phase shifts induced by reflection at their boundaries. The parameter $q$, called the coupling coefficient, is related to the transmission ($T$) of wave energy through the evanescent zone between those cavities and can take a value between 0 and 1. Although first derived in the context of large evanescent zones (hence weak coupling), the resonance condition in Eq.~\eqref{eq:resonance} is general, as proven by \citet{Takata2016}, and $q$ can be written as:
\begin{equation}
	q=\frac{1-\sqrt{1-T^2}}{1+\sqrt{1-T^2}}\mathrm{.}
	\label{eq:qT}
\end{equation}
When $q \ll 1$, the wave is mainly trapped in one of the two cavities with properties close to those of pure p- or g-modes. As $q$ increases, the energy of the wave is distributed amongst the two regions. The final frequency of the mixed mode and its dominant character depends not only on $\Theta_\mathrm{g}$ and $\Theta_\mathrm{p}$, but also on the properties of the evanescent zone, through $q$.

In the framework of the large evanescent zone solved by \citet{Shibahashi1979}, which hereafter will be called the weak-coupling approximation, the coupling coefficient is defined as:
\begin{equation}
	q_\mathrm{w}=\frac{1}{4} \exp(-2\int_{r_1}^{r_2} |k_r|dr)=\frac{1}{4} \exp(-2\pi X)\approx \frac{T^2}{4}\mathrm{,}
	\label{eq:qw}
\end{equation}
\noindent with $X=1/\pi \int_{r_1}^{r_2} |k_r|dr$, and $r_1$, $r_2$ being the limits of the evanescent zone (Fig.~\ref{fig:propdiagramexplanation}). As can be seen from Eq.~\eqref{eq:qw}, $q_\mathrm{w}$ can reach a maximum of 0.25. However, by fitting the asymptotic relation for mixed modes, \citet{Mosser2017} found that over 40\% of the \Kepler red giants have $q$ significantly larger than 0.25.

The limitations of this coupling-coefficient prescription may be due to approximations made in its derivation: the Cowling approximation and the weak-coupling approximation. The Cowling approximation is not valid in the case of dipole modes, in particular dipole mixed p- and g-modes \citep[e.g.~appendix~A.1 from][]{Pincon2020}. Moreover, there are phases of evolution, such as subgiant branch (SGB) and CHeB phases in which the turning points ($r_1$, $r_2$) may become very close in terms of oscillation wavenumber, violating one of the assumptions underlying the weak-coupling approximation.

Takata's work \citep{Takata2005,Takata2006,Takata2016,Takata2016-2} has made it possible to study mixed dipole modes without using these approximations. \citet{Takata2005,Takata2006} simplified the problem by a change of dependent variables, and identified a first integral specific to dipolar oscillations. This reduces the problem to a second-order system of ordinary differential equations while keeping the perturbation to the gravitational potential. On the other hand, \citet{Takata2016,Takata2016-2} performed an asymptotic analysis of the system and provides a JWKB solution for the dipole mixed modes in the case of a very narrow evanescent zone. The system of equations is similar in form to that obtained using the Cowling approximation.
Also, the equivalent dispersion relation has the same form as Eq.~\eqref{eq:dispersionrelation}, but with the critical frequencies replaced by the corresponding reduced characteristic frequencies:
\begin{equation}\label{eq:N_red}
	\Nred = \frac{ N}{J}
\end{equation}
and
\begin{equation}\label{eq:S_red}
	\Sred = J S_{\ell=1}\mathrm{,}
\end{equation}
where $J$ includes the effects of perturbations to the gravitational potential, and is related to the density concentration through:
\begin{equation}\label{eq:J}
	J = 1 - \frac{4\pi}{3} \frac{\rho r^3}{m}= 1 -\frac{\rho}{\langle\rho\rangle}_r\mathrm{,}
\end{equation}
where $\rho$ is the local density, $m$ the mass coordinate, and $\langle\rho\rangle_r$ the mean density of a sphere of radius $r$.

The terms $P$ and $Q$ appearing in the system of equations \citep[Eqs.~20 and 21 of][]{Takata2016} are now:
\begin{equation}\label{eq:P}
	P = 2J\left(1 - \frac{\omega^2}{\Sred^2}\right)
\end{equation}
and
\begin{equation}\label{eq:Q}
	Q = J\left(1 - \frac{\Nred^2}{\omega^2}\right)\mathrm{.}
\end{equation}

The local radial wavenumber in the asymptotic approximation can be written as $k_r^2=-PQ/r^2$. The boundaries of the evanescent zone, $r_1$ and $r_2$, are defined by the zeros of $P(r)$ and $Q(r)$ respectively (note that $r_1$ and $r_2$ are switched in \citealt{Pincon2020} compared to \citealt{Takata2016} and the Takata order is used in this work):
\begin{equation}\label{eq:r1}
	P(r=r_1) = 0 \Leftrightarrow \Sred(r_1) = \omega
\end{equation}{and}
\begin{equation}\label{eq:r2}
	Q(r=r_2) = 0 \Leftrightarrow \Nred(r_2) = \omega.
\end{equation}
Due to $J$, the new turning points could differ from those in the Cowling approximation, slightly modifying the location of the intermediate evanescent zone. Figure~\ref{fig:propdiagramexplanation} shows a propagation diagram of a 1\msol star in the RGB. It shows how \Nred behaves as $S_\ell$ and how \Sred behaves as $N$ in the core of the star and behave like their non-reduced counterparts in the outer regions.

The resonance condition for dipole mixed modes obtained in \citet{Takata2016} for the case of very narrow evanescent zones, has the same functional shape as that from \citet{Shibahashi1979} in the framework of the Cowling approximation with a large evanescent zone. The phases in the p- and g-mode cavities ($\Theta_p,\,\Theta_g$) may differ from those in previous works \citep[see][]{Takata2016,Pincon2019}, but more important for the present paper is the expression for $X$, and hence the transmission coefficient $T$. The $X$ of \citet{Takata2016} contains a term that includes the properties of stratification in the evanescent zone via the \Nred and \Sred gradients, in addition to the integral of the radial wavenumber in the evanescent zone. Thus, for red giants, this formulation allows us to link the transmission rate to the properties of the stellar structure in the region above the hydrogen-burning shell.

\citet{Pincon2020} studied the ability to extract information about properties of the evanescent zone (radial extent and density stratification) from the coupling coefficient $q$ observed in subgiants and RGB stars. To follow an analytical approach, they adopted simplified structure models in which \Nred and \Sred follow power-laws. They follow the same power-law when the evanescent zone has a radiative stratification ($\dd\ln \Nred/\dd\ln r=\dd\ln\Sred/\dd\ln r=\beta$). When the evanescent zone has a convective stratification, \Sred follows a power law whilst \Nred is 0.

\subsection{Computation of the coupling coefficient $q$ using stellar models: approach and limitations }\label{sec:Method2}
In this work we numerically computed the coupling coefficient, $q$, following the weak and strong approximations in stellar models from the end of the main sequence to the end of central helium-burning. We covered different masses and chemical compositions, since these parameters affect both the radial extent and stratification of the corresponding evanescent zones.

We followed the approach described in \citet{Takata2016}, which is summarized in Appendix D.1 of \citet{Pincon2020}, for the solution of the oscillation problem in the evanescent zone. First, a new spatial variable is defined by
\begin{equation}\label{eq:s}
	s = \ln \left(\frac{r}{r_0}\right),
\end{equation}
where
\begin{equation}\label{eq:r0}
	r_0 = \sqrt{r_1 r_2}.
\end{equation}
In this new reference system the centre of the evanescent zone $r_0$ becomes $s=0$, and the boundaries of the evanescent zone $r_1$ and $r_2$ in terms of $s$ then become $s=s_0$ and $s=-s_0$ respectively, where
\begin{equation}\label{eq:s0}
	s_0 = (\ln r_1 -\ln r_2)/2.
\end{equation}
Both $s = 0$ and $r_0$ refer to the same location in the star in different coordinate systems, and $s_0$ can be negative.

Using Equation~(61) of \citet{Takata2016}, the expression for $X$ in the coupling coefficient is:
\begin{equation}\label{eq:X}
	X=\frac{1}{\pi}\int_{-|s_0|}^{|s_0|} \kappa(s)\sqrt{{s_0}^2-s^2}\, \dd s + \frac{\mathcal{G}^2_{s=0}}{2\kappa_{s=0}} = \frac{I}{\pi}+\frac{\mathcal{G}^2_{s=0}}{2\kappa_{s=0}}\mathrm{.}
\end{equation}

\noindent The first term is equivalent to that in Eq.~\eqref{eq:qw}, which is the integral of the wavenumber in the evanescent zone, with
\begin{equation}\label{eq:kappa}
	\kappa = \sqrt{\frac{PQ}{{s_0}^2 - s^2}},
\end{equation}
and reduces to
\begin{equation}
    I = \int_{-|s_0|}^{|s_0|} \sqrt{PQ} \dd s.
\end{equation}
The second term is the stratification term, with $\mathcal{G}_{s=0}$ related to the gradients of \Nred and \Sred at the centre of the evanescent zone:
\begin{equation}\label{eq:G}
	\mathcal{G} = \frac{1}{4}\frac{\dd}{\dd s}\left[\ln\left(\frac{P}{Q}\frac{s + s_0}{s_0 - s}\right)\right] - \frac{1}{2}\left(\Nu - \A - J\right)\mathrm{,}
\end{equation}
where
\begin{equation}\label{eq:Nu}
	\mathcal{V} = \frac{2J}{\Sred^2}\frac{g}{r},
\end{equation}
and
\begin{equation}\label{eq:A}
	\mathcal{A} = J \frac{\Nred^2 r }{g}.
\end{equation}
It is possible to split the gradient part of $\mathcal{G}$ into three parts:
\begin{equation} \label{eq:Gsplit}
 	\frac{\dd}{\dd s}\left[\ln\left(\frac{P}{Q}\frac{s + s_0}{s_0 - s}\right)\right] = \frac{\dd \ln P}{\dd s} - \frac{\dd \ln Q}{\dd s} + \frac{\dd}{\dd s} \ln\left(\frac{s + s_0}{s_0 - s}\right),
\end{equation}
which will become convenient later.

Hence, using the detailed structure of stellar models over the course of their evolution, we computed the coupling coefficient at frequencies close to the central value of their expected oscillation domain ($\omega=2\pi\nu_{\rm max}$). The coupling coefficient $q$, and its dependence on $\omega$, if any, determines the properties of the dipole oscillation spectra for different stellar parameters. The coupling coefficient $q$ is defined as:
\begin{equation}\label{eq:q}
	q = \frac{1 - \sqrt{1 - e^{-2\pi X}}}{1 + \sqrt{1 - e^{-2\pi X}}}.
\end{equation}
The expression for $X$ (and hence for $q$) in Eq.~\eqref{eq:X} of Takata's asymptotic solution for the propagation of dipole-mixed modes in a intermediate evanescent zone, has been obtained assuming a series of hypotheses, outside of which its validity cannot be guaranteed. These assumptions are: the extent of the evanescent zone is very small relative to the local wavelength; the structure of the star does not vary rapidly in the EZ; the functions $P$ and $Q$ have no zeros inside the EZ; and the formalism applies only to the case of a single EZ.

\begin{figure}
	\centering
	\includegraphics[width=\linewidth]{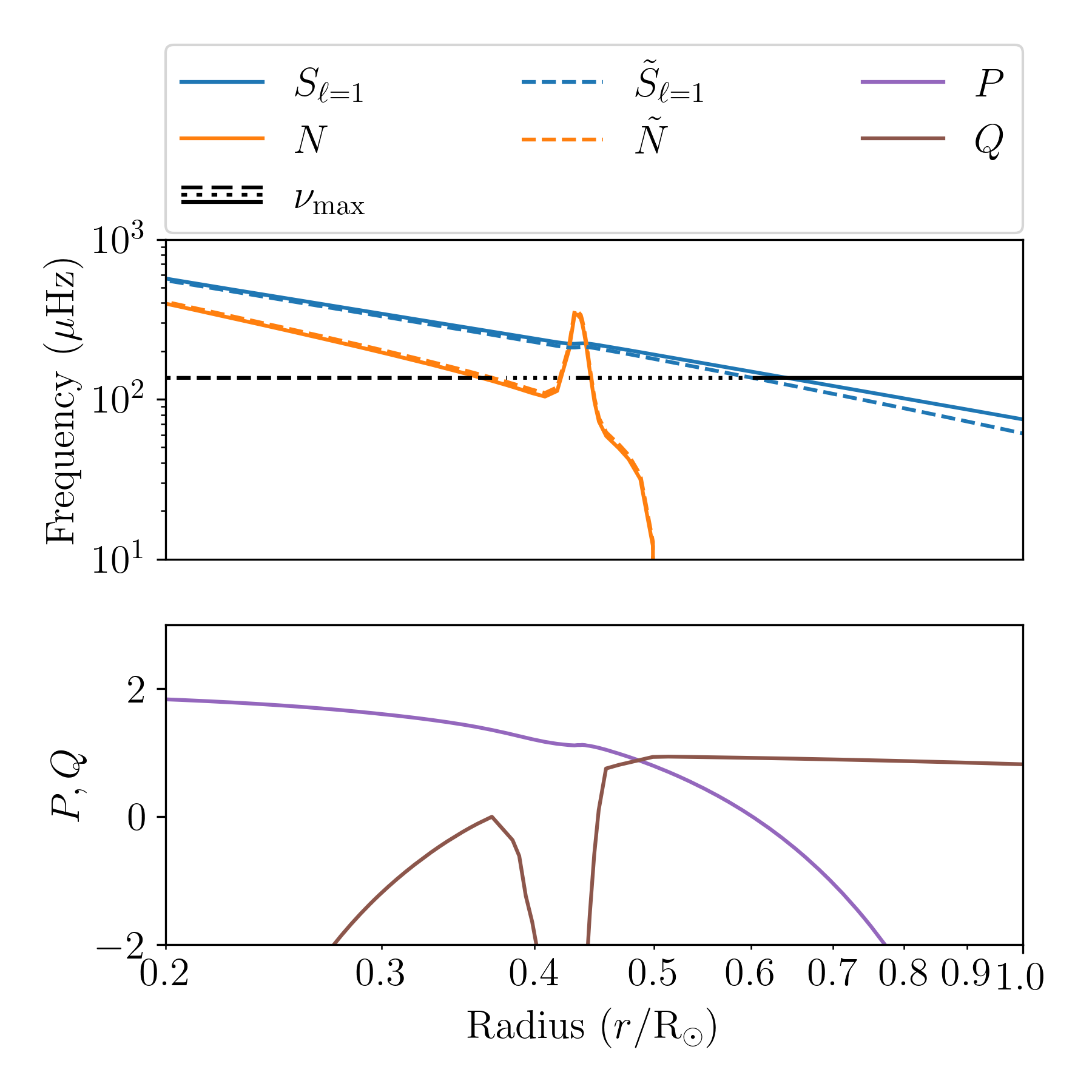}
	\caption{Similar to Fig.~\ref{fig:propdiagramexplanation} but zoomed in around the evanescent zone and showcasing an example where the strong-coupling prescription is not valid. This model is 150~Myr earlier in the RGB than the propagation diagram shown in Fig.~\ref{fig:propdiagramexplanation}.}
	\label{fig:propdiagramexplanationbad}
\end{figure}

As discussed in \citet{Pincon2020}, as the star evolves from the subgiant phase, the characteristics of the evanescent region change, transitioning from being very thin with a radiative thermal stratification to being thick and located at the bottom end of the convective envelope. For these two scenarios, the asymptotic solutions from Takata and Shibahashi apply, respectively, when computing the parameter $q$. Currently, there is no validated expression for the coupling in the transition between these two situations. One could follow the approach in \citet{Pincon2020}, which consists of a combination of type-a and type-b solutions. However, this is not the only issue in this evolutionary phase. In fact, as illustrated by the propagation diagram in Fig.~\ref{fig:propdiagramexplanationbad}, the chemical composition gradient left by the first dredge-up (FDU) appears as a spike in \Nred. That feature can transit through the EZ as \numax decreases due to the envelope's expansion and hence introduce a rapid variation in the physical quantities inside EZ. Whether this gradient appears as a large spike or as a discontinuity depends on the numerical treatment of the evolution of the CE and on the effect of non-standard mixing processes, such as undershooting below the CE. \cite{Jiang2022} has presented a detailed study on the effect of this spike travelling through the EZ on the value of $q$. In Fig.~\ref{fig:propdiagramexplanationbad} we see how the spike left by the FDU, even if large, can introduce a zero in the function $Q$ inside the EZ, violating one crucial assumption in the derivation of Eq.~\eqref{eq:X}. It can also cause another EZ to appear, violating another assumption underlying Takata's formalism.

Therefore, we decided to perform the numerical estimation of $q$ for models for which either $q_w$ or $q_s$ can be computed safely, that is,
we used the strong prescription when the evanescent zone is less than 20\% convective in terms of $s$ and Takata's weak prescription when it is more than 80\% convective in $s$. These thresholds were selected such that we avoid sudden changes in $q$ even though the EZ does not change notably. We defined this fraction of the evanescent zone which is convective ($f_\mathrm{CZ}$) as:
\begin{equation} \label{eq:fCZ}
	f_\mathrm{CZ} = \frac{|s_0| - s_\mathrm{bCZ}}{2 |s_0|},
\end{equation}
where $s_\mathrm{bCZ}$ is the location of the bottom of the convective envelope in terms of $s$ and is defined as:
\begin{equation}
	s_\mathrm{bCZ} = \ln \left(\frac{r_\mathrm{bCZ}}{r_0}\right).
\end{equation}
If $0.2 < f_\mathrm{CZ} < 0.8$ or the spike in \Nred is in the evanescent zone we do not compute the coupling coefficient and therefore we leave it undefined. The final coupling equation becomes:
\begin{equation} \label{eq:qall}
	q =
	\begin{cases}
		\begin{aligned}
			q_s = \frac{1 - \sqrt{1 - e^{-2\pi X}}}{1 + \sqrt{1 - e^{-2\pi X}}}
		\end{aligned}&f_\mathrm{CZ} \leq 0.2,\\
		\begin{aligned}
			q_w = \frac{1}{4} \exp(-2I)
		\end{aligned} &f_\mathrm{CZ} \geq 0.8,\\
		\mathrm{undefined} & \mathrm{otherwise,}\\
	\end{cases}
\end{equation}

\noindent where $X$ and $I$ are defined in Eq.~\eqref{eq:X}. For the weak coupling we used $X$ defined by \citet{Takata2016}, which includes the perturbation to the gravitational potential.
These limits on $f_\mathrm{CZ}$ are less restrictive than $0 < f_\mathrm{CZ} < 1$ which corresponds to the transition phase between type-a and type-b EZs defined in \cite{Pincon2020}.
Nevertheless, the resulting behaviour of Eq.~\eqref{eq:qall} follows observational one and is consistent with \citet{Jiang2022} as this results in the use of the weak-coupling prescription when $q \lesssim 0.12$. We cannot however compare our modelled coupling coefficients directly to \citet{Jiang2022} since we include a diffusive overshooting below the convective envelope in our models, leading to a glitch with finite width.

In this work we computed the coefficient $q$ based on the structure of numerically computed stellar models. Some numerical issues can appear in the computation of terms in Eq.~\ref{eq:X} due, for instance, to the finite number of mesh points defining the EZ, or due to numerical noise in the evaluation of derivatives.  We deal with these issues as described in Appendix \ref{sec:calcq}.

To complement this approach we also considered the possibility of computing $q_s$ assuming that  $\Sred$ and $\Nred$ follow simple power-laws in $r$. We followed two different approaches: either assuming that the two characteristic frequencies follow the same power-law (parallel approximation), as in \citet{Takata2016} and \citet{Pincon2020}, or considering that $\Sred$ and $\Nred$ may be described by different power-laws (non-parallel approximation). The bias in $q$ introduced by these approximations with respect to the complete numerical computation is analysed in Sect.~\ref{sec:testing}. In Table~\ref{table:prescriptions} we present the domains of applicability of our various prescriptions.
We are aware that the work of \citet{Takata2016} referred to red giant stars with two cavities and an intermediate evanescent region, and therefore did not consider stars in the central helium burning phase, which develop a convective core and therefore another evanescent region. However this second EZ is far from the evanescent zone of interest here (that above the H-burning shell) and we assume its effect on the value of $q$ can be neglected as it affects the phase and not the properties of the evanescent zone. %

\begin{table*}
	\caption{Summary of applicability of coupling prescriptions. \label{table:prescriptions}}
	\centering
	\begin{tabular}{llll}
		\hline
		\hline
		Type & Prescription & Evanescent zone properties & Evolutionary phases \\
		\hline
		Weak & \citet{Shibahashi1979} & Type-b, large, convective & RGB, AGB \\
		Weak & \citet{Takata2016} & Type-b, large, convective & RGB, AGB \\
		Strong & \citet{Takata2016} & Type-a, small, radiative & SGB, RC \\
		Strong & Parallel & Type-a, small, radiative & SGB, RC \\
		Strong & Non-parallel & Type-a, small, radiative & SGB, RC \\
		Intermediate & -- & Partially radiative and convective& E-RGB, He-flashes, E-AGB \\
		\hline
	\end{tabular}
    \tablefoot{Type-a and type-b evanescent zones are defined following \citet{Pincon2019}.}
\end{table*}

We have implemented the above equations in the stellar evolution code \MESA v11701 \citep{Paxton2011, Paxton2013, Paxton2015, Paxton2018, Paxton2019} and the implementation details can be found in Appendix \ref{sec:calcq}. A small grid of models was run to explore the behaviour of $q$ in different conditions. We considered five different metallicities ([Fe/H]), and we select five key values for the stellar mass to cover very low mass stars (0.7~\msol), typical stellar mass in the observed populations (1.0~\msol, 1.5~\msol), mass close to the transition mass (2.3~\msol), and typical intermediate mass star (3~\msol).
Table \ref{table:parameters_standard} shows some of our key model parameters. We calibrated our initial metallicity using ${(Z/X)_{\mathrm{surf}}=0.0178}$ from \citep{Serenelli2009}. The initial metallicity, helium abundance, and mixing-length parameter were calibrated on the Sun, so that a 1~\msol model has ${(Z/X)_{\mathrm{surf}}=0.0178}$ \citep{Serenelli2009}, ${\log_{10} (L/L_\odot)=0 \pm 10^{-5}}$, and ${\log_{10} (R/R_\odot)=0 \pm 10^{-5}}$ at the solar age of $4.567 \pm 5 \times 10^{-3}$ Gyr \citep{Connelly2012}. The inlists and \texttt{run\_star\_extras.f} used are available online \footnote{\url{https://doi.org/10.5281/zenodo.13861465}}.
\begin{table}
	\caption{Key parameters used in the simulations.\label{table:parameters_standard}}
	\centering
	\small
	\begin{tabular}{llr}
		\hline \hline
		Parameter & Symbol & Values \\
		\hline
		Mass & $m_i$ &  0.7, 1.0, 1.5, 2.3, 3.0 \\
		Metallicity & \FeH & -1.0, -0.5, 0.0, 0.25, 0.4 \\
		\hline
		solar initial metallicity & Z$_\odot$ & 0.0145 \\
		solar initial helium abundance & Y$_\odot$ & 0.263 \\
		Mixing-length parameter & $\alpha_{\mathrm{MLT}}$ & 1.702\\
		Enrichment law & $\frac{\Delta \mathrm{Y}}{\Delta \mathrm{Z}}$ & 1.007 \\
		Undershooting & $f_\mathrm{US}$ & 0.02\\
		Overshooting & $\alpha_\mathrm{OS,CHeB}$ & 0.5\\
		\hline
	\end{tabular}
\end{table}

\section{Results}\label{sec:Results}

\subsection{Evolutionary state}
\label{sec:evstate}
\begin{figure}
	\centering
	\includegraphics[width=\linewidth]{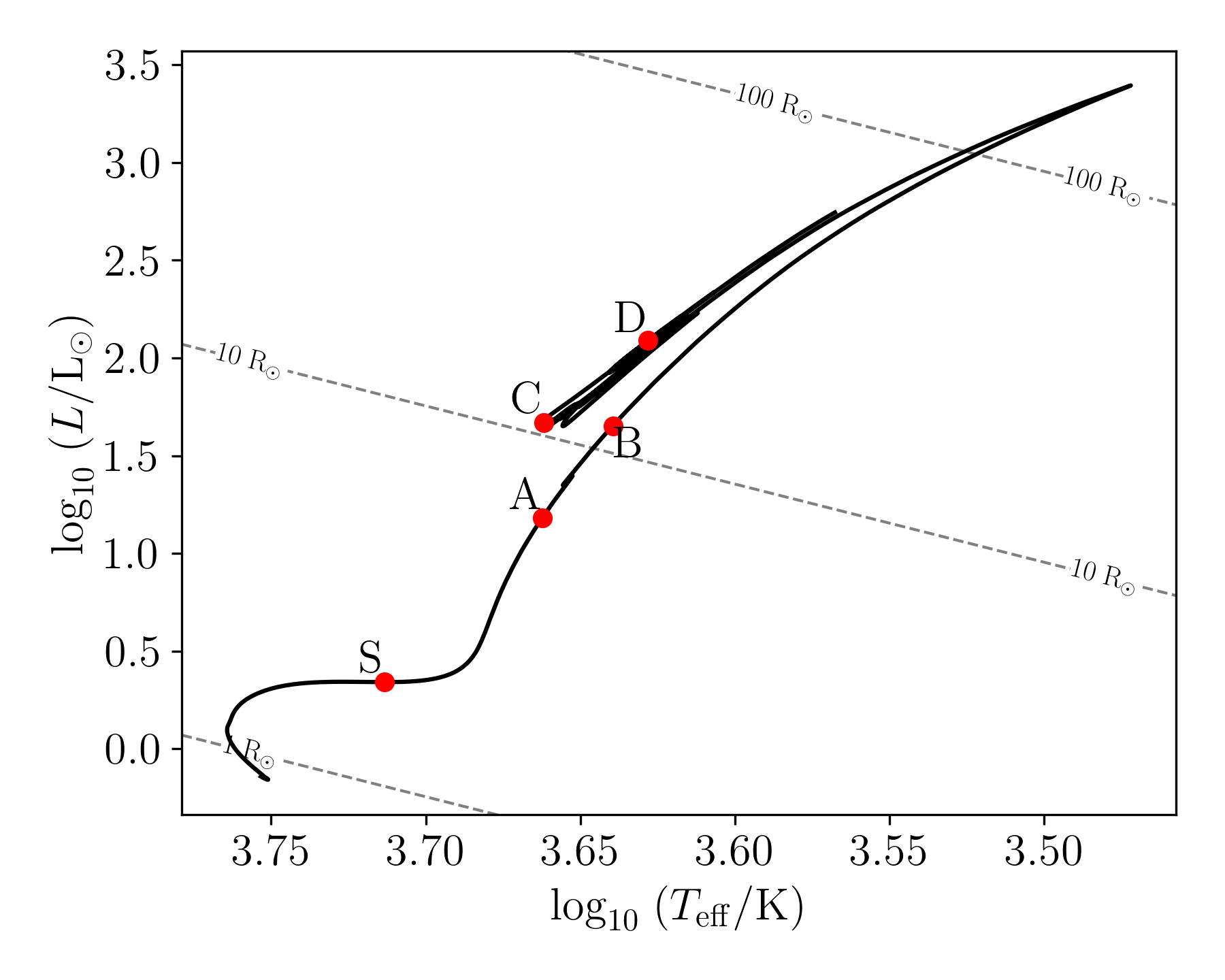}
	\caption{Locations of models A--D and S in the HRD for a 1\msol \FeH = 0 star shown as red dots. Lines of constant radius are shown as the dashed grey lines.}
	\label{fig:hrdproplabels}
\end{figure}

First, we look at how the coupling between the p- and g-mode cavities changes in different evolutionary phases for a modelled star of a given mass and initial chemical composition.
We consider five models (S, A-D) during the evolution of a 1~\msol solar metallicity star:
\begin{enumerate}[(A)]
\setcounter{enumi}{18}
    \item SGB star with $\numax \simeq 950~\uHz$.
\setcounter{enumi}{0}
    \item RGB at the effective temperature of the RC.
    \item RGB at the luminosity of the RC (post-RGB bump).
    \item Middle of core helium-burning (CHeB) in terms of time (at $Y_\mathrm{C} = 0.52$).
    \item Early-asymptotic giant branch (E-AGB) at approximately the effective temperature of model B (degenerate CO core with shell helium-burning).
\end{enumerate}
Models A and B are similar to the YR and ER models in \citet{Pincon2020} respectively.
Figure~\ref{fig:hrdproplabels} shows the location of these models in the HRD as red points.

\begin{figure}
    \centering
    \includegraphics[width=\linewidth]{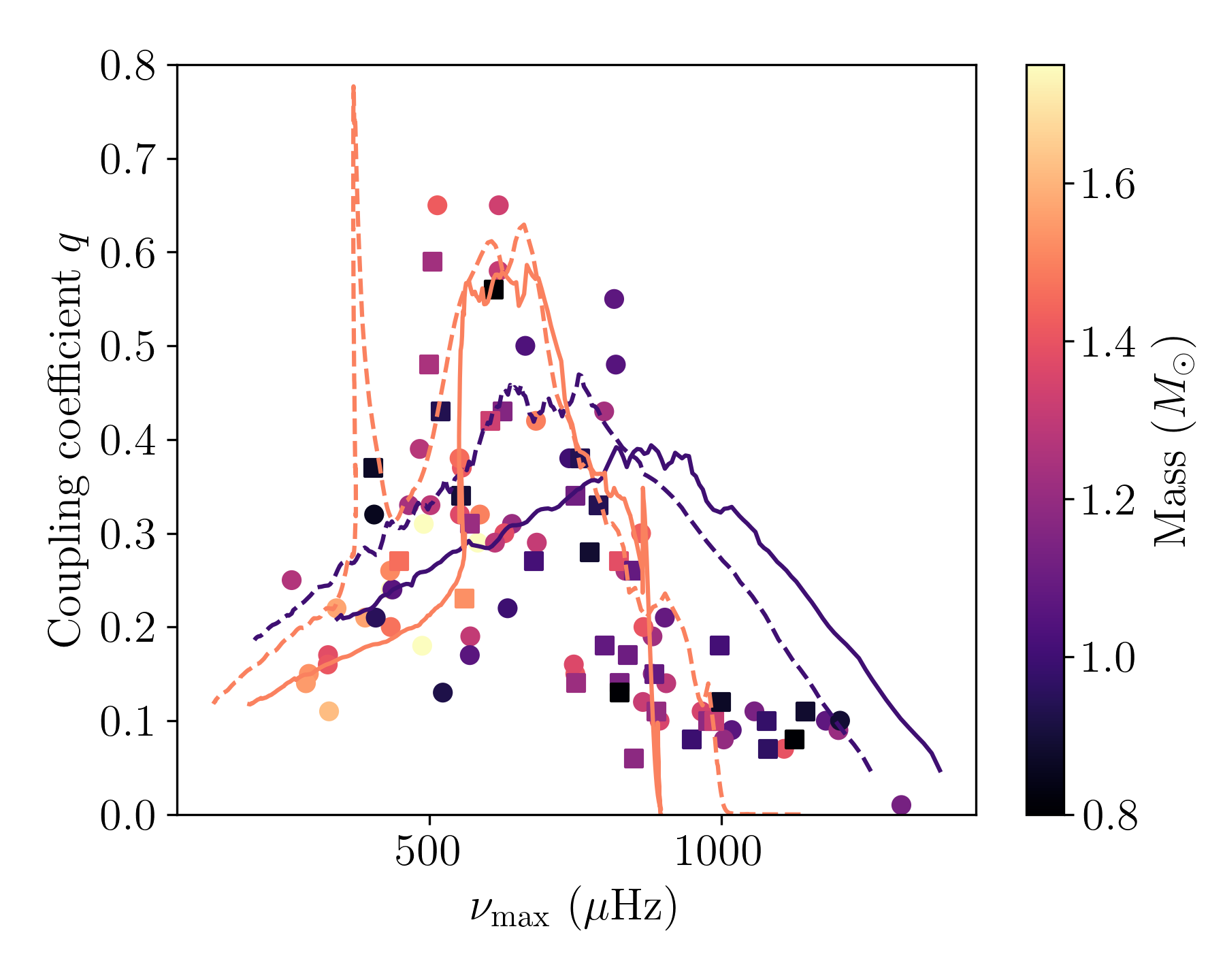}
    \caption{Mixed mode coupling $q$ of SGB stars as a function of \numax. Observed couplings from \citet{Appourchaux2020} are shown as squares, and from Mosser (2024, priv.~comm.) are shown as circles. Mass is calculated using the asteroseismic scaling relations \citep{Kjeldsen1995}, and is indicated using colour. Model tracks with masses 1~\msol and 1.5~\msol with $\FeH=-0.5$ are shown as dashed lines, and with $\FeH=0.0$ as solid lines.  Stars evolve from right to left.}
    \label{fig:qapproximationcompareobs}
\end{figure}

\begin{figure}
    \centering
    \includegraphics[width=\linewidth]{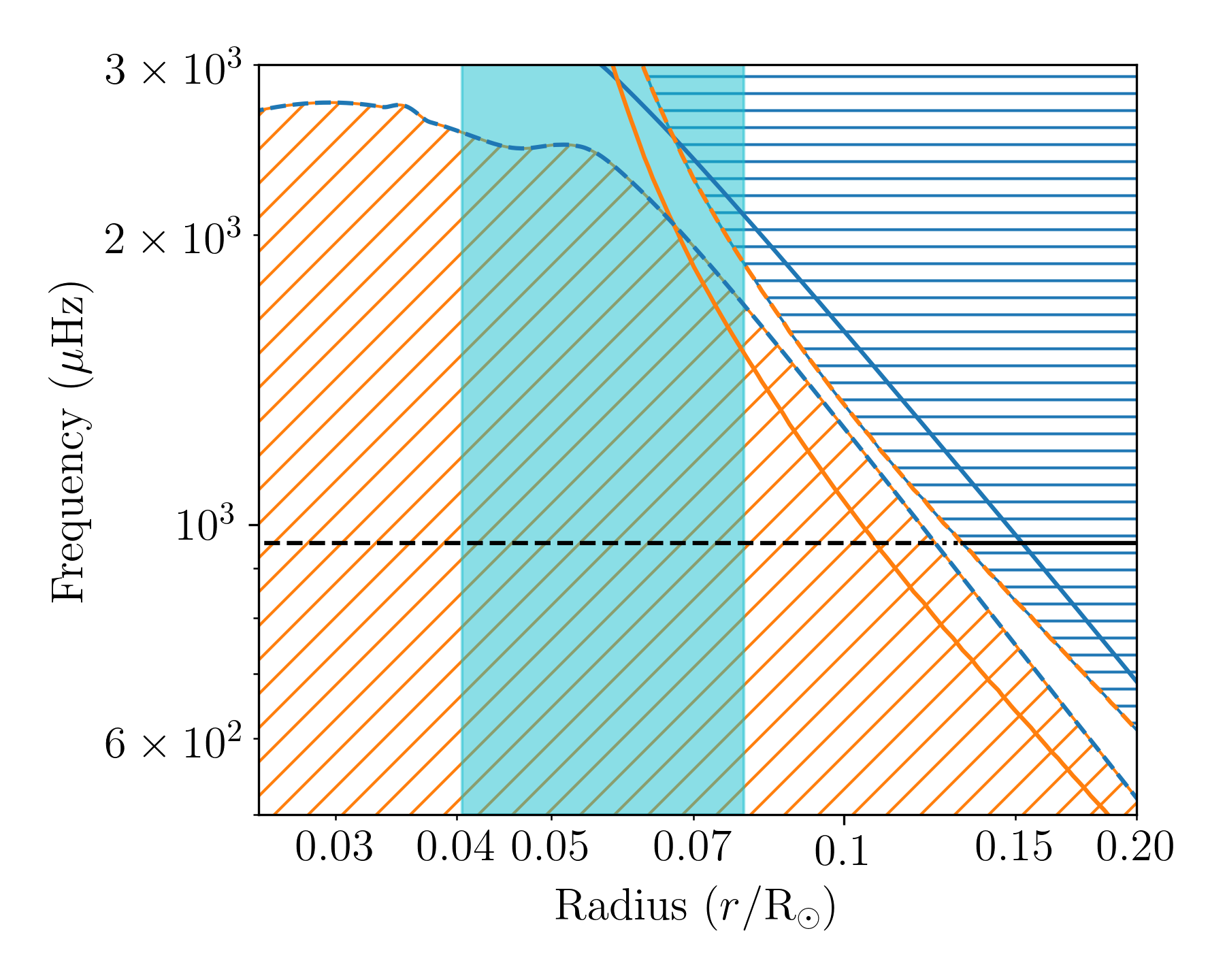}
    \caption{Propagation diagram of model S, zoomed in on the H-burning shell and evanescent zone. The Lamb $(S)$ and reduced Lamb $(\Sred)$ frequencies are shown as blue and green lines, the \brunt $(N)$ and reduced \brunt $(\Nred)$ frequencies are shown as orange and red lines. \numax is shown as a horizontal black line where the dashed part is inside the g-cavity, dotted part in the EZ, and solid part in the p-cavity.}
    \label{fig:propdiagramsgb}
\end{figure}

Although our study focusses primarily on RC stars, we briefly discuss the case of SGB stars, which are also characterised by a narrow evanescent regions and hence strong coupling.
Figure~\ref{fig:qapproximationcompareobs} shows the coupling coefficient as a function of \numax in the SGB. Observed coupling coefficients are from \citet{Appourchaux2020} and unpublished data from \citet[][2024, priv.~comm.]{Mosser2017}. Although we notice that the structure may have more than one evanescent zone (above and below the g-mode cavity) and thus the Takata prescription is of limited validity, the model-predicted coupling is able to qualitatively reproduce the features of the data, including the local maximum of $q$ related to the size of the evanescent region vanishing when $r_1$ and $r_2$ cross (Fig.~\ref{fig:propdiagramsgb}).

\begin{figure}
    \centering
    \includegraphics[width=\linewidth]{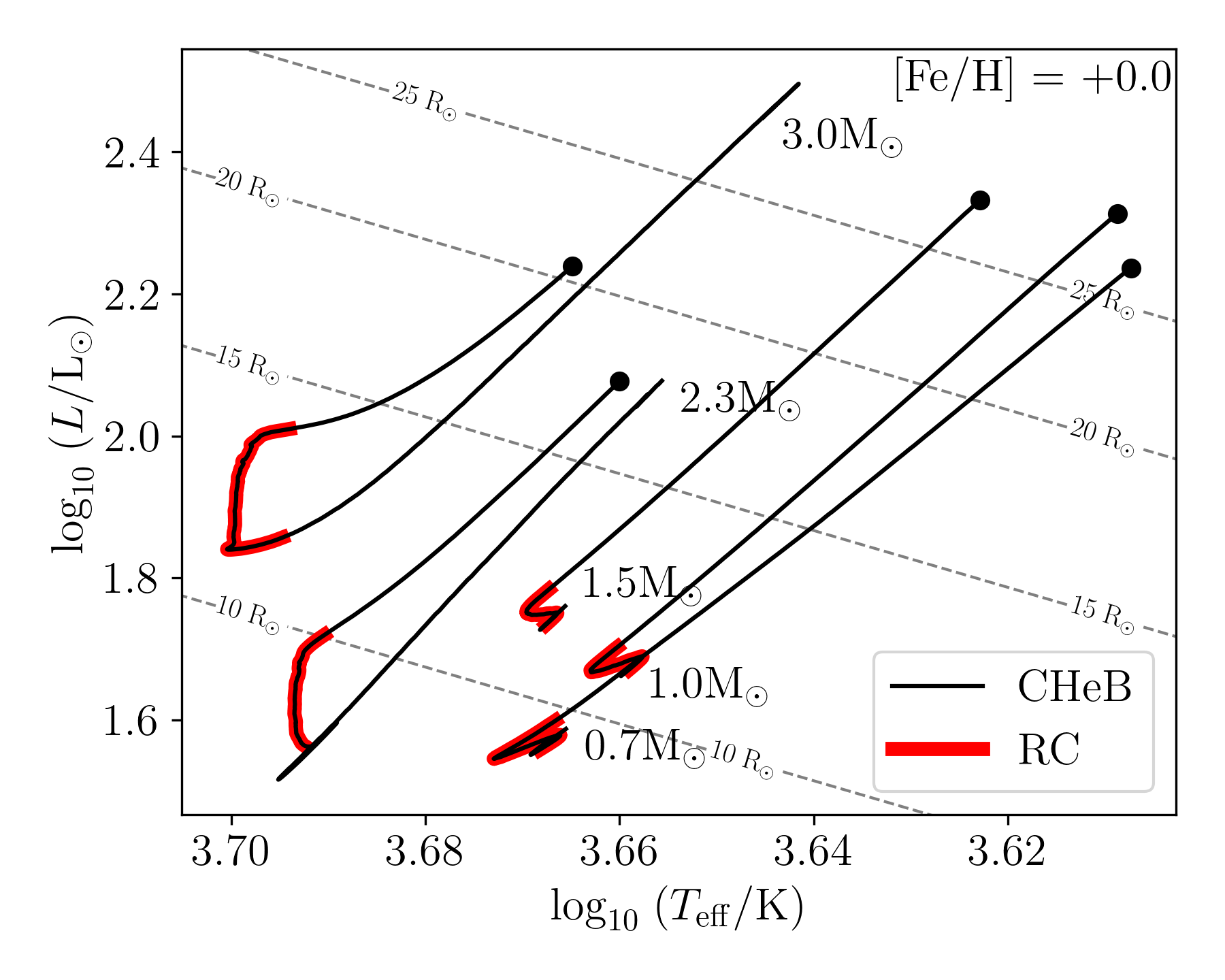}
    \caption{Tracks in the HRD of five 0.7--3 \msol stars with \FeH = 0 during CHeB. The RC phase is highlighted in red. Lines of constant radius are shown as grey dashed lines. Black dots indicate the end of CHeB.}
    \label{fig:chebvsrc}
\end{figure}

We define models as being in the CHeB phase if they satisfy the following three conditions simultaneously. They:
\begin{enumerate}[(1)]
    \item have a convective core,
    \item have a central hydrogen mass fraction less than $10^{-6}$, and
    \item have a central helium mass fraction ($Y_\mathrm{C}$) greater than $10^{-6}$.
\end{enumerate}
The RC phase we define as a sub-phase of CHeB using an additional fourth condition which must be satisfied along with those for the CHeB:
\begin{enumerate}[(1)]
    \setcounter{enumi}{3}
    \item have a central helium mass fraction between $0.95 \times Y_\mathrm{{C,max}} \leq Y_\mathrm{C} \leq 0.1$,
\end{enumerate}
where $Y_\mathrm{{C,max}}$ is the maximum central helium mass fraction reached during evolution before CHeB. This additional constraint is included so that the portion of the evolutionary track used in the Hertzsprung-Russell diagram (HRD) lies mainly in the same area as the observed RC (and the secondary clump in the more massive models).
Figure~\ref{fig:chebvsrc} shows the placement in the HRD of evolutionary tracks between 0.7 and 3\msol with solar metallicity during CHeB and with our RC definition highlighted in red.

\begin{figure}
    \centering
    \includegraphics[width=\linewidth]{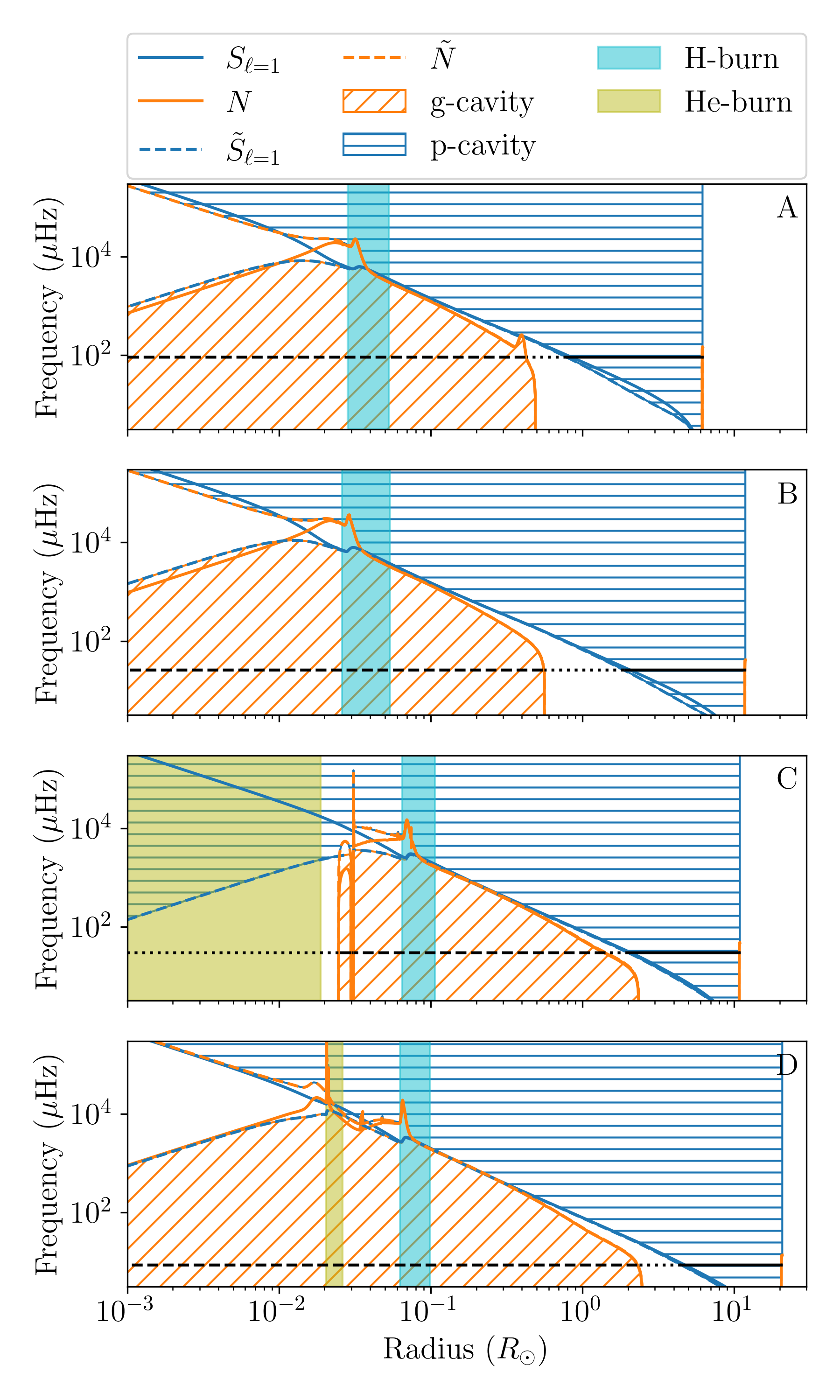}
    \caption{Propagation diagrams of models A--D. The Lamb $(S)$ and reduced Lamb $(\Sred)$ frequencies are shown as blue and green lines, the \brunt $(N)$ and reduced \brunt $(\Nred)$ frequencies are shown as orange and red lines. \numax is shown as a black line where the dashed part is inside the g-cavity, dotted part in the EZ, and solid part in the p-cavity. }
    \label{fig:prop}
\end{figure}

\begin{figure*}
	\centering
	\includegraphics[width=17cm]{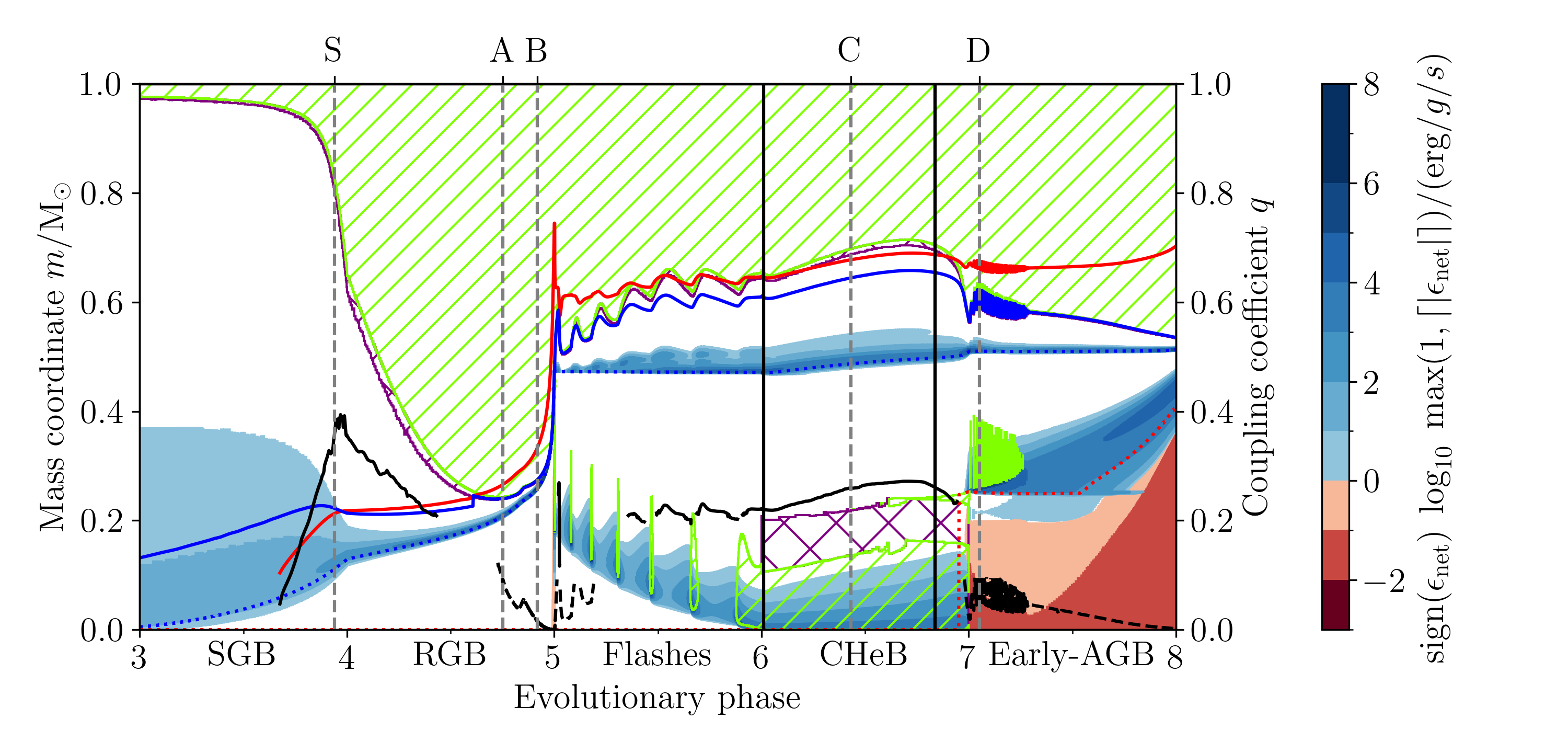}
	\caption{Kippenhahn diagram of a 1~\msol star with solar metallicity. The x-axis shows the stellar age with different scales between evolutionary phases. The SGB is shown from 3--4, the ascent up the RGB from 4--5, the helium flashes from 5--6, CHeB from 6--7, and early-AGB from 7--8. Convection is shown by the green //-hashed regions. Convective overshoot is shown by the purple cross-hashed region. The colour scale shows the net nuclear energy generation rate. The blue dashed line shows the boundary between the helium core and hydrogen envelope. The red dashed line shows the boundary between the helium core and the CO core. Strong coupling is shown as a solid black line and weak coupling as a dashed black line. The boundaries of the evanescent zone, $r_1$ and $r_2$, are shown as the red and blue lines respectively. The two vertical black lines in the CHeB phase show the start and end of the RC. The locations of models A--D and S are shown as vertical dashed grey lines.}
	\label{fig:evokipp0011}
\end{figure*}

Figure~\ref{fig:prop} shows the propagation diagrams of models A--D. Models A, B, and D have type-b
evanescent zones and therefore we used the weak-coupling prescription from \citet{Takata2016} described in Table \ref{table:prescriptions} in Sect.~\ref{sec:Method2}. Model C has a type-a evanescent zone and therefore we used the strong-coupling prescription for this model. The Kippenhahn diagram in Fig.~\ref{fig:evokipp0011} shows the coupling in the various evolutionary phases from the SGB onward with the locations of models A--D shown along the top. During the SGB, the star is in the strong-coupling regime (solid black line in Fig.~\ref{fig:evokipp0011}) as the evanescent zone is fully radiative (type-a evanescent zone, following \citealt{Pincon2020} and Table \ref{table:prescriptions}). As the star evolves along the RGB, it encounters the glitch in \Nred below the convective envelope and also enters the transition regime (the gap between the solid and dashed black lines). The star continues up the RGB and its evanescent zone becomes type-b. Therefore, the coupling is in the weak regime (dashed black lines) and decreases as the star evolves further. When the evanescent zone is of type-b, its bottom edge closely follows the bottom of the convective envelope. As the star continues up the RGB, we encounter models A and B (top two panels in Fig.~\ref{fig:prop}), which are structurally similar. However, model A has not yet gone through the RGB-bump (RGBb) whilst model B has. In the coupling, the RGBb is also visible as a temporary increase in the coupling coefficient (between A and B in Fig.~\ref{fig:evokipp0011}). The temporary increase in coupling coefficient occurs for the same reason as the RGBb: the hydrogen-burning shell encounters the composition gradient left behind by the first dredge-up. This causes the bottom of the convective envelope to recede, increasing $r_2$ whilst leaving $r_1$ relatively untouched, therefore reducing the size of the evanescent zone. The hydrogen-burning shell is visible as the blue shaded region at around $3 \times 10 ^{-2}$ \rsol in Fig.~\ref{fig:prop} A and B, and at around 0.2\msol in Fig.~\ref{fig:evokipp0011}. Model A has a coupling coefficient of 0.091, whilst the coupling coefficient of model B is around a fifth of model A, at 0.016. This difference is due to the increase in size of the evanescent zone, with $r_1 - r_2 = 0.34~\rsol$ in model A ($\sim 5$\% of stellar radius) and $r_1 - r_2 = 1.32~\rsol$ in model B (11\% of $R$). Additionally, the star has almost doubled in size expanding from 6.2~\rsol in model A to 11.9~\rsol in model B.

Once CHeB has started, the central density has decreased by almost 2 orders of magnitude compared to the RGB-tip. As a consequence, the hydrogen-burning shell advances to around $6 \times 10 ^{-2}$~\rsol ($0.48~\msol$) and the evanescent zone is of type-a as seen in panel C of Fig.~\ref{fig:prop} and in Fig.~\ref{fig:evokipp0011}. The evanescent zone is just below the convective envelope and stays there until the end of the RC phase. This is in contrast to models A and B where the evanescent zone is almost entirely (model A) or entirely (model B) contained in the convective envelope. Therefore, the density distribution leads to the evanescent zone being smaller than in models A and B, resulting in a stronger coupling of approximately 0.26.

When the star has consumed 95\% of its central helium content, the central regions of the star start to contract and the envelope expands. This contraction occurs because the support from helium-burning decreases. This in turn leads to a deeper convective envelope and, as a consequence, the convective boundary enters the evanescent zone again.

\begin{figure}
	\centering
	\includegraphics[width=\linewidth]{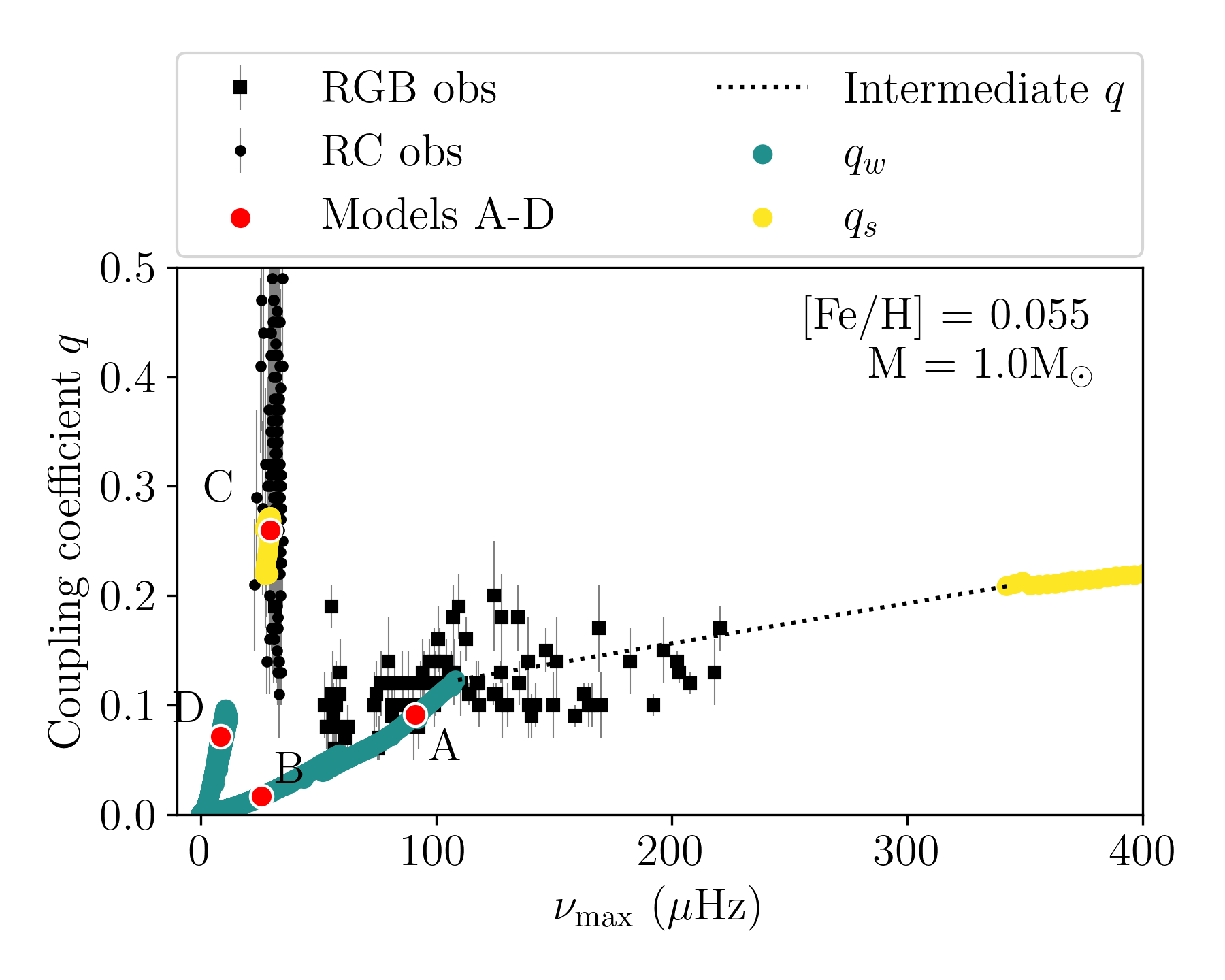}
	\caption{Track of a 1~\msol star with solar metallicity showing the coupling coefficient $q$ as a function of \numax. Yellow points indicate strong coupling, cyan points indicate weak coupling. Observed coupling and \deltanu \citep{Vrard2016,Mosser2017} in stars with masses between 0.9\msol and 1.1\msol and metallicities between -0.070 and 0.180 are shown as black dots and squares, with dots having {$\Delta\Pi_1 \leq 130~\mathrm{s}$ } and squares $\Delta \Pi_1 > 130~\mathrm{s}$. The red points show the locations of models A--D. }
	\label{fig:deltanuqref}
\end{figure}

During the early-AGB (model D) the hydrogen-burning shell is at 0.51~\msol. Helium burning in the star takes place in a shell located at approximately 0.25~\msol, following a contraction of the central regions to an increased density contrast. The coupling in this phase is weak, as for an RGB model of similar $T_\mathrm{eff}$, yet with a coefficient of 0.05--0.10, significantly larger than that on the RGB (see models D and B in Fig.~\ref{fig:deltanuqref}). The evanescent zone during this phase is of type-b and starts, as in model B, just inside the lower boundary of the convective envelope.

Figure~\ref{fig:deltanuqref} shows the coupling coefficient versus \numax for a 1\msol star with solar metallicity.
The four evolutionary snapshots, models A--D, described above are shown as red dots. The RC is clearly visible as the high-$q$ concentration of black dots at $\deltanu \sim 4~\uHz$. The large spread in coupling coefficients inferred from the observed spectra compared to the model values becomes apparent and is likely due to the effect that buoyancy glitches have on deriving $q$ \citep{Vrard2016,Mosser2017,Mosser2018}. The computed weak coupling underestimates the strength of the observed coupling as shown in Fig.~\ref{fig:deltanuqref}. However, since small values of $q$ are harder to measure from the observed oscillations spectra (Sect. 3.4 in \citealt{Mosser2017}) part of this apparent discrepancy, and the lack of observed $q$ below around 0.05, may be ascribed to an observational bias. Additionally, the absence of modelled coupling between $ 10 \lesssim \deltanu/\uHz \lesssim 24$ is visible due to the limitations in the coupling prescriptions described in \ref{sec:Method2}, namely the star being in the transition regime and splitting of the evanescent zone due to the spike in \Nred.

\subsection{Verification of the strong, parallel, and non-parallel coupling prescriptions} \label{sec:testing}
In this section we explore the effect of using additional approximations when numerically calculating the coupling coefficient compared to the Takata prescription.
As illustrated in \citet{Pincon2019} and described in Appendix 3 of \citet{Takata2016}, the strong-coupling prescription can also be approximated by assuming \Sred and \Nred follow the same power-law. This simplifies the calculation of $\mathcal{G}$ somewhat. In this case Eq.~\eqref{eq:G} can be replaced by the following equation:
\begin{equation}\label{eq:Dapprox}
	\mathcal{G} = -{\beta_{\Sred}} \left(\frac{\alpha}{1 - \alpha} + \frac{1}{\ln \alpha}\right) - \frac{1}{2}\left(\Nu - \A - J\right),
\end{equation}
where
\begin{equation}\label{eq:beta}
	{\beta_{\Sred}} \equiv - \left(\frac{\dd \ln \Sred}{\dd \ln r}\right)_{r_0},
\end{equation}
and
\begin{equation}\label{eq:alpha}
	\alpha = \left(\frac{r_2}{r_1}\right)^{\beta_{\Sred}}.
\end{equation}
This method of calculating the gradient term is referred to as the parallel approximation in this work.

As an additional test, we removed the assumption from the parallel approximation that \Nred and \Sred follow the same power-law, but still follow power-laws, which results in the following equation for $\mathcal{G}$.
\begin{equation}\label{eq:GnoP}
	\mathcal{G} = - {\beta_{\Sred}} \left( \frac{\alpha^{\gamma} \gamma - \alpha
		\left(\alpha^{\gamma} \gamma + \alpha^{\gamma} - 1\right)}{2
		\left(\alpha - 1\right) \left(\alpha^{\gamma} - 1\right)} +
	\frac{1}{\ln{\alpha}}\right) -
	\frac{1}{2}\left(\Nu - \A - J\right),
\end{equation}
where
\begin{equation}
	\gamma = \frac{\beta_{\Nred}}{\beta_{\Sred}}{\rm ,}
\end{equation}
and
\begin{equation}\label{eq:betan}
	{\beta_{\Nred}} \equiv - \left(\frac{\dd \ln \Nred}{\dd \ln r}\right)_{r_0}.
\end{equation}
Equation~\eqref{eq:GnoP} reduces to Eq.~\eqref{eq:Dapprox} when \Nred and \Sred are parallel (i.e.~$\gamma=1$). We refer to this as the non-parallel approximation.

\begin{figure}
	\centering
	\includegraphics[width=\linewidth]{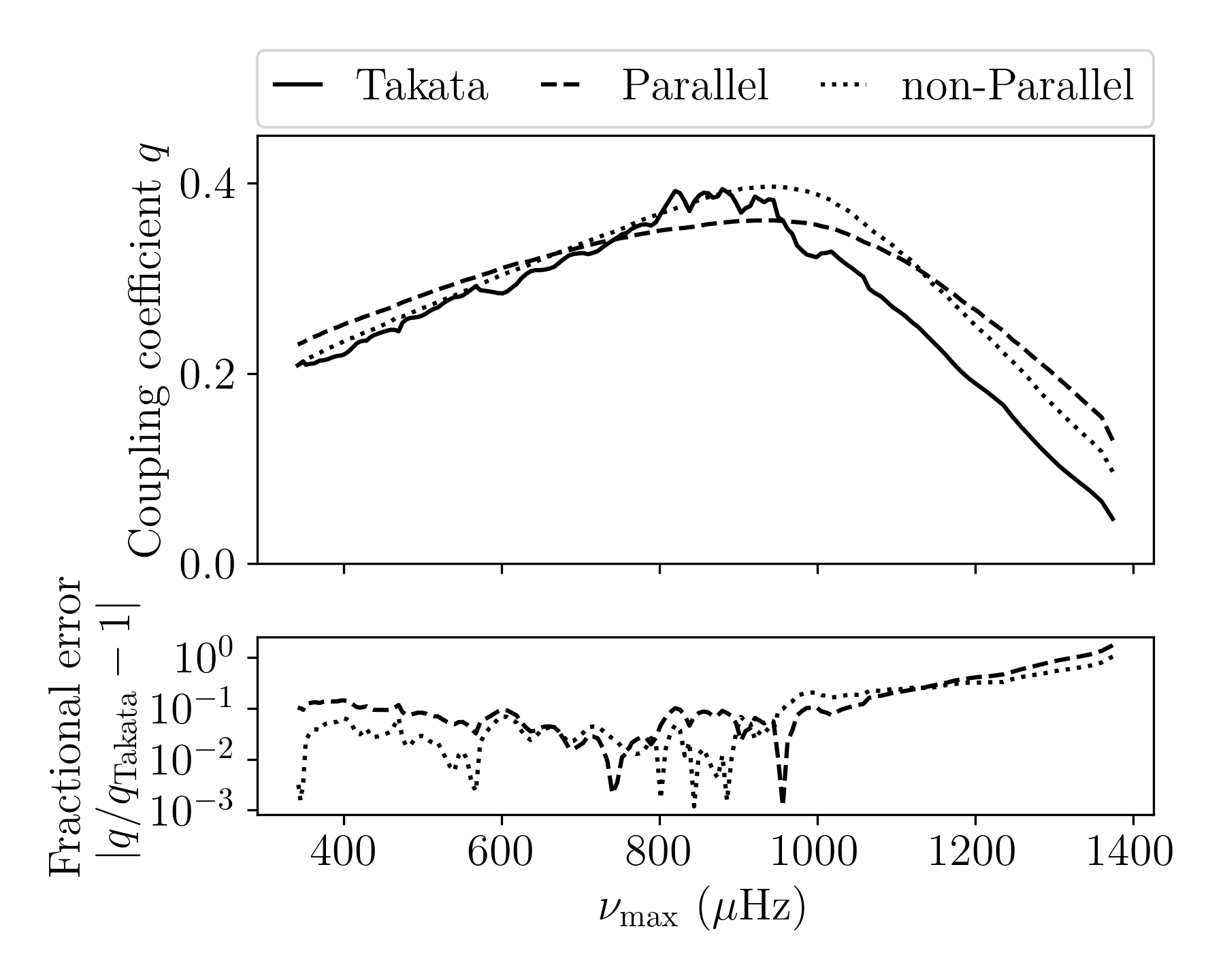}
	\caption{Modelled coupling coefficients $q$ during the SGB and early RGB (top) and fractional error compared to the Takata prescription (bottom) of a 1~\msol solar metallicity star as a function of \numax. The Takata prescription is shown as a solid black line, the parallel approximation as a dashed black line, and the non-parallel approximation as a dotted black line.}
	\label{fig:qapproximationcompare0}
\end{figure}

We quantitatively compare the coupling computed following these different assumptions early on in the RGB and SGB, as well as during the RC.
Figure~\ref{fig:qapproximationcompare0} shows the modelled coupling coefficient computed using the Takata prescription, parallel approximation, and non-parallel approximation as a function of \numax during the SGB and early RGB of a 1~\msol solar metallicity star in the top panel, and the fractional error in the bottom panel. We stop at $\numax=300~\uHz$ as below this frequency the Takata prescription encounters the limitations described in Sect.~\ref{sec:Method2} and is therefore not valid. During the SGB ($\numax \lesssim 1~\mathrm{mHz}$) both the parallel and non-parallel approximations overestimate $q$, with the parallel approximation overestimating by $\sim 0.1$, as seen in the bottom panel of Fig.~\ref{fig:qapproximationcompare0}. On the RGB below $\numax$ of 800~\uHz the non-parallel approximation outperforms the parallel approximation with a typical error of approximately 5\%, whereas the parallel approximation has typical errors of 5--15\%. However, the Takata prescription is more sensitive to numerical noise in $\mathcal{G}$ when the evanescent zone is very thin and its boundaries cross (i.e.~$r_1 \approx r_2$), resulting in a typical uncertainty of around 3\% during the SGB and early RGB.

\begin{figure}
	\centering
	\includegraphics[width=\linewidth]{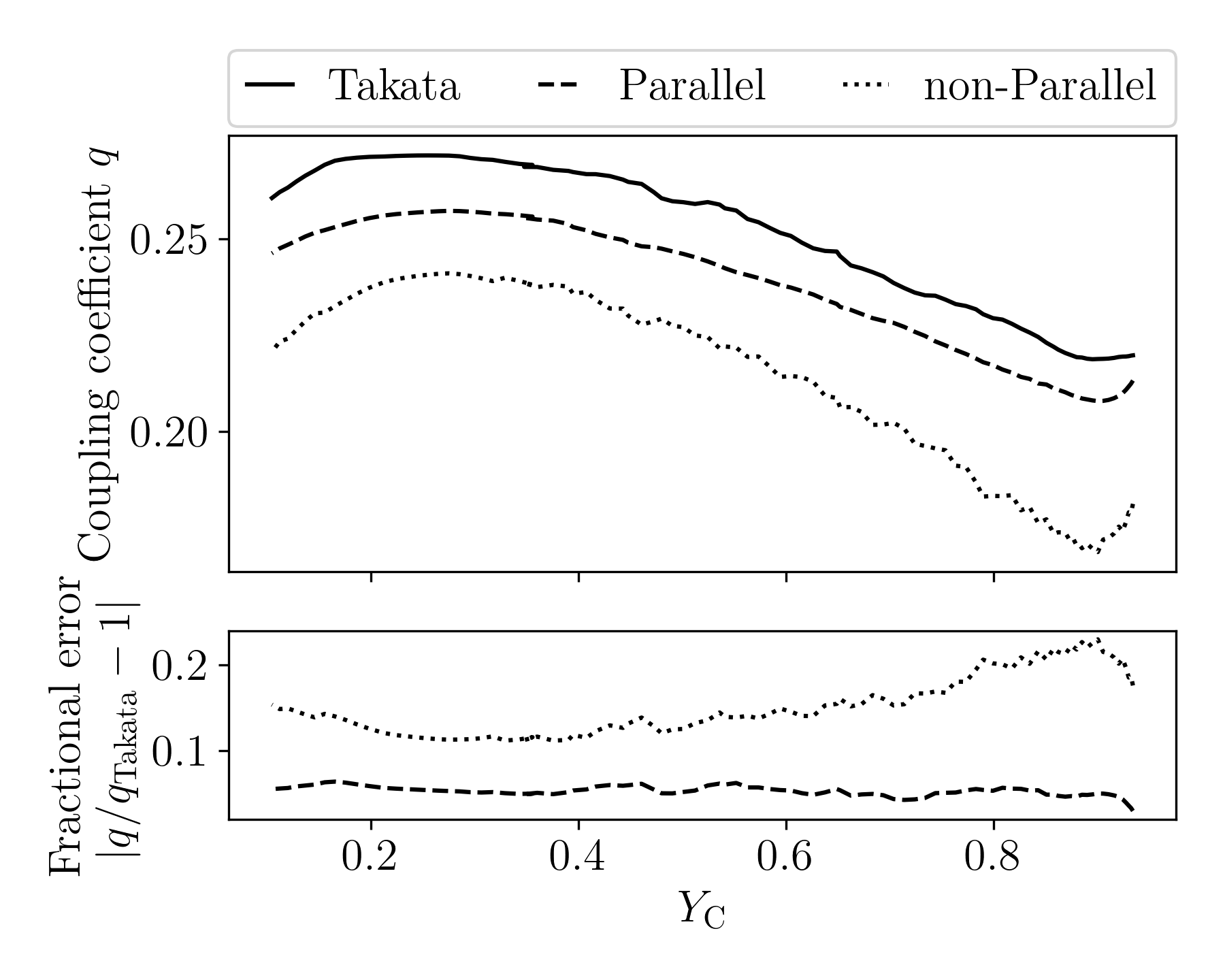}
	\caption{As Fig. \ref{fig:qapproximationcompare0} but during the RC and as a function of $Y_\mathrm{C}$.}
	\label{fig:qapproximationcompare1}
\end{figure}

\begin{figure}
	\centering
	\includegraphics[width=\linewidth]{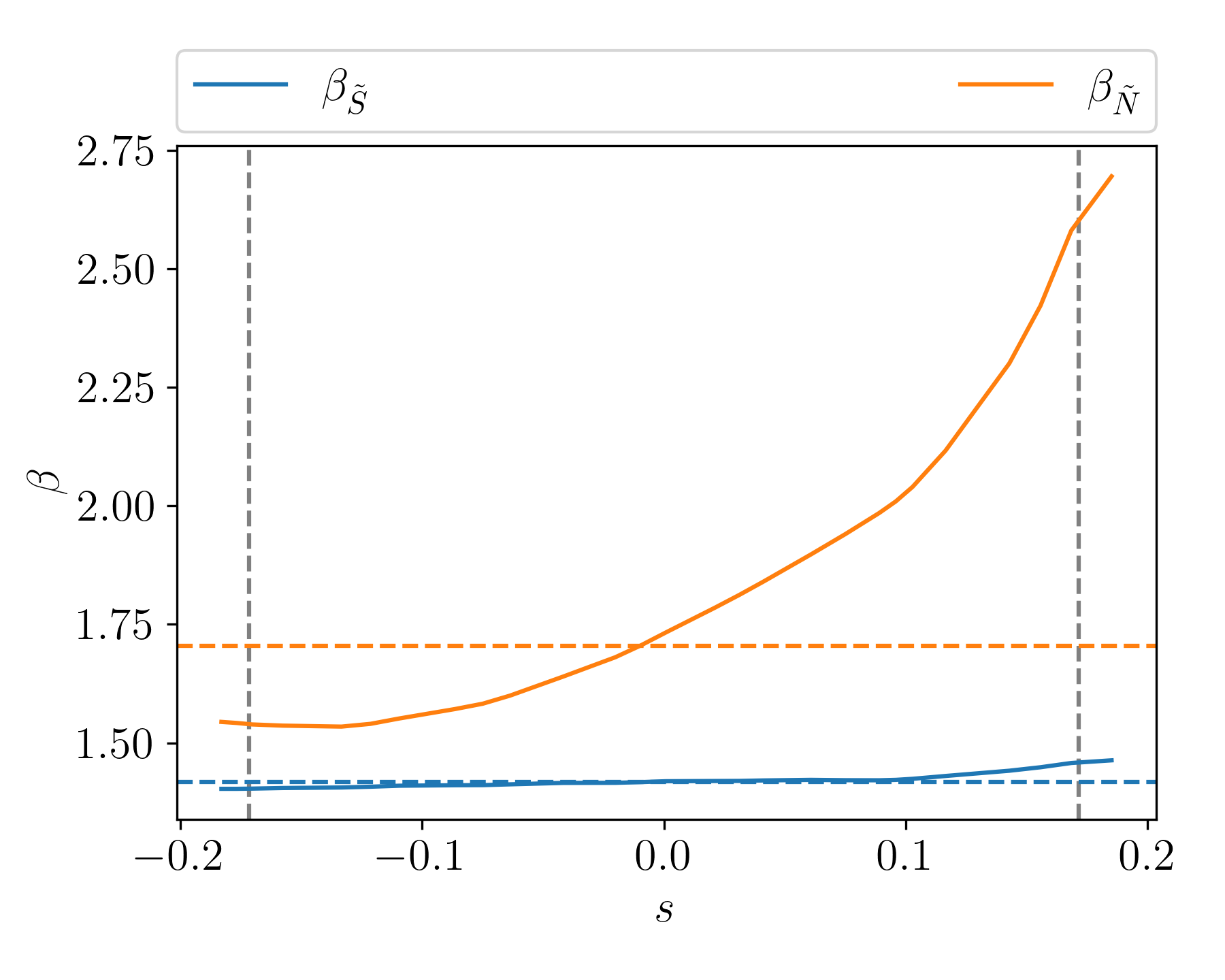}
	\caption{Power-law exponents $\beta_{\Sred}$ and $\beta_{\Nred}$ as a function of $s$ of model C are shown as solid blue and orange lines respectively. The values of $\beta_{\Sred}$ and $\beta_{\Nred}$ used in the power-laws are shown as the horizontal dashed blue and orange lines respectively. The vertical grey dashed lines show the boundaries of the evanescent zone, $\pm s_0$.}
	\label{fig:qapproxbeta1}
\end{figure}

Similar to Fig.~\ref{fig:qapproximationcompare0}, the top panel in Fig.~\ref{fig:qapproximationcompare1} shows the modelled coupling coefficient computed using the Takata prescription, parallel approximation, and non-parallel approximation of a 1~\msol solar metallicity star in the RC. The bottom panel shows the fractional error compared to the Takata prescription. Both the parallel and non-parallel approximations underestimate the coupling coefficient in the RC. However, the parallel approximation performs better as its error is typically around 5\%, whereas the non-parallel approximation underestimates $q$ by 8--14\%.
In this case, the non-parallel approximation performs worse than the parallel approximation as the evanescent zone is near the bottom of the convective envelope. In this region \Nred steepens and it is not well described by a single power-law as $\beta_{\Nred}$ increases, as seen in Fig.~\ref{fig:qapproxbeta1} in model C. This increasing $\beta_{\Nred}$ causes $Q(s=0)$ to be overestimated, which makes the $\dd \ln Q / \dd s$ part of $\mathcal{G}$ to be too small (Eq.~\ref{eq:Gsplit}), increasing $\mathcal{G}$ and therefore decreasing $q$. During the SGB and early-RGB the evanescent zone is far from the bottom of the convective envelope (e.g.~Fig.~\ref{fig:evokipp0011}). This means that \Nred is well approximated by a power-law with a constant exponent and does not encounter this issue.

\subsection{Dependence on oscillation frequency}
\begin{figure}
	\centering
	\includegraphics[width=\linewidth]{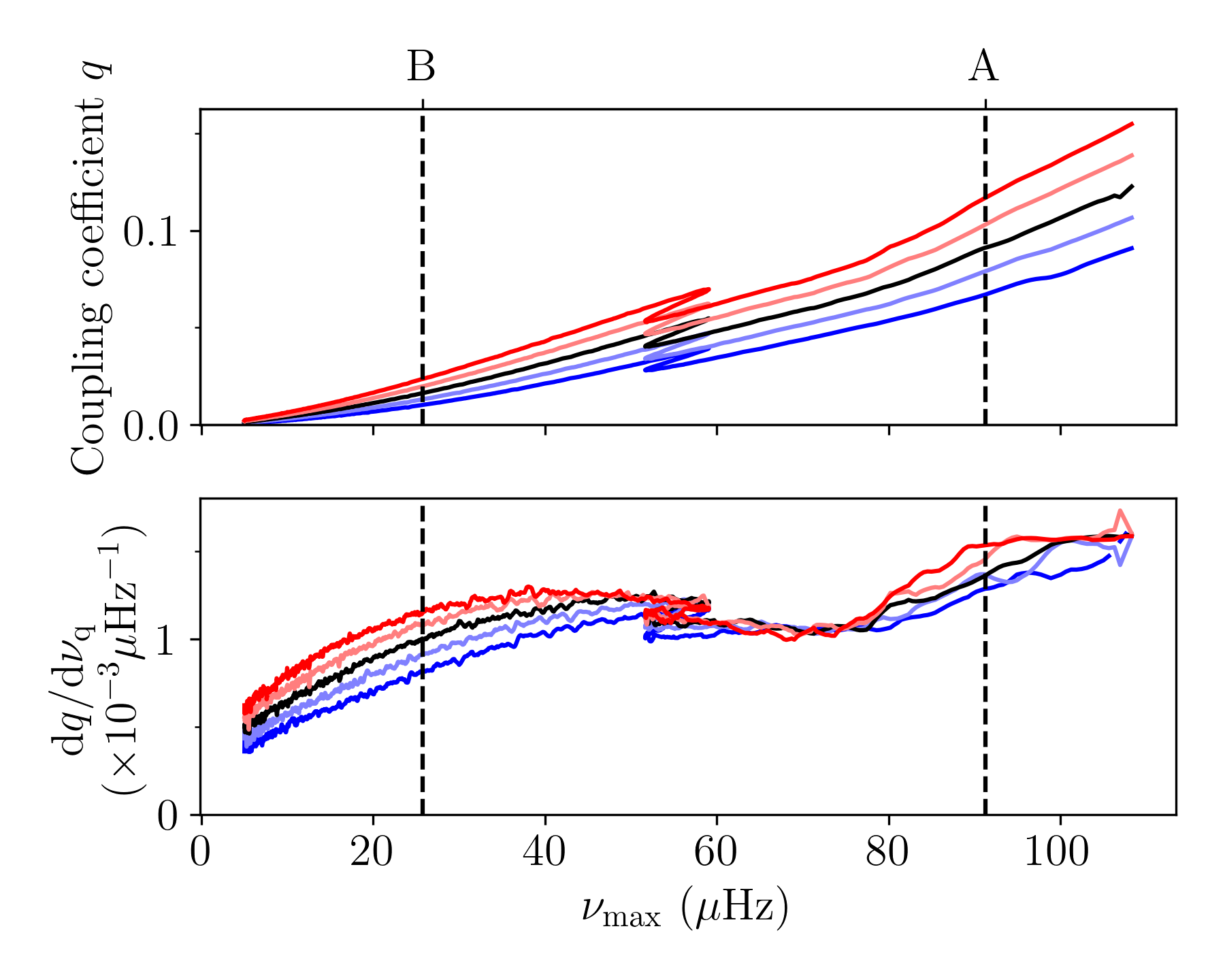}
	\caption{Coupling (top) and coupling gradient (bottom) as a function of \numax in a 1~\msol star with solar metallicity. The black line in each panel shows the values at \numax. The red line in each panel shows the values at $\numax + 2\deltanu$, light red at $\numax + \deltanu$, light blue at $\numax - 2\deltanu$, and blue at $\numax - 2\deltanu$. The vertical dashed lines show the locations of models A and B.}
	\label{fig:rgbdqdnu0011}
\end{figure}

Up to now, we have only studied $q$ at \numax. However, stars showing solar-like oscillations have a frequency spectrum with a Gaussian-like power envelope, characterized by $\sigma = 0.66 \; \numax^{0.88}$ \citep{Mosser2012a}.
For a solar metallicity star with $M=1$~\msol and $R\simeq 10$~\rsol (either in RGB or RC) we expect an oscillation spectrum with $\sigma \simeq 1.5\deltanu$. Therefore, it is necessary to study how $q$ changes as a function of frequency for a given model. We label these frequencies $\nu_\mathrm{q}$ to signify that these frequencies are not related to changes in oscillation frequency due to evolution of the star.
Figure~\ref{fig:rgbdqdnu0011} displays the coupling coefficient, calculated using Eq.~\eqref{eq:qall}, for models along the RGB evolutionary track of a 1~\msol star with solar metallicity. For each model, $q$ was computed at five different frequencies, covering a range of $\pm 2 \deltanu$ around \numax. The first model for which $q$ is shown has $\numax \sim 110~\uHz$ ($\deltanu \sim 10~\uHz$), and is slightly younger than model A.

\citet{Jiang2020} and \citet{Pincon2020} found that, for a given model on the RGB,  the coupling coefficient increases as a function of mode frequency $\nu_\mathrm{q}$. Our models show a similar behaviour, as illustrated in the bottom panel of Fig.~\ref{fig:rgbdqdnu0011} and reported in Table \ref{tab:rgbdqdnu}.
We notice that, since $q$ is sensitive to the detailed behaviour of \Nred with frequency, in stars before the RGB bump $q=q(\nu_\mathrm{q})$ can potentially be used to devise observational tests of the thermal stratification near the boundary of the convective region, resulting, for example, from different envelope-overshooting prescriptions. A further, quantitative exploration of this sensitivity is, however, beyond the scope of this paper.

Table \ref{tab:rgbdqdnu} shows the mean coupling coefficient $\langle q\rangle$ at \numax, mean coupling gradient \dqdnu, mean \numax, and mean \deltanu for models during the temporary drop in $L$ in the RGBb. This ensures that $f_\mathrm{CZ} \simeq 0$ and that using the weak coupling approximation is valid. We define the mean coupling gradient as:
\begin{equation}
	\left\langle\frac{\dd q}{\dd\nu_\mathrm{q}}\right\rangle = \frac{1}{\tau_{\mathrm{RGBb}}}\int_{\mathrm{RGBb}}\frac{q(\numax + \deltanu) - q(\numax - \deltanu)}{2 \deltanu}\dd t,
\end{equation}
where we define the integration domain as the star being in the RGB with an increasing effective temperature. Within one \deltanu frequency interval, there can be a difference of between 10--20\% in $q$. This variation of $q$, due to different $\nu_\mathrm{q}$, should be taken into account when determining $q$ from observations. We also find that before the RGB bump \dqdnu does not vary much as a function of $\nu_\mathrm{q}$ as seen by the constant \dqdnu in the bottom panel of Fig.~\ref{fig:rgbdqdnu0011} in the 60--80~\uHz region. However, after the bump this is no longer the case as all models have steeper \dqdnu at higher $\nu_\mathrm{q}$ with $\langle\dd^2 q/\dd\nu_\mathrm{q}^2 \rangle \simeq 10 - 100 \times 10^{-6} ~\uHz^{-2}$ at around half the bump's \numax. We also observe that \dqdnu decreases with increasing \FeH, which was also seen by \citet{Jiang2020}.
\begin{table}
	\caption{Modelled mean coupling coefficient and mean coupling coefficient gradient at \numax during the RGBb descent.
		\label{tab:rgbdqdnu}}
	\centering
	\small
	
\begin{tabular}{lccccc}\\
\hline \hline \\
Mass & Initial [Fe/H] & $\langle q \rangle$ & $\langle \dd q/\dd\nu_\mathrm{q} \rangle$ & $\langle\numax\rangle$ & $\langle\deltanu\rangle$ \\\
(\msol)&  &  & $(\times 10^{-3} \uHz^{-1})$ & (\uHz)& (\uHz) \\
\hline \\
0.70 & -1.00 & 0.033 &  2.00 $\pm$ 0.02 & 23.5 & 3.55 \\
1.00 & -1.00 & 0.030 &  2.14 $\pm$ 0.09 & 19.6 & 2.83 \\
1.50 & -1.00 & 0.028 &  2.32 $\pm$ 0.09 & 16.3 & 2.23 \\
\hline
0.70 & -0.50 & 0.039 &  1.43 $\pm$ 0.03 & 39.6 & 5.17 \\
1.00 & -0.50 & 0.036 &  1.58 $\pm$ 0.06 & 32.1 & 4.05 \\
1.50 & -0.50 & 0.034 &  1.76 $\pm$ 0.04 & 26.1 & 3.14 \\
\hline
0.70 &  0.00 & 0.053 &  1.03 $\pm$ 0.02 & 71.3 & 7.91 \\
1.00 &  0.00 & 0.047 &  1.17 $\pm$ 0.03 & 55.0 & 5.98 \\
1.50 &  0.00 & 0.042 &  1.31 $\pm$ 0.02 & 42.7 & 4.48 \\
\hline
0.70 &  0.25 & 0.061 &  0.88 $\pm$ 0.01 & 93.3 & 9.60 \\
1.00 &  0.25 & 0.052 &  1.01 $\pm$ 0.03 & 68.9 & 7.02 \\
1.50 &  0.25 & 0.045 &  1.15 $\pm$ 0.03 & 52.2 & 5.17 \\
\hline
0.70 &  0.40 & 0.065 &  0.81 $\pm$ 0.01 & 105.6 & 10.47 \\
1.00 &  0.40 & 0.055 &  0.95 $\pm$ 0.01 & 76.4 & 7.55 \\
1.50 &  0.40 & 0.047 &  1.09 $\pm$ 0.01 & 56.8 & 5.48 \\
\hline
\end{tabular}
 \end{table}

\begin{figure}
	\centering
	\includegraphics[width=\linewidth]{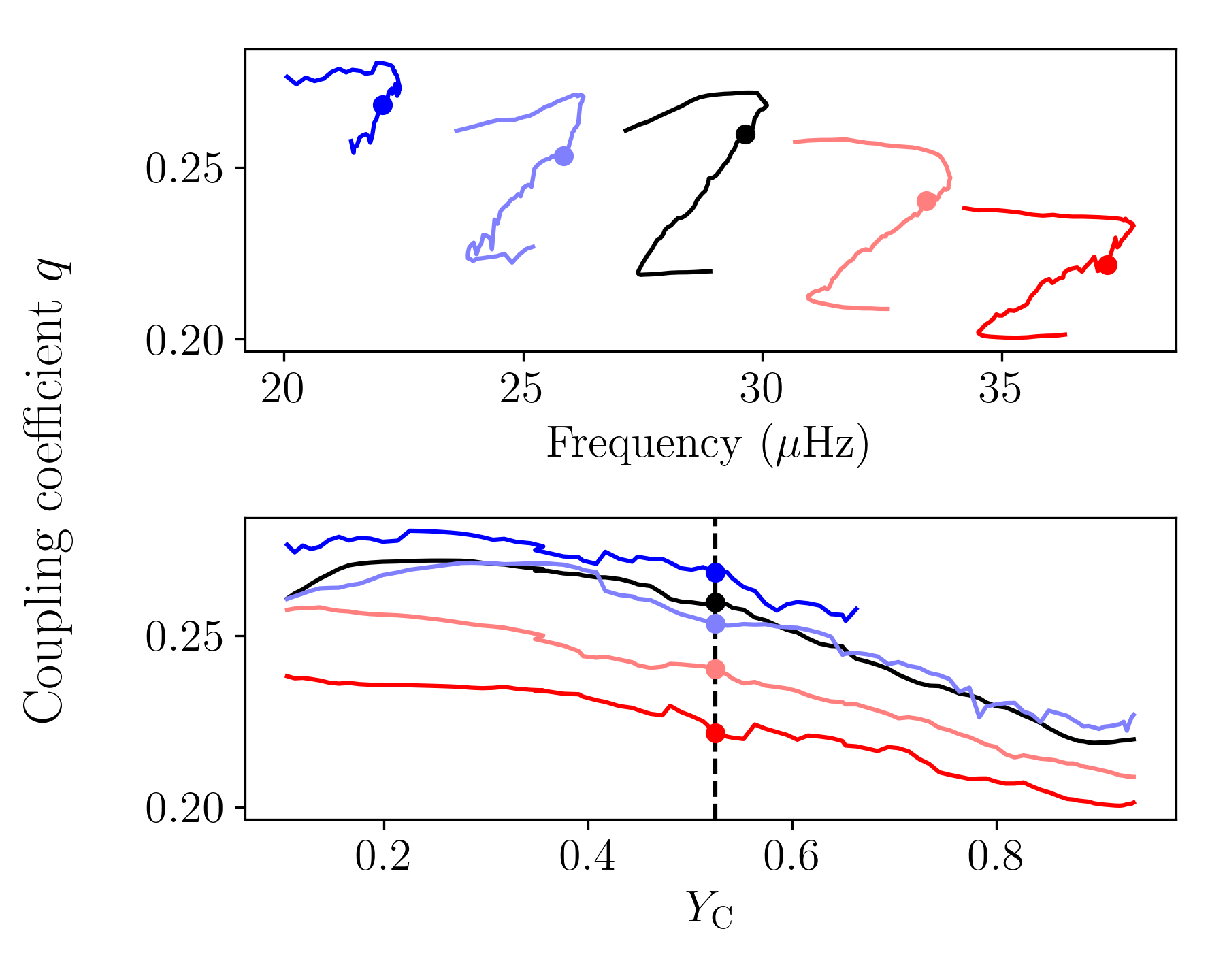}
	\caption{Strong-coupling coefficients as a function of frequency in the RC (top), and coupling coefficient as a function of central helium mass fraction (bottom). The black line corresponds to $\nu_\mathrm{q} = \numax$, the coloured lines correspond to $\nu_\mathrm{q}$ offset from \numax in steps of \deltanu, with dark red being $+2\;\deltanu$, light red $+1\;\deltanu$, light blue $-1\;\deltanu$, and dark blue $-2\;\deltanu$.}
	\label{fig:dqdnu_all}
\end{figure}

\begin{figure}
	\centering
	\includegraphics[width=\linewidth]{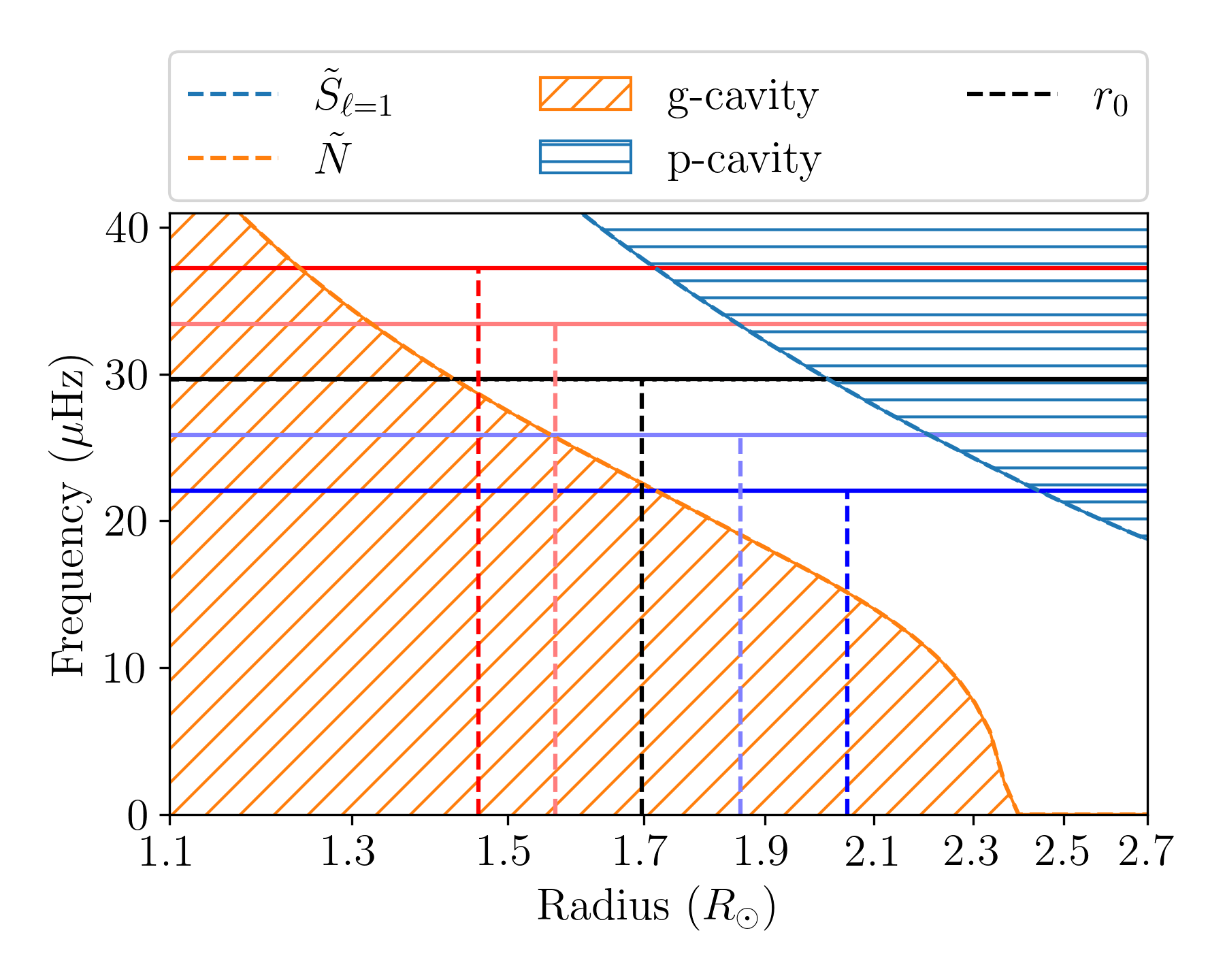}
	\caption{Propagation diagram of a 1\msol solar metallicity star in the RC with $Y_\mathrm{C} \simeq 0.5$. The black horizontal line corresponds to $\nu_\mathrm{q} = \numax$, and the coloured horizontal lines correspond to $\nu_\mathrm{q}$ offset from \numax in steps of \deltanu. The vertical dashed lines correspond to $r_0$ ($s=0$) at the respective frequency.}
	\label{fig:dqdnu_prop}
\end{figure}

Figure~\ref{fig:dqdnu_all} shows the coupling as a function of $\nu_\mathrm{q}$ in models of 1~\msol and solar metallicity  during the central helium-burning phase. Table \ref{tab:rcdqdnu} shows the range of $q$ within $\pm 2$\deltanu of \numax when the central helium abundance is between $0.4 \leq Y_\mathrm{C} \leq 0.6$, as well as the fraction of $X$ which comes from the integral term in  Eq. \eqref{eq:X}.
RC models show a dependence of $q$ on $\nu_\mathrm{q}$, however, due to the simpler structure of type-a evanescent zones (see Fig.~\ref{fig:dqdnu_prop}), the frequency dependence is weaker compared to that on the RGB. In RC models, changing the frequency by one \deltanu typically only results in a change in $q$ of 1--10\% (see Table \ref{tab:rcdqdnu}).
\begin{table}
	\caption{Modelled coupling coefficients in the RC.
		\label{tab:rcdqdnu}}
	\centering
	\small
	
\begin{tabular}{lccccc}\\
\hline \hline \\
Mass & Initial [Fe/H] & $q$-range\tablefootmark{a} & $\langle X_I/X\rangle$ & $\langle\numax\rangle$ & $\langle\deltanu\rangle$ \\
(\msol)&  &  &  & (\uHz)& (\uHz) \\
\hline \\
0.70 & -1.00 & 0.59 -- 0.72 &  0.37 & 39.8 & 5.56 \\
1.00 & -1.00 & 0.40 -- 0.54 &  0.24 & 33.3 & 4.36 \\
1.50 & -1.00 & 0.43 -- 0.48 &  0.18 & 38.3 & 4.39 \\
2.30 & -1.00 & 0.55 -- 0.61 &  0.98 & 43.6 & 4.49 \\
3.00 & -1.00 & 0.44 -- 0.49 &  0.88 & 44.1 & 4.56 \\
\hline
0.70 & -0.50 & 0.53 -- 0.63 &  0.24 & 32.8 & 4.66 \\
1.00 & -0.50 & 0.31 -- 0.40 &  0.51 & 30.5 & 3.98 \\
1.50 & -0.50 & 0.26 -- 0.29 &  0.39 & 38.9 & 4.33 \\
2.30 & -0.50 & 0.32 -- 0.38 &  0.27 & 71.7 & 6.29 \\
3.00 & -0.50 & 0.67 -- 0.74 &  0.84 & 34.5 & 3.46 \\
\hline
0.70 &  0.00 & 0.39 -- 0.42 &  0.61 & 29.9 & 4.20 \\
1.00 &  0.00 & 0.23 -- 0.27 &  0.67 & 29.7 & 3.79 \\
1.50 &  0.00 & 0.18 -- 0.21 &  0.49 & 39.3 & 4.26 \\
2.30 &  0.00 & 0.19 -- 0.21 &  0.37 & 91.2 & 7.34 \\
3.00 &  0.00 & 0.24 -- 0.26 &  0.35 & 63.5 & 5.26 \\
\hline
0.70 &  0.25 & 0.34 -- 0.36 &  0.68 & 29.7 & 4.13 \\
1.00 &  0.25 & 0.20 -- 0.23 &  0.68 & 29.8 & 3.76 \\
1.50 &  0.25 & 0.17 -- 0.18 &  0.53 & 39.6 & 4.22 \\
2.30 &  0.25 & 0.17 -- 0.19 &  0.40 & 81.0 & 6.61 \\
3.00 &  0.25 & 0.19 -- 0.20 &  0.39 & 72.6 & 5.74 \\
\hline
0.70 &  0.40 & 0.32 -- 0.33 &  0.76 & 29.6 & 4.07 \\
1.00 &  0.40 & 0.19 -- 0.21 &  0.68 & 29.5 & 3.70 \\
1.50 &  0.40 & 0.16 -- 0.16 &  0.55 & 39.1 & 4.16 \\
2.30 &  0.40 & 0.16 -- 0.18 &  0.42 & 80.1 & 6.51 \\
3.00 &  0.40 & 0.17 -- 0.18 &  0.41 & 72.9 & 5.71 \\
\hline
\end{tabular}
     \tablefoot{Means and $q$-range are taken over $0.4 \leq Y_\mathrm{C} \leq 0.6$.\\
    \tablefoottext{a}{Range of $q$ within 2 \deltanu of \numax.}}
\end{table}
On the other hand, below around 20~\uHz the oscillation modes enter the intermediate-coupling regime. If we were to evaluate the coupling using the available strong-coupling approximation, we would find a sudden decrease of the coupling with decreasing frequency. By numerically calculating the modes' eigenfunctions and inertiae, however, it becomes apparent that the coupling of these low frequency cases broadly follows the behaviour of the coupling calculated at other frequencies. This suggests that the abrupt change in the apparent behaviour of $q$ is likely related to the limitations of the analytical approximation, which should not be used beyond its domain of applicability.


\subsection{Dependence on mass and metallicity, and comparison with observations} \label{sec:Observations}

To compare our models with observations, we use the \citet{Mosser2017} catalogue, which contains the measured coupling coefficients $q$ of over 5000 red giant stars.
To select stars in the CHeB phase, we use only stars with ${\Delta P > 130 ~\mathrm{s}}$ (see e.g.~\citealt{Mosser2014}).
For spectroscopic parameters and associated uncertainties we use SDSS-DR14 \citep{SDSS-DR14}. Finally, we use the full set of inferred masses from \citet{Miglio2021} including stars with less reliable age estimates. Stars common to these three data sets result in just under 2500 objects.

\begin{figure}
	\centering
	\includegraphics[width=\linewidth]{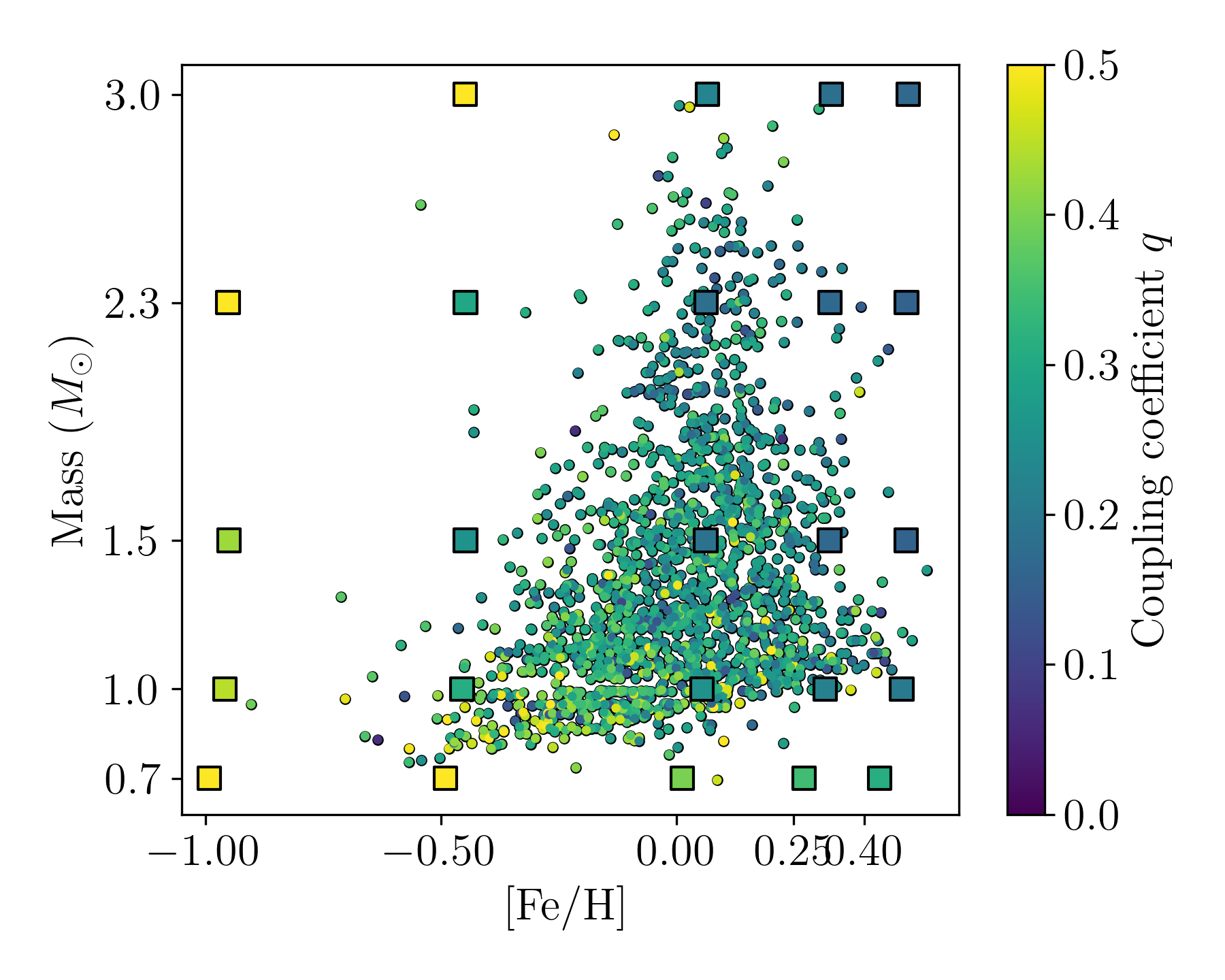}
	\caption{Modelled masses and \FeH are shown as coloured squares. Observed masses and \FeH are shown as dots. The colour scale shows the mean coupling coefficient in the RC ($\langle q \rangle_\mathrm{RC}$) for models and observed coupling coefficient for observations.}
	\label{fig:massfehqrcbinned}
\end{figure}

Figure~\ref{fig:massfehqrcbinned} shows the modelled mean coupling in the red clump as square markers and the observed coupling as coloured dots. In models, for a given mass and metallicity, the mean coupling in the RC is defined as:
\begin{equation}
	\langle q\rangle_\mathrm{RC} = \frac{\int_\mathrm{RC} q \dd t}{\tau_\mathrm{RC}},
\end{equation}
where $\tau_\mathrm{RC}$ is the lifetime in the RC and the integral is taken over the time spent in the RC.
A typical observational uncertainty of $q$ is around 20\%. In the RC, we see that the observed stars are concentrated in the lower-mass, higher-metallicity region. Figure~\ref{fig:massfehqrcbinned} further shows that there are hardly any observed stars in the top left and bottom right regions of the plot. This lack of stars is expected from Galactic chemical evolution, that is, a lack of young (more massive) metal-poor stars and old (less massive) metal-rich stars.

Overall, we notice that the observed data show, as also predicted by the models, an anti-correlation between mass or metallicity and coupling, with stronger coupling coefficients being more prevalent at lower masses and lower metallicities. We explore and discuss the origin of these correlations in more detail in what follows (see in particular Sec. \ref{sec:rhpcontr}). We will limit ourselves to commenting on the main trends, recalling that a thorough comparison between observed and model-predicted coupling should take into account several potential sources of systematic biases associated with the procedure to estimate $q$ from observations (e.g.~the presence of acoustic glitches, buoyancy glitches, rotational splitting, and assuming that $q$ is not a function of $\nu_\mathrm{q}$, see e.g.~\citealt{Mosser2018}).

\begin{figure}
	\centering
	\includegraphics[width=\linewidth]{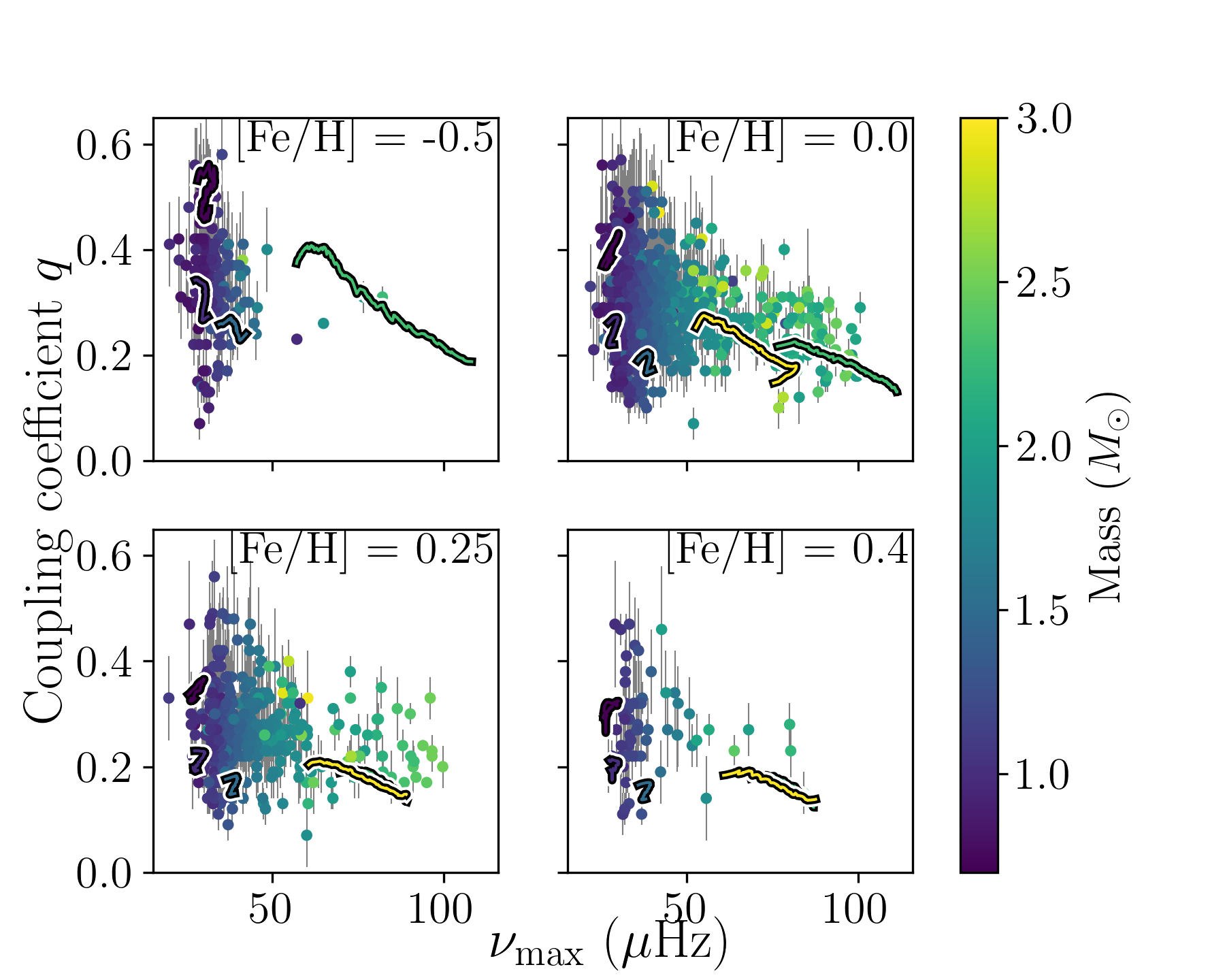}
	\caption{Coupling coefficient ($q$) at \numax versus \numax in the RC. The colour scale shows the stellar mass. The points are observed coupling and \deltanu and the tracks are from modelled stars. From left to right, top to bottom, the model \FeH are -0.5, 0.0, 0.25, and 0.4. Similarly, the ranges for the observed \FeH are -0.75 $\le$ \FeH $\leq$ -0.25, -0.25 $\le$ \FeH $\leq$ 0.125, 0.125 $\le$ \FeH $\leq$ 0.325, 0.325 $\le$ \FeH.}
	\label{fig:deltanuq}
\end{figure}

Figure~\ref{fig:deltanuq} shows coupling coefficient $q$ vs.~\numax for models and observations of RC stars. In general, our models agree for the bulk of observed $q$ with observations being on average 1.5-$\sigma$ away from a model when taking the quadratic mean. The modelled masses cover the range in observed \numax. However, there are some cases where the observed $q$ lies far away from a corresponding model, for example, the group of points with low $q$ at a $\numax \simeq 30~\uHz$, some of which could be stars leaving the RC with a coupling lower than the bulk of RC stars (as also noticed in Sec. \ref{sec:evstate}).

\begin{figure}
	\centering
	\includegraphics[width=\linewidth]{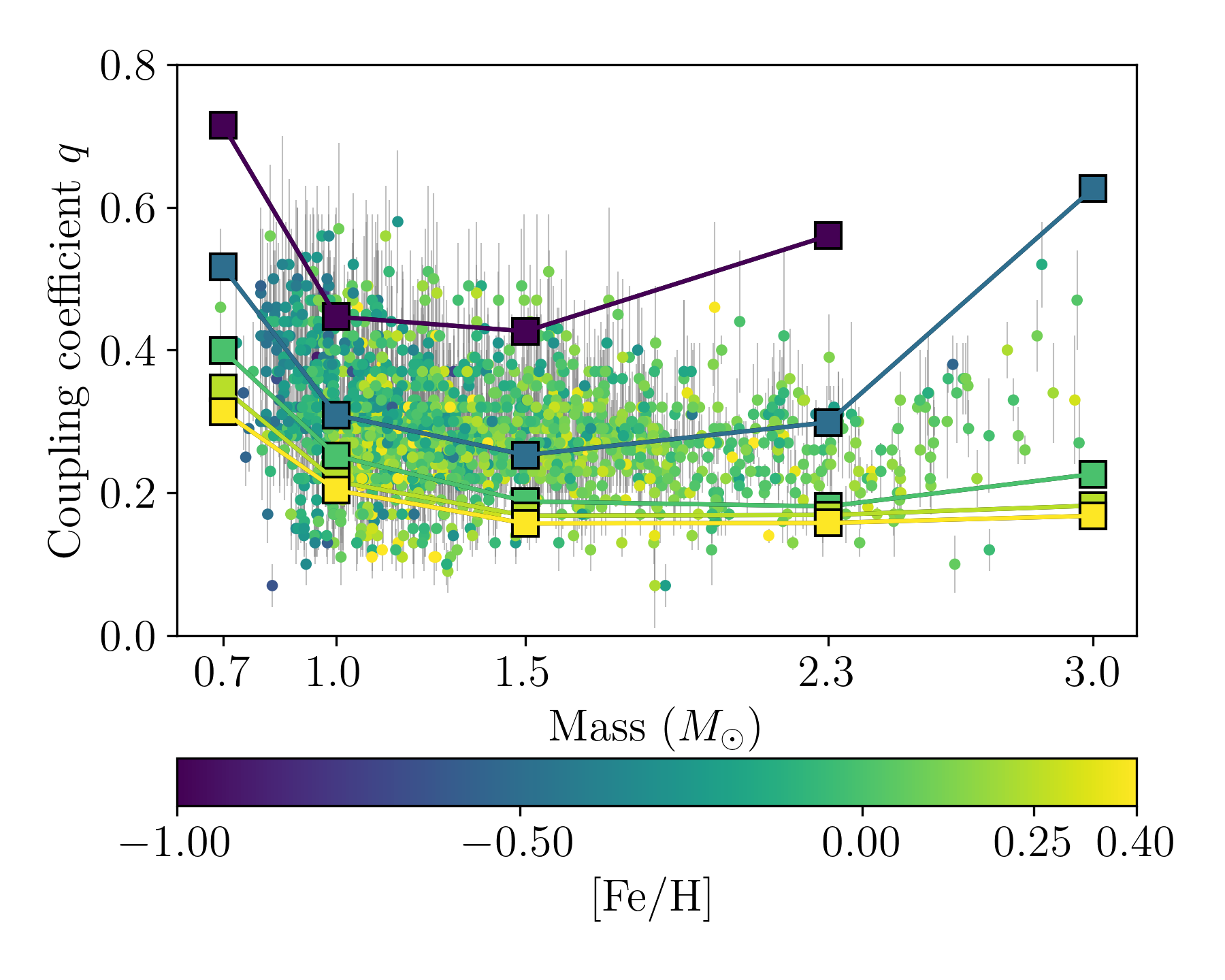}
	\caption{Mean coupling coefficient in the RC ($\langle q \rangle_\mathrm{RC}$) for models (squares) and observed coupling coefficient for observations (dots) as a function of mass. Each set of models with the same initial \FeH is connected by a solid line. \FeH is shown using the colour scale.} \label{fig:massqrc}
\end{figure}

Figure~\ref{fig:massqrc} shows the models' mean coupling in the RC (Sect.~\ref{sec:Method2}) as a function of mass at different metallicities together with the observed coupling and metallicity of RC stars. The mean RC coupling is strongest at lower masses and increases when $M \gtrsim 2~\msol$. The observed coupling coefficients follow this trend as well, with decreasing $q$ at low masses and increasing $q$ at high masses.

\begin{figure}
	\centering
	\includegraphics[width=\linewidth]{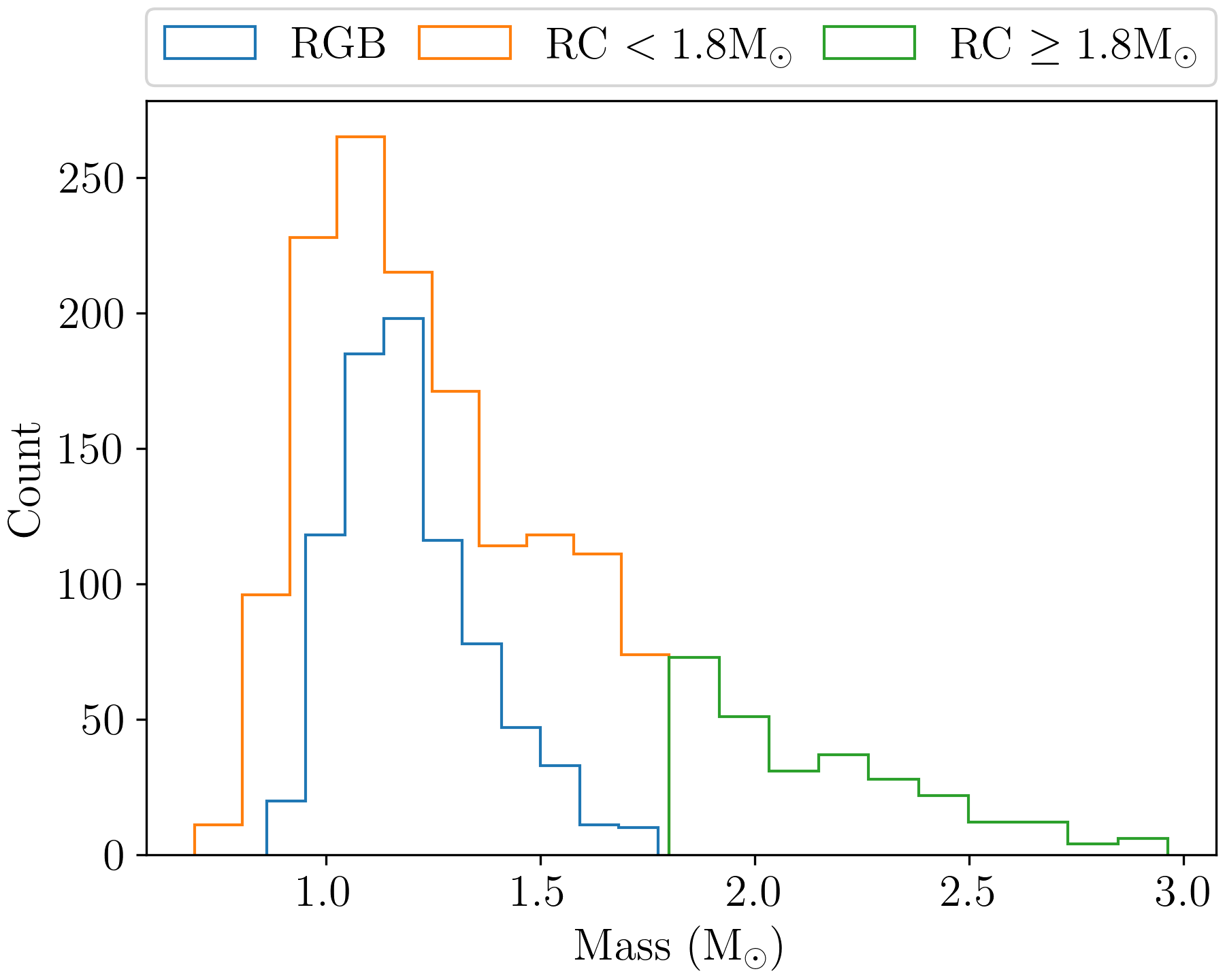}
	\caption{Stellar mass distribution of RGB stars (blue), RC stars with masses ${M < 1.8~\msol}$ (orange), and RC stars with masses ${M \geq 1.8~\msol}$ (green). }
	\label{fig:fehqmasshist}
\end{figure}

To determine the dependence of $q$ on \FeH we split the observations into three groups and fit a linear model with Gaussian intrinsic scatter with variance $\epsilon^2$ given by
\begin{equation}
	q = m \FeH + b + \mathcal{N}(0, \epsilon^2).
 \label{eq:linfit}
\end{equation}
This model is fit to each group using flat priors in $\arctan(m)$, $b$, and $\ln \epsilon$ with \texttt{emcee} \citep{DFM2013}, a Markov chain Monte Carlo (MCMC) ensemble sampler.
These groups are RGB stars ($\Delta P < 130$~s), RC stars with masses ${M < 1.8~\msol}$ and RC stars with masses ${M \geq 1.8~\msol}$. These groups contain 816, 1403, and 276 stars, respectively. Figure~\ref{fig:fehqmasshist} shows the stellar mass distributions of these groups. The mean masses of stars in each group are $1.14 \pm 0.06~\msol$, $1.09 \pm 0.11~\msol$, and $2.01 \pm 0.13~\msol$.

\begin{figure}
	\centering
	\includegraphics[width=\linewidth]{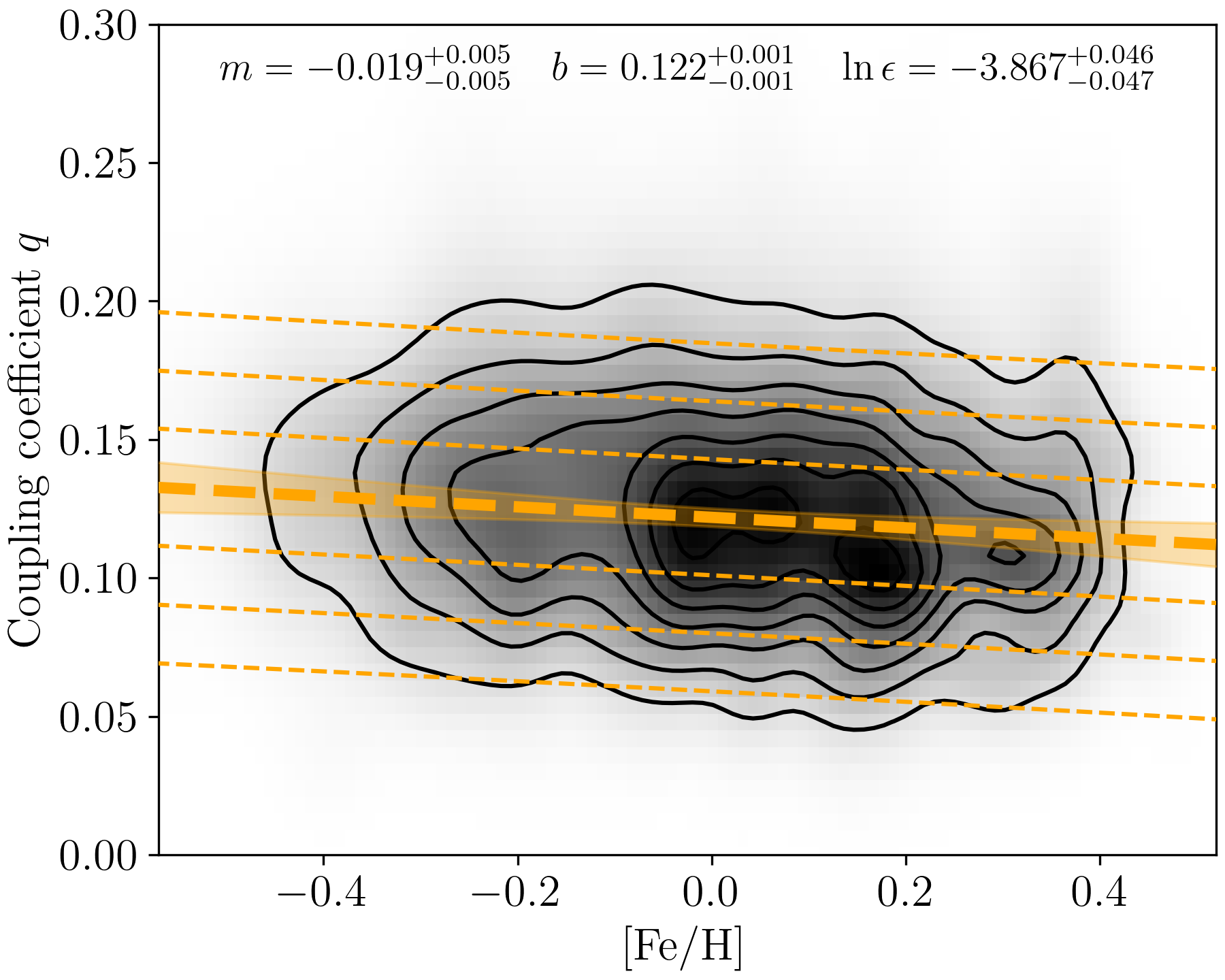}
	\caption{KDE of observed \FeH and $q$ for RGB stars with $\Delta P < 130~\mathrm{s}$. The thick orange dashed line shows the fit to the observed data, with the parameters of the fit shown at the top of the panel and Table \ref{tab:dqdfeh}. The orange shaded region shows the 3-$\sigma$ confidence interval on $m$ and $b$, whilst the thin orange dashed lines show the maximum-likelihood intrinsic relation in steps of 1-$\sigma$.}
	\label{fig:fehqkde0}
\end{figure}

\begin{figure}
	\centering
	\includegraphics[width=\linewidth]{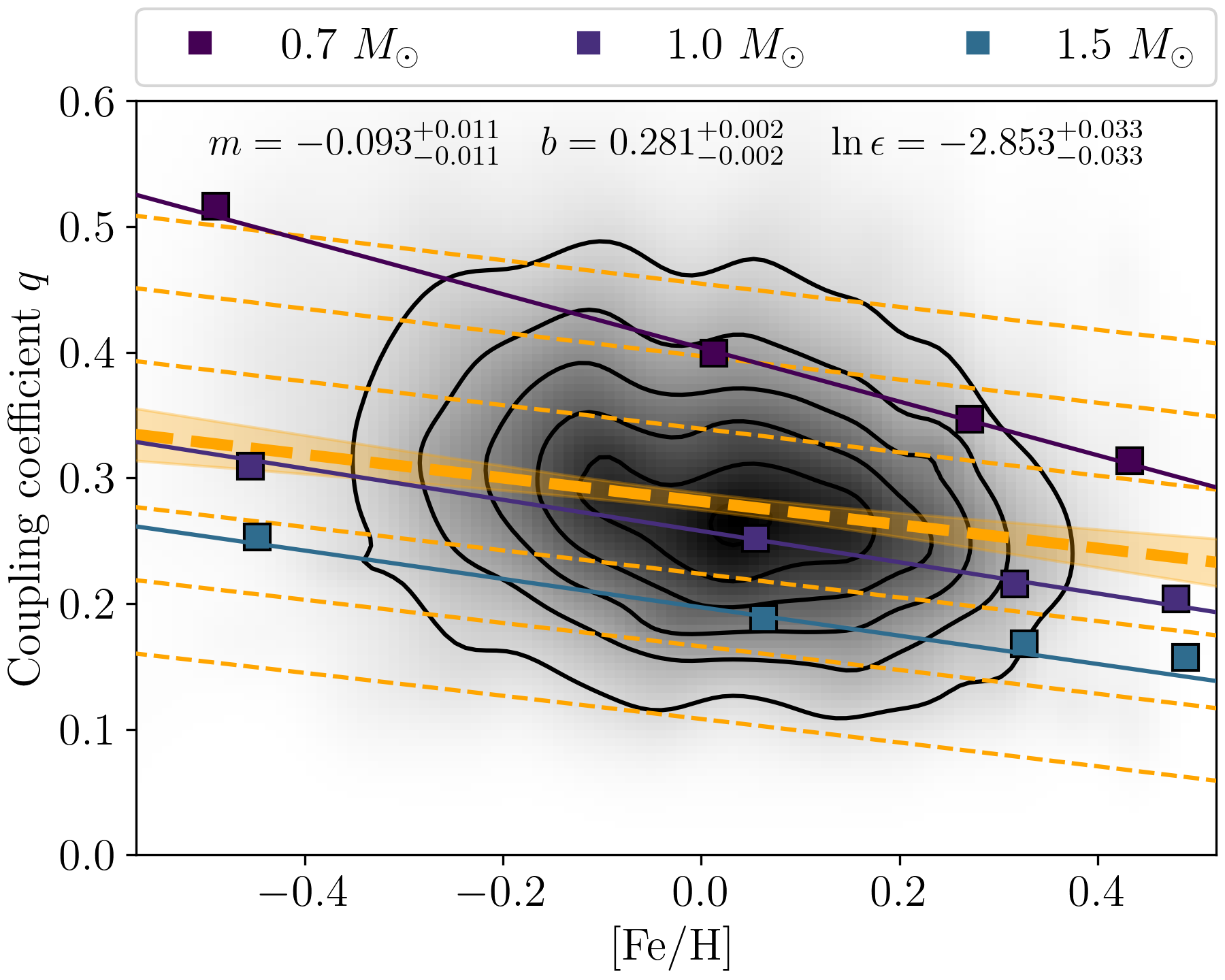}
	\caption{As Fig.~\ref{fig:fehqkde0} but for RC stars with masses $M < 1.8~\msol$. Modelled \FeH and $\langle q\rangle_\mathrm{RC}$ are shown for different masses as coloured squares. The solid coloured lines show the fits to our RC models. The parameters of these fits are shown in Table \ref{tab:moddqdfeh}.}
	\label{fig:fehqkde1}
\end{figure}

\begin{figure}
	\centering
	\includegraphics[width=\linewidth]{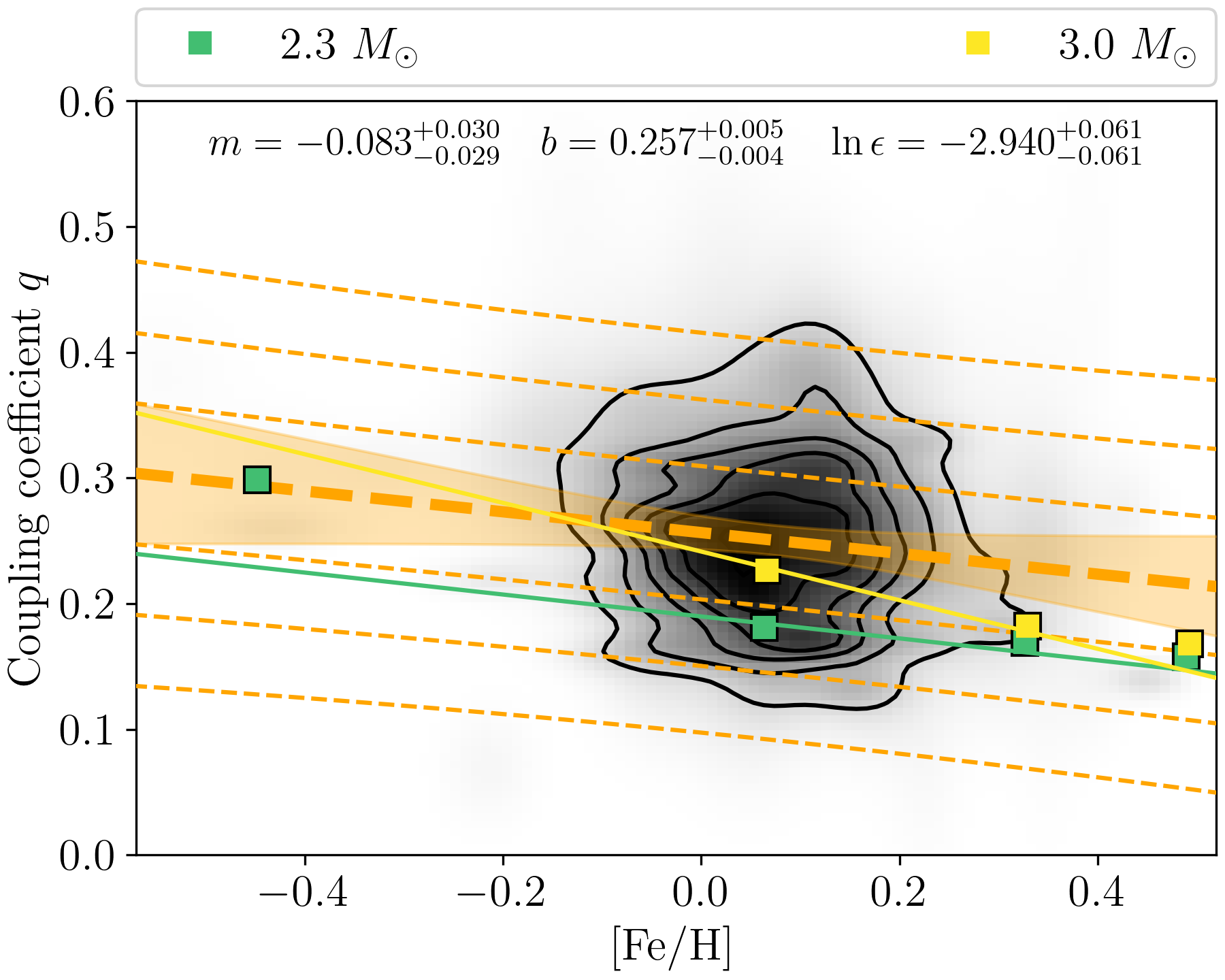}
	\caption{As Fig.~\ref{fig:fehqkde1} but for RC stars with masses $M \geq 1.8~\msol$.}
	\label{fig:fehqkde2}
\end{figure}

Figures~\ref{fig:fehqkde0}, \ref{fig:fehqkde1}, and \ref{fig:fehqkde2} show normalised 2-d Kernel Density Estimates (KDE) of observed \FeH and $q$ of RGB stars, RC stars with masses ${M < 1.8~\msol}$, and RC stars with masses ${M \geq 1.8~\msol}$ respectively. The kernel of each observation is a 2-d Gaussian with the observed uncertainties being the bandwidths in each direction. In Figs.~\ref{fig:fehqkde1} and \ref{fig:fehqkde2}, modelled stars are overlaid as squares with different initial masses labelled by colour. A linear fit to these points, weighted by the observed \FeH distribution, is shown by the solid lines. The parameters of these fits are shown in Table \ref{tab:moddqdfeh}. The posterior distributions of the fitted parameters are reported in Appendix \ref{sec:posteriors} for completeness. For RC stars the model slopes are consistent with each other, implying no dependence on mass of the slope. This is likely due to the lack of stars with masses around 3~\msol and 0.7~\msol, as seen in Fig.~\ref{fig:fehqmasshist}. In the lower-mass RC case the modelled slope is steeper than the observed slopes, but is consistent to within 2.3-$\sigma$. The slope of the higher mass RC case is consistent with observations. For RGB stars the fits to modelled coupling are excluded as a large portion of the models would be in the transition regime between strong and weak coupling. However, qualitatively they still show a similar dependence on \FeH. The y-intercepts are significantly different between the modelled and observed coupling when neglecting the intrinsic scatter. However, including the intrinsic scatter and treating it as an additional uncertainty on $b$ resolves this difference. When comparing the observed and modelled $q$, modelled $q$ tends to be underestimated \citep[][and e.g.~Fig.~\ref{fig:deltanuqref}]{Ong2023,Kuszlewicz2023}. Additionally, there are systematics which are unaccounted for as the modelled coupling coefficient and the observed coupling coefficient are determined through different methods.

For completeness we tested how changing the initial helium abundance in models affects the coupling, finding effects of the order of $10 \%$ or less for changes in $Y_{\rm init}$ of 0.02 (see Appendix \ref{sec:helium} for details).

\begin{table*}
	\caption{Parameters of the fits to observed \FeH and $q$.
		\label{tab:dqdfeh}}
	\centering
	\small
	\begin{tabular}{llcccc}
    \hline \hline 
    Phase& & $\langle M \rangle$ & $m$ & $b$ & $\ln\epsilon$\\
         & (\msol) & (\msol)             &     &     &           \\
    \hline
RC  &$   < 1.8$ & $1.09 \pm 0.22$ & $-0.093_{-0.011}^{+0.011}$ & $0.281_{-0.002}^{+0.002}$ & $-2.853_{-0.033}^{+0.033}$ \\
    &$\geq 1.8$ & $2.01 \pm 0.26$ & $-0.083_{-0.029}^{+0.030}$ & $0.257_{-0.004}^{+0.005}$ & $-2.940_{-0.061}^{+0.061}$ \\
    \hline
RGB &           & $1.14 \pm 0.14$ & $-0.019_{-0.005}^{+0.005}$ & $0.122_{-0.001}^{+0.001}$ & $-3.867_{-0.047}^{+0.046}$ \\
    \hline
\end{tabular} \end{table*}
\begin{table}
	\caption{Parameters of fits to \FeH and $q$ of our RC models.
		\label{tab:moddqdfeh}}
	\centering
	\small
	\begin{tabular}{llcc}
        \hline \hline 
        Phase & Mass & $m$ & $b$  \\
        & (\msol) & & \\
        \hline
RC     & 0.7 & $-0.214 \pm 0.009$ & $0.404 \pm 0.002$ \\
    & 1.0 & $-0.124 \pm 0.008$ & $0.258 \pm 0.001$ \\
    & 1.5 & $-0.113 \pm 0.014$ & $0.197 \pm 0.003$ \\
    & 2.3 & $-0.087 \pm 0.050$ & $0.190 \pm 0.009$ \\
    & 3.0 & $-0.194 \pm 0.083$ & $0.241 \pm 0.015$ \\

        \hline
    \end{tabular} \end{table}

\subsection{Coupling as a proxy of the core-envelope density contrast}
\label{sec:rhpcontr}
The main reason for the dependence of coupling on mass and metallicity can be ascribed to the different density contrast between the helium core and the convective envelope.
Large coupling is an indication of shallow and, on average, denser convective envelopes. This is seen both in the low-mass end \citep[e.g.][]{Matteuzzi2023}, giving us a signature of partially stripped stars (e.g.~red horizontal branch stars) and when mass increases beyond ${\sim 2 ~\msol}$  at solar metallicity.

\begin{figure}
	\centering
	\includegraphics[width=\linewidth]{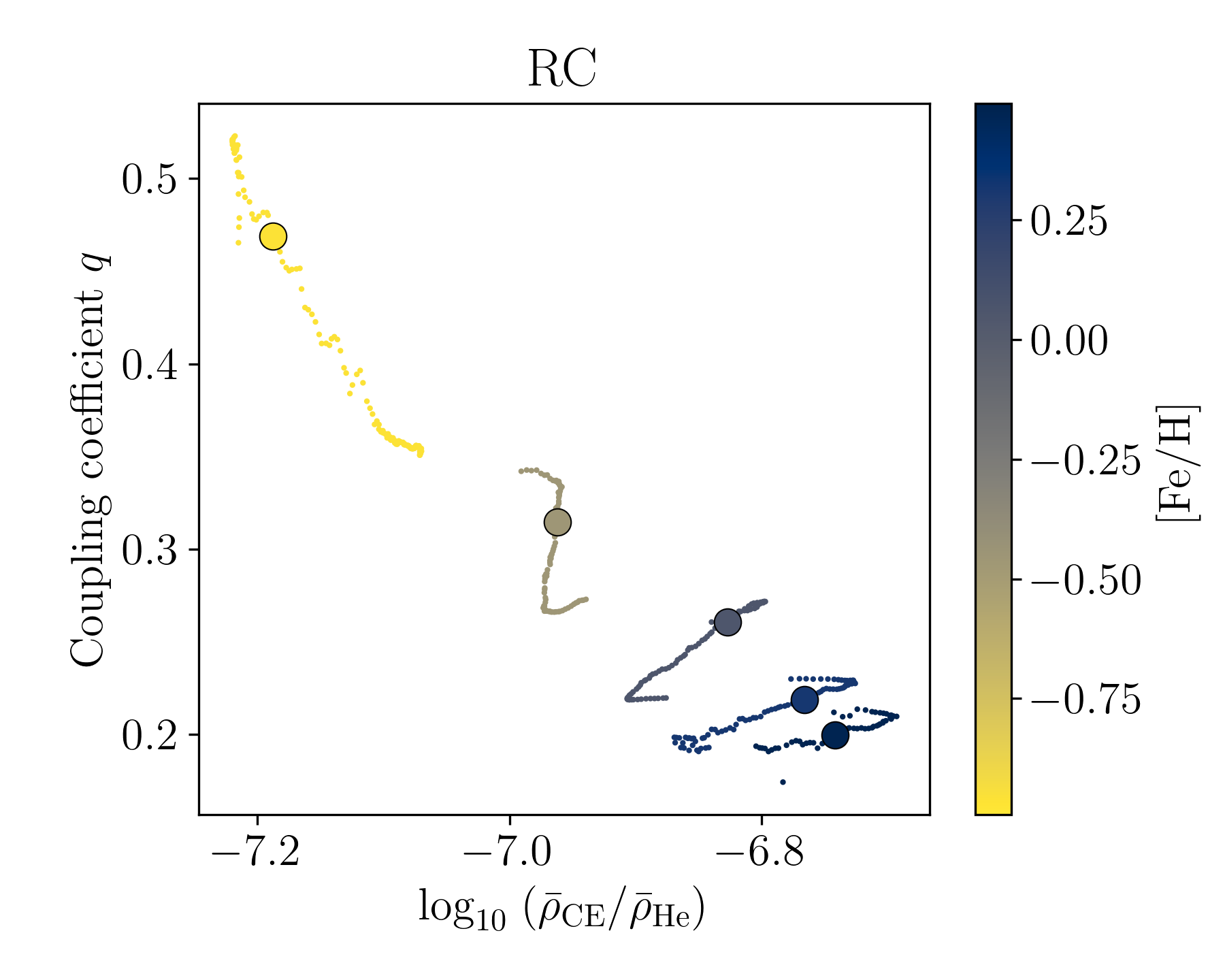}
	\caption{Coupling coefficient $q$ versus the logarithm of the ratio of the mean convective envelope density ($\bar{\rho}_\mathrm{CE}$) to the mean helium core density ($\bar{\rho}_\mathrm{He}$) in 1\msol stars. The large dots show the coupling and density contrast around $Y_\mathrm{c} = 0.5$. The colour scale shows the \FeH.}
	\label{fig:rhocontrast}
\end{figure}

\begin{figure}
	\centering
	\includegraphics[width=\linewidth]{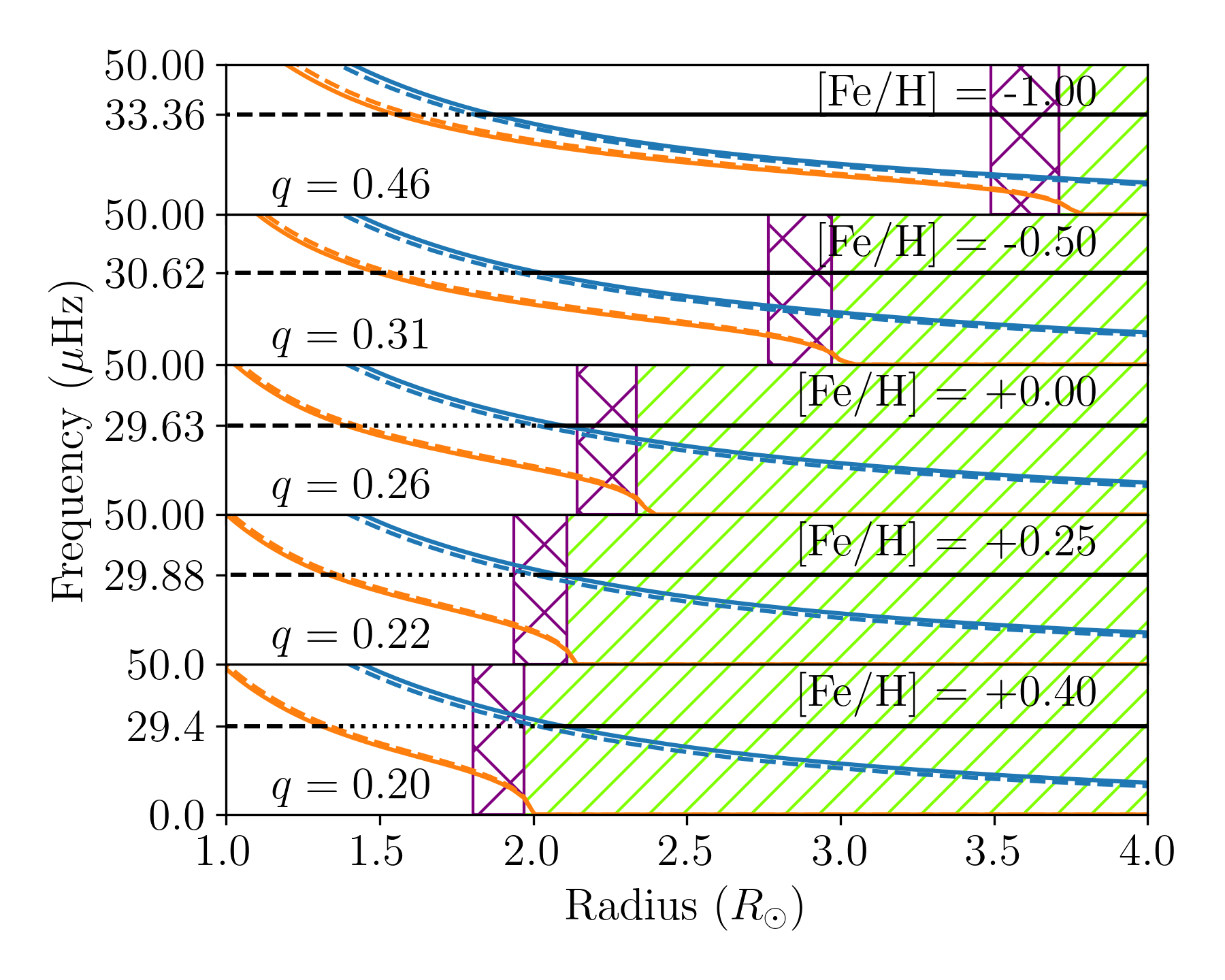}
	\caption{Propagation diagrams of 1~\msol RC models with $Y_\mathrm{C} \simeq 0.5$ showing the Lamb $(S)$ and reduced Lamb $(\Sred)$ frequencies as blue solid and dashed lines respectively. The \brunt $(N)$ and reduced \brunt $(\Nred)$ frequencies are shown as orange solid and dashed lines respectively. \numax is shown as a horizontal black line. The green hashed regions show where the star is undergoing convection, and the purple cross-hashed region shows where convective overshoot is occurring.}
	\label{fig:evanescentlocr}
\end{figure}

Figure~\ref{fig:rhocontrast} shows this strong relationship between the contrast and the coupling in these models. To first order, this relationship between density contrast and coupling is the main component driving the changes in the coupling. In models with ${M \lesssim 2~\msol}$ the mean helium-core density ($\bar{\rho}_\mathrm{He}$) remains approximately constant throughout the RC, which leaves the changes in the mean convective envelope density ($\bar{\rho}_\mathrm{CE}$) to drive the changes in density contrast.
Changes in the convective envelope caused by different metallicities can be seen in Fig.~\ref{fig:evanescentlocr}. It shows part of a propagation diagram of a 1~\msol star, at various metallicities in the RC, when the central helium mass fraction is approximately 0.5.
The size of the evanescent zone is smaller in low-metallicity stars and as metallicity increases, the upper boundary moves outwards, whilst the inner boundary moves towards the centre of the star but at a slower rate. As the metallicity increases, so does the opacity. This both inflates the star and gives it a deeper convective envelope.

\begin{figure}
	\centering
	\includegraphics[width=\linewidth]{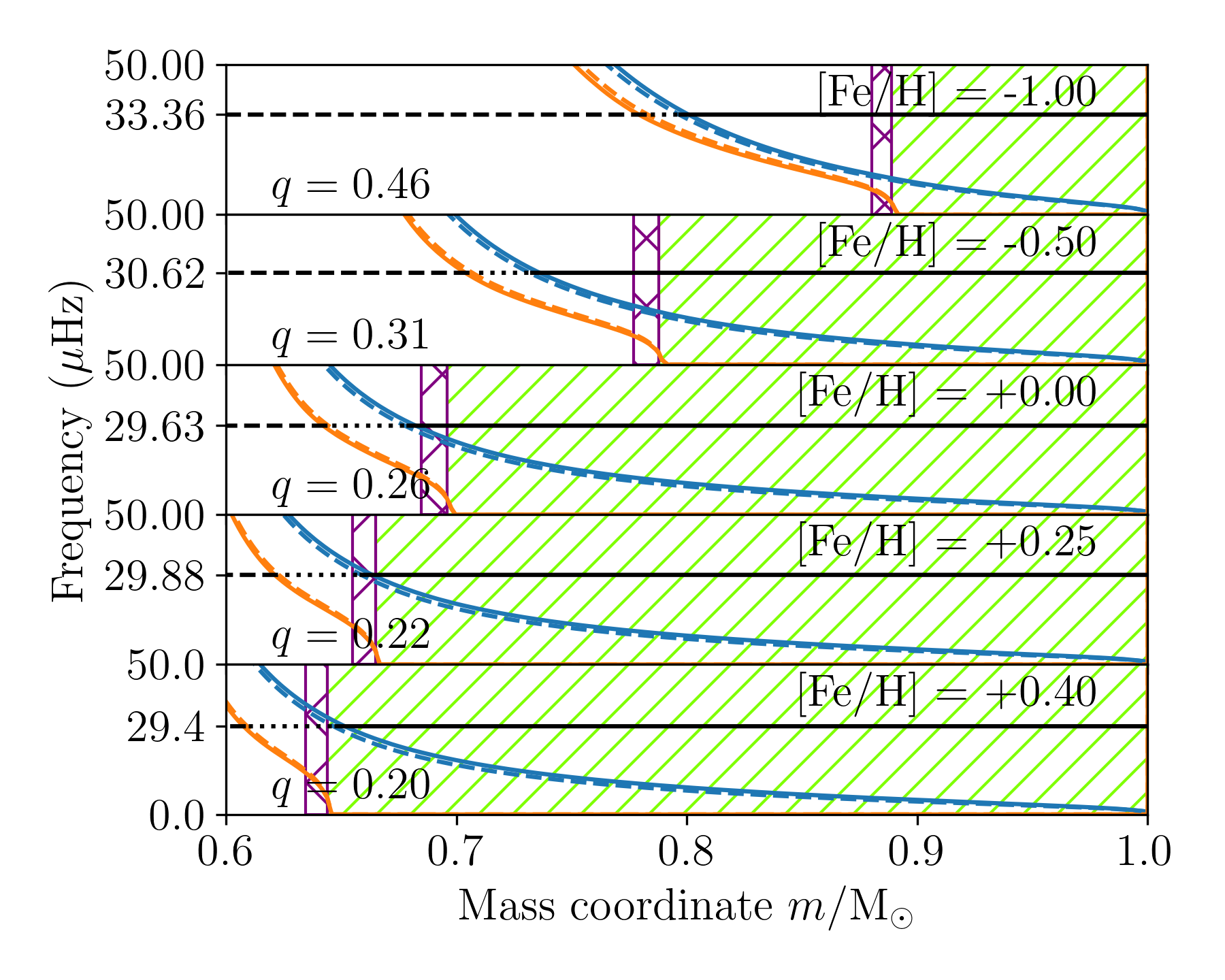}
	\caption{As Fig.~\ref{fig:evanescentlocr} but as a function of mass.}
	\label{fig:evanescentlocmass}
\end{figure}
Figure~\ref{fig:evanescentlocmass} is similar to Fig.~\ref{fig:evanescentlocr} but shows the propagation diagram as a function of mass coordinate instead of radius. The location of the evanescent zone moves deeper into the star as the metallicity increases and as the convective envelope becomes more massive, going from a mass coordinate of 0.8~\msol to around 0.6~\msol at the highest metallicity. These two effects, shown in Figs. \ref{fig:evanescentlocr} and \ref{fig:evanescentlocmass}, account for the relationship between density contrast and coupling seen in Fig.~\ref{fig:rhocontrast}.

The combination of the relationships in Figs.~\ref{fig:massfehqrcbinned}, \ref{fig:massqrc}, \ref{fig:fehqkde0}, \ref{fig:fehqkde1}, \ref{fig:fehqkde2}, and \ref{fig:rhocontrast} shows that the coupling coefficient is an indirect probe to the depth of the convective envelope. A shallower convective envelope leads to a smaller density contrast, and hence to a larger coupling.

\section{Summary and prospects} \label{sec:SummaryConcl}
In this work we established through detailed stellar modelling that the coupling coefficient describing the interaction between p- and g-modes depends on stellar global parameters, evolutionary stage, and structural properties. Crucially, we have checked the formulation by \citet{Takata2016} in RC stars by comparing the modelled coupling coefficients to observed coupling coefficients. Our main conclusions are as follows:
\begin{itemize}
	\item We have shown that both mass and metallicity play a significant role in mixed-mode coupling in the RC. They both affect the density contrast between the core and envelope, and thus affect the depth of the convective envelope, and therefore also determine the location and size of the evanescent zone. Additionally, as shown in Figs.~\ref{fig:fehqkde0}, \ref{fig:fehqkde1}, and \ref{fig:fehqkde2}, there is an anti-correlation between metallicity and coupling in both the RGB and in the RC. We also find that coupling decreases as a function of mass in low-mass models ($M \lesssim 1.8~\msol$), and increases as a function of mass for high-mass models ($M \gtrsim 1.8~\msol$ ). We thus expect stars with highest coupling to be either very low-mass core-He burning stars (e.g.~red-horizontal branch or partially stripped stars, see \citealt{Matteuzzi2023}) or luminous, core-He burning massive (${M \gtrsim 2~\msol}$) stars.

	\item These model-predicted trends in \FeH and $q$ (Figs.~\ref{fig:fehqkde0}, \ref{fig:fehqkde1}, and \ref{fig:fehqkde2}) broadly agree with observations.
 However, a note of caution should be given about the possible presence of glitches, which may affect the determination of the mean period spacings, which in turn may affect coupling coefficients \citep{Vrard2016,Mosser2017}. Additionally, concerning model-predicted values of $q$, there is some ambiguity about what value $q$ has in the transition regime between the strong and weak cases as there is no coupling prescription for the intermediate regime, for example, during the RGB.

	\item We find the coupling coefficient to be frequency dependent \citep[see][for the case of RGB stars]{Jiang2020}. Current measurements of coupling from the data only give an average value \citep{Mosser2017}, introducing a potential source of systematic error when comparing observed to model-predicted coupling. The frequency dependence is particularly strong in the RGB (panels A and B in Fig.~\ref{fig:prop}), opening up the possibility of a detailed mapping of chemical and thermal stratification of the convective boundary region if the frequency dependence of the coupling coefficient can be measured. Within one \deltanu frequency interval, there can be a difference of between 10--20\% in $q$. This variation in $q$ due to different $\nu_\mathrm{q}$ should be taken into account when determining $q$ from observations, and when comparing to model predictions.

	\item We have also introduced an additional approximation to calculate the gradient term for the coupling based on the work done by \citet{Takata2016} and \citet{Pincon2019} and the fact that the \brunt and Lamb frequencies do not share the same power-law exponent (non-parallel approximation). This works well in type-a evanescent zones on the early RGB with errors of around 5\% using the non-parallel approximation compared to 10\% when using the parallel approximation.%

\end{itemize}

The relationship between the coupling coefficient and the internal structure of stars highlighted in this work provides a foundation for further study in which one can also potentially identify structures with non-standard core-mass to envelope-mass ratios, providing a potential way to identify, for example, a direct signature of mass transfer products.

\begin{acknowledgements}
    We thank R.G. Izzard and W.J. Chaplin for their helpful comments and thorough reading of the manuscript.
	W.E.v.R, A.M., and J.M. acknowledge support from the European Research Council Consolidator Grant funding scheme (project ASTEROCHRONOMETRY, G.A. n. 772293, \url{https://www.asterochronometry.eu}).
	This work has made extensive use of the python packages \texttt{NumPy} \citep[][\url{https://numpy.org}]{numpy}, \texttt{matplotlib} \citep[][\url{https://matplotlib.org}]{matplotlib},
	\texttt{scipy} \citep[][\url{https://scipy.org}]{scipy}, and
	\texttt{Astropy} \citep[][\url{https://www.astropy.org}]{astropy:2013, astropy:2018}.

\end{acknowledgements}

\bibliographystyle{aa}
\bibliography{my_bib}

\begin{thebibliography}{65}
\expandafter\ifx\csname natexlab\endcsname\relax\def\natexlab#1{#1}\fi

\bibitem[{{Abolfathi} {et~al.}(2018){Abolfathi}, {Aguado}, {Aguilar}, {Allende
  Prieto}, {Almeida}, {Ananna}, {Anders}, {Anderson}, {Andrews}, {Anguiano},
  {Arag{\'o}n-Salamanca}, {Argudo-Fern{\'a}ndez}, {Armengaud}, {Ata},
  {Aubourg}, {Avila-Reese}, {Badenes}, {Bailey}, {Balland}, {Barger},
  {Barrera-Ballesteros}, {Bartosz}, {Bastien}, {Bates}, {Baumgarten},
  {Bautista}, {Beaton}, {Beers}, {Belfiore}, {Bender}, {Bernardi}, {Bershady},
  {Beutler}, {Bird}, {Bizyaev}, {Blanc}, {Blanton}, {Blomqvist}, {Bolton},
  {Boquien}, {Borissova}, {Bovy}, {Bradna Diaz}, {Brandt}, {Brinkmann},
  {Brownstein}, {Bundy}, {Burgasser}, {Burtin}, {Busca}, {Ca{\~n}as},
  {Cano-D{\'\i}az}, {Cappellari}, {Carrera}, {Casey}, {Cervantes Sodi}, {Chen},
  {Cherinka}, {Chiappini}, {Choi}, {Chojnowski}, {Chuang}, {Chung}, {Clerc},
  {Cohen}, {Comerford}, {Comparat}, {Correa do Nascimento}, {da Costa},
  {Cousinou}, {Covey}, {Crane}, {Cruz-Gonzalez}, {Cunha}, {da Silva Ilha},
  {Damke}, {Darling}, {Davidson}, {Dawson}, {de Icaza Lizaola}, {de la
  Macorra}, {de la Torre}, {De Lee}, {de Sainte Agathe}, {Deconto Machado},
  {Dell'Agli}, {Delubac}, {Diamond-Stanic}, {Donor}, {Downes}, {Drory}, {du Mas
  des Bourboux}, {Duckworth}, {Dwelly}, {Dyer}, {Ebelke}, {Davis Eigenbrot},
  {Eisenstein}, {Elsworth}, {Emsellem}, {Eracleous}, {Erfanianfar},
  {Escoffier}, {Fan}, {Fern{\'a}ndez Alvar}, {Fernandez-Trincado}, {Fernando
  Cirolini}, {Feuillet}, {Finoguenov}, {Fleming}, {Font-Ribera}, {Freischlad},
  {Frinchaboy}, {Fu}, {G{\'o}mez Maqueo Chew}, {Galbany}, {Garc{\'\i}a
  P{\'e}rez}, {Garcia-Dias}, {Garc{\'\i}a-Hern{\'a}ndez}, {Garma Oehmichen},
  {Gaulme}, {Gelfand}, {Gil-Mar{\'\i}n}, {Gillespie}, {Goddard}, {Gonz{\'a}lez
  Hern{\'a}ndez}, {Gonzalez-Perez}, {Grabowski}, {Green}, {Grier}, {Gueguen},
  {Guo}, {Guy}, {Hagen}, {Hall}, {Harding}, {Hasselquist}, {Hawley}, {Hayes},
  {Hearty}, {Hekker}, {Hernandez}, {Hernandez Toledo}, {Hogg},
  {Holley-Bockelmann}, {Holtzman}, {Hou}, {Hsieh}, {Hunt}, {Hutchinson},
  {Hwang}, {Jimenez Angel}, {Johnson}, {Jones}, {J{\"o}nsson}, {Jullo}, {Khan},
  {Kinemuchi}, {Kirkby}, {Kirkpatrick}, {Kitaura}, {Knapp}, {Kneib},
  {Kollmeier}, {Lacerna}, {Lane}, {Lang}, {Law}, {Le Goff}, {Lee}, {Li}, {Li},
  {Lian}, {Liang}, {Lima}, {Lin}, {Long}, {Lucatello}, {Lundgren}, {Mackereth},
  {MacLeod}, {Mahadevan}, {Maia}, {Majewski}, {Manchado}, {Maraston},
  {Mariappan}, {Marques-Chaves}, {Masseron}, {Masters}, {McDermid}, {McGreer},
  {Melendez}, {Meneses-Goytia}, {Merloni}, {Merrifield}, {Meszaros}, {Meza},
  {Minchev}, {Minniti}, {Mueller}, {Muller-Sanchez}, {Muna}, {Mu{\~n}oz},
  {Myers}, {Nair}, {Nandra}, {Ness}, {Newman}, {Nichol}, {Nidever},
  {Nitschelm}, {Noterdaeme}, {O'Connell}, {Oelkers}, {Oravetz}, {Oravetz},
  {Ort{\'\i}z}, {Osorio}, {Pace}, {Padilla}, {Palanque-Delabrouille},
  {Palicio}, {Pan}, {Pan}, {Parikh}, {P{\^a}ris}, {Park}, {Peirani},
  {Pellejero-Ibanez}, {Penny}, {Percival}, {Perez-Fournon}, {Petitjean},
  {Pieri}, {Pinsonneault}, {Pisani}, {Prada}, {Prakash}, {Queiroz}, {Raddick},
  {Raichoor}, {Barboza Rembold}, {Richstein}, {Riffel}, {Riffel}, {Rix},
  {Robin}, {Rodr{\'\i}guez Torres}, {Rom{\'a}n-Z{\'u}{\~n}iga}, {Ross},
  {Rossi}, {Ruan}, {Ruggeri}, {Ruiz}, {Salvato}, {S{\'a}nchez}, {S{\'a}nchez},
  {Sanchez Almeida}, {S{\'a}nchez-Gallego}, {Santana Rojas}, {Santiago},
  {Schiavon}, {Schimoia}, {Schlafly}, {Schlegel}, {Schneider}, {Schuster},
  {Schwope}, {Seo}, {Serenelli}, {Shen}, {Shen}, {Shetrone}, {Shull}, {Silva
  Aguirre}, {Simon}, {Skrutskie}, {Slosar}, {Smethurst}, {Smith}, {Sobeck},
  {Somers}, {Souter}, {Souto}, {Spindler}, {Stark}, {Stassun}, {Steinmetz},
  {Stello}, {Storchi-Bergmann}, {Streblyanska}, {Stringfellow}, {Su{\'a}rez},
  {Sun}, {Szigeti}, {Taghizadeh-Popp}, {Talbot}, {Tang}, {Tao}, {Tayar},
  {Tembe}, {Teske}, {Thakar}, {Thomas}, {Tissera}, {Tojeiro}, {Tremonti},
  {Troup}, {Urry}, {Valenzuela}, {van den Bosch}, {Vargas-Gonz{\'a}lez},
  {Vargas-Maga{\~n}a}, {Vazquez}, {Villanova}, {Vogt}, {Wake}, {Wang},
  {Weaver}, {Weijmans}, {Weinberg}, {Westfall}, {Whelan}, {Wilcots}, {Wild},
  {Williams}, {Wilson}, {Wood-Vasey}, {Wylezalek}, {Xiao}, {Yan}, {Yang},
  {Ybarra}, {Y{\`e}che}, {Zakamska}, {Zamora}, {Zarrouk}, {Zasowski}, {Zhang},
  {Zhao}, {Zhao}, {Zheng}, {Zheng}, {Zhou}, {Zhu}, {Zinn}, \&
  {Zou}}]{SDSS-DR14}
{Abolfathi}, B., {Aguado}, D.~S., {Aguilar}, G., {et~al.} 2018, \apjs, 235, 42

\bibitem[{Anderson {et~al.}(1999)Anderson, Bai, Bischof, Blackford, Demmel,
  Dongarra, Du~Croz, Greenbaum, Hammarling, McKenney, \& Sorensen}]{LAPACK}
Anderson, E., Bai, Z., Bischof, C., {et~al.} 1999, {LAPACK} Users' Guide, 3rd
  edn. (Philadelphia, PA: Society for Industrial and Applied Mathematics)

\bibitem[{{Appourchaux}(2020)}]{Appourchaux2020}
{Appourchaux}, T. 2020, \aap, 642, A226

\bibitem[{{Astropy Collaboration} {et~al.}(2018){Astropy Collaboration},
  {Price-Whelan}, {Sip{\H{o}}cz}, {G{\"u}nther}, {Lim}, {Crawford}, {Conseil},
  {Shupe}, {Craig}, {Dencheva}, {Ginsburg}, {Vand erPlas}, {Bradley},
  {P{\'e}rez-Su{\'a}rez}, {de Val-Borro}, {Aldcroft}, {Cruz}, {Robitaille},
  {Tollerud}, {Ardelean}, {Babej}, {Bach}, {Bachetti}, {Bakanov}, {Bamford},
  {Barentsen}, {Barmby}, {Baumbach}, {Berry}, {Biscani}, {Boquien}, {Bostroem},
  {Bouma}, {Brammer}, {Bray}, {Breytenbach}, {Buddelmeijer}, {Burke},
  {Calderone}, {Cano Rodr{\'\i}guez}, {Cara}, {Cardoso}, {Cheedella}, {Copin},
  {Corrales}, {Crichton}, {D'Avella}, {Deil}, {Depagne}, {Dietrich}, {Donath},
  {Droettboom}, {Earl}, {Erben}, {Fabbro}, {Ferreira}, {Finethy}, {Fox},
  {Garrison}, {Gibbons}, {Goldstein}, {Gommers}, {Greco}, {Greenfield},
  {Groener}, {Grollier}, {Hagen}, {Hirst}, {Homeier}, {Horton}, {Hosseinzadeh},
  {Hu}, {Hunkeler}, {Ivezi{\'c}}, {Jain}, {Jenness}, {Kanarek}, {Kendrew},
  {Kern}, {Kerzendorf}, {Khvalko}, {King}, {Kirkby}, {Kulkarni}, {Kumar},
  {Lee}, {Lenz}, {Littlefair}, {Ma}, {Macleod}, {Mastropietro}, {McCully},
  {Montagnac}, {Morris}, {Mueller}, {Mumford}, {Muna}, {Murphy}, {Nelson},
  {Nguyen}, {Ninan}, {N{\"o}the}, {Ogaz}, {Oh}, {Parejko}, {Parley}, {Pascual},
  {Patil}, {Patil}, {Plunkett}, {Prochaska}, {Rastogi}, {Reddy Janga},
  {Sabater}, {Sakurikar}, {Seifert}, {Sherbert}, {Sherwood-Taylor}, {Shih},
  {Sick}, {Silbiger}, {Singanamalla}, {Singer}, {Sladen}, {Sooley},
  {Sornarajah}, {Streicher}, {Teuben}, {Thomas}, {Tremblay}, {Turner},
  {Terr{\'o}n}, {van Kerkwijk}, {de la Vega}, {Watkins}, {Weaver}, {Whitmore},
  {Woillez}, {Zabalza}, \& {Astropy Contributors}}]{astropy:2018}
{Astropy Collaboration}, {Price-Whelan}, A.~M., {Sip{\H{o}}cz}, B.~M., {et~al.}
  2018, \aj, 156, 123

\bibitem[{{Astropy Collaboration} {et~al.}(2013){Astropy Collaboration},
  {Robitaille}, {Tollerud}, {Greenfield}, {Droettboom}, {Bray}, {Aldcroft},
  {Davis}, {Ginsburg}, {Price-Whelan}, {Kerzendorf}, {Conley}, {Crighton},
  {Barbary}, {Muna}, {Ferguson}, {Grollier}, {Parikh}, {Nair}, {Unther},
  {Deil}, {Woillez}, {Conseil}, {Kramer}, {Turner}, {Singer}, {Fox}, {Weaver},
  {Zabalza}, {Edwards}, {Azalee Bostroem}, {Burke}, {Casey}, {Crawford},
  {Dencheva}, {Ely}, {Jenness}, {Labrie}, {Lim}, {Pierfederici}, {Pontzen},
  {Ptak}, {Refsdal}, {Servillat}, \& {Streicher}}]{astropy:2013}
{Astropy Collaboration}, {Robitaille}, T.~P., {Tollerud}, E.~J., {et~al.} 2013,
  \aap, 558, A33

\bibitem[{{Auvergne} {et~al.}(2009){Auvergne}, {Bodin}, {Boisnard}, {Buey},
  {Chaintreuil}, {Epstein}, {Jouret}, {Lam-Trong}, {Levacher}, {Magnan},
  {Perez}, {Plasson}, {Plesseria}, {Peter}, {Steller}, {Tiph{\`e}ne}, {Baglin},
  {Agogu{\'e}}, {Appourchaux}, {Barbet}, {Beaufort}, {Bellenger}, {Berlin},
  {Bernardi}, {Blouin}, {Boumier}, {Bonneau}, {Briet}, {Butler}, {Cautain},
  {Chiavassa}, {Costes}, {Cuvilho}, {Cunha-Parro}, {de Oliveira Fialho},
  {Decaudin}, {Defise}, {Djalal}, {Docclo}, {Drummond}, {Dupuis}, {Exil},
  {Faur{\'e}}, {Gaboriaud}, {Gamet}, {Gavalda}, {Grolleau}, {Gueguen},
  {Guivarc'h}, {Guterman}, {Hasiba}, {Huntzinger}, {Hustaix}, {Imbert},
  {Jeanville}, {Johlander}, {Jorda}, {Journoud}, {Karioty}, {Kerjean},
  {Lafond}, {Lapeyrere}, {Landiech}, {Larqu{\'e}}, {Laudet}, {Le Merrer},
  {Leporati}, {Leruyet}, {Levieuge}, {Llebaria}, {Martin}, {Mazy}, {Mesnager},
  {Michel}, {Moalic}, {Monjoin}, {Naudet}, {Neukirchner}, {Nguyen-Kim},
  {Ollivier}, {Orcesi}, {Ottacher}, {Oulali}, {Parisot}, {Perruchot},
  {Piacentino}, {Pinheiro da Silva}, {Platzer}, {Pontet}, {Pradines},
  {Quentin}, {Rohbeck}, {Rolland}, {Rollenhagen}, {Romagnan}, {Russ}, {Samadi},
  {Schmidt}, {Schwartz}, {Sebbag}, {Smit}, {Sunter}, {Tello}, {Toulouse},
  {Ulmer}, {Vandermarcq}, {Vergnault}, {Wallner}, {Waultier}, \&
  {Zanatta}}]{CoRoT}
{Auvergne}, M., {Bodin}, P., {Boisnard}, L., {et~al.} 2009, \aap, 506, 411

\bibitem[{{Beck} {et~al.}(2011){Beck}, {Bedding}, {Mosser}, {Stello}, {Garcia},
  {Kallinger}, {Hekker}, {Elsworth}, {Frandsen}, {Carrier}, {De Ridder},
  {Aerts}, {White}, {Huber}, {Dupret}, {Montalb{\'a}n}, {Miglio}, {Noels},
  {Chaplin}, {Kjeldsen}, {Christensen-Dalsgaard}, {Gilliland}, {Brown},
  {Kawaler}, {Mathur}, \& {Jenkins}}]{Beck2011}
{Beck}, P.~G., {Bedding}, T.~R., {Mosser}, B., {et~al.} 2011, Science, 332, 205

\bibitem[{{Beck} {et~al.}(2012){Beck}, {Montalban}, {Kallinger}, {De Ridder},
  {Aerts}, {Garc{\'\i}a}, {Hekker}, {Dupret}, {Mosser}, {Eggenberger},
  {Stello}, {Elsworth}, {Frandsen}, {Carrier}, {Hillen}, {Gruberbauer},
  {Christensen-Dalsgaard}, {Miglio}, {Valentini}, {Bedding}, {Kjeldsen},
  {Girouard}, {Hall}, \& {Ibrahim}}]{Beck2012}
{Beck}, P.~G., {Montalban}, J., {Kallinger}, T., {et~al.} 2012, \nat, 481, 55

\bibitem[{{Bedding} {et~al.}(2010){Bedding}, {Huber}, {Stello}, {Elsworth},
  {Hekker}, {Kallinger}, {Mathur}, {Mosser}, {Preston}, {Ballot}, {Barban},
  {Broomhall}, {Buzasi}, {Chaplin}, {Garc{\'\i}a}, {Gruberbauer}, {Hale}, {De
  Ridder}, {Frandsen}, {Borucki}, {Brown}, {Christensen-Dalsgaard},
  {Gilliland}, {Jenkins}, {Kjeldsen}, {Koch}, {Belkacem}, {Bildsten}, {Bruntt},
  {Campante}, {Deheuvels}, {Derekas}, {Dupret}, {Goupil}, {Hatzes}, {Houdek},
  {Ireland }, {Jiang}, {Karoff}, {Kiss}, {Lebreton}, {Miglio}, {Montalb{\'a}n},
  {Noels}, {Roxburgh}, {Sangaralingam}, {Stevens}, {Suran}, {Tarrant}, \&
  {Weiss}}]{Bedding2010}
{Bedding}, T.~R., {Huber}, D., {Stello}, D., {et~al.} 2010, \apjl, 713, L176

\bibitem[{{Bedding} {et~al.}(2011){Bedding}, {Mosser}, {Huber},
  {Montalb{\'a}n}, {Beck}, {Christensen-Dalsgaard}, {Elsworth}, {Garc{\'\i}a},
  {Miglio}, {Stello}, {White}, {De Ridder}, {Hekker}, {Aerts}, {Barban},
  {Belkacem}, {Broomhall}, {Brown}, {Buzasi}, {Carrier}, {Chaplin}, {di Mauro},
  {Dupret}, {Frandsen}, {Gilliland }, {Goupil}, {Jenkins}, {Kallinger},
  {Kawaler}, {Kjeldsen}, {Mathur}, {Noels}, {Silva Aguirre}, \&
  {Ventura}}]{Bedding2011}
{Bedding}, T.~R., {Mosser}, B., {Huber}, D., {et~al.} 2011, \nat, 471, 608

\bibitem[{Bjorn \& Čertík(2021)}]{bjodah2021}
Bjorn \& Čertík, O. 2021

\bibitem[{{Borucki} {et~al.}(2010){Borucki}, {Koch}, {Basri}, {Batalha},
  {Brown}, {Caldwell}, {Caldwell}, {Christensen-Dalsgaard}, {Cochran},
  {DeVore}, {Dunham}, {Dupree}, {Gautier}, {Geary}, {Gilliland}, {Gould},
  {Howell}, {Jenkins}, {Kondo}, {Latham}, {Marcy}, {Meibom}, {Kjeldsen},
  {Lissauer}, {Monet}, {Morrison}, {Sasselov}, {Tarter}, {Boss}, {Brownlee},
  {Owen}, {Buzasi}, {Charbonneau}, {Doyle}, {Fortney}, {Ford}, {Holman},
  {Seager}, {Steffen}, {Welsh}, {Rowe}, {Anderson}, {Buchhave}, {Ciardi},
  {Walkowicz}, {Sherry}, {Horch}, {Isaacson}, {Everett}, {Fischer}, {Torres},
  {Johnson}, {Endl}, {MacQueen}, {Bryson}, {Dotson}, {Haas}, {Kolodziejczak},
  {Van Cleve}, {Chandrasekaran}, {Twicken}, {Quintana}, {Clarke}, {Allen},
  {Li}, {Wu}, {Tenenbaum}, {Verner}, {Bruhweiler}, {Barnes}, \&
  {Prsa}}]{Kepler}
{Borucki}, W.~J., {Koch}, D., {Basri}, G., {et~al.} 2010, Science, 327, 977

\bibitem[{{Bossini} {et~al.}(2015){Bossini}, {Miglio}, {Salaris},
  {Pietrinferni}, {Montalb{\'a}n}, {Bressan}, {Noels}, {Cassisi}, {Girardi}, \&
  {Marigo}}]{Bossini2015}
{Bossini}, D., {Miglio}, A., {Salaris}, M., {et~al.} 2015, \mnras, 453, 2290

\bibitem[{{Chaplin} \& {Miglio}(2013)}]{Chaplin2013}
{Chaplin}, W.~J. \& {Miglio}, A. 2013, \araa, 51, 353

\bibitem[{{Connelly} {et~al.}(2012){Connelly}, {Bizzarro}, {Krot}, {Nordlund},
  {Wielandt}, \& {Ivanova}}]{Connelly2012}
{Connelly}, J.~N., {Bizzarro}, M., {Krot}, A.~N., {et~al.} 2012, Science, 338,
  651

\bibitem[{{Cowling}(1941)}]{Cowling1941}
{Cowling}, T.~G. 1941, \mnras, 101, 367

\bibitem[{{Cunha} {et~al.}(2015){Cunha}, {Stello}, {Avelino},
  {Christensen-Dalsgaard}, \& {Townsend}}]{Cunha2015}
{Cunha}, M.~S., {Stello}, D., {Avelino}, P.~P., {Christensen-Dalsgaard}, J., \&
  {Townsend}, R.~H.~D. 2015, \apj, 805, 127

\bibitem[{{Deheuvels} {et~al.}(2020){Deheuvels}, {Ballot}, {Eggenberger},
  {Spada}, {Noll}, \& {den Hartogh}}]{Deheuvels2020}
{Deheuvels}, S., {Ballot}, J., {Eggenberger}, P., {et~al.} 2020, \aap, 641,
  A117

\bibitem[{{Deheuvels} {et~al.}(2016){Deheuvels}, {Brand{\~a}o}, {Silva
  Aguirre}, {Ballot}, {Michel}, {Cunha}, {Lebreton}, \&
  {Appourchaux}}]{Deheuvels2016}
{Deheuvels}, S., {Brand{\~a}o}, I., {Silva Aguirre}, V., {et~al.} 2016, \aap,
  589, A93

\bibitem[{{Deheuvels} {et~al.}(2012){Deheuvels}, {Garc{\'\i}a}, {Chaplin},
  {Basu}, {Antia}, {Appourchaux}, {Benomar}, {Davies}, {Elsworth}, {Gizon},
  {Goupil}, {Reese}, {Regulo}, {Schou}, {Stahn}, {Casagrande},
  {Christensen-Dalsgaard}, {Fischer}, {Hekker}, {Kjeldsen}, {Mathur}, {Mosser},
  {Pinsonneault}, {Valenti}, {Christiansen}, {Kinemuchi}, \&
  {Mullally}}]{Deheuvels2012}
{Deheuvels}, S., {Garc{\'\i}a}, R.~A., {Chaplin}, W.~J., {et~al.} 2012, \apj,
  756, 19

\bibitem[{{Di Mauro}(2016)}]{DiMauro2016}
{Di Mauro}, M.~P. 2016, in Frontier Research in Astrophysics II (FRAPWS2016),
  29

\bibitem[{{Di Mauro} {et~al.}(2016){Di Mauro}, {Ventura}, {Cardini}, {Stello},
  {Christensen-Dalsgaard}, {Dziembowski}, {Patern{\`o}}, {Beck}, {Bloemen},
  {Davies}, {De Smedt}, {Elsworth}, {Garc{\'\i}a}, {Hekker}, {Mosser}, \&
  {Tkachenko}}]{DiMauro2016paper}
{Di Mauro}, M.~P., {Ventura}, R., {Cardini}, D., {et~al.} 2016, \apj, 817, 65

\bibitem[{{Eggenberger} {et~al.}(2012){Eggenberger}, {Montalb{\'a}n}, \&
  {Miglio}}]{Eggenberger2012}
{Eggenberger}, P., {Montalb{\'a}n}, J., \& {Miglio}, A. 2012, \aap, 544, L4

\bibitem[{Foreman-Mackey(2016)}]{corner}
Foreman-Mackey, D. 2016, The Journal of Open Source Software, 1, 24

\bibitem[{{Foreman-Mackey} {et~al.}(2013){Foreman-Mackey}, {Hogg}, {Lang}, \&
  {Goodman}}]{DFM2013}
{Foreman-Mackey}, D., {Hogg}, D.~W., {Lang}, D., \& {Goodman}, J. 2013, \pasp,
  125, 306

\bibitem[{Fornberg(1988)}]{fornberg_generation_1988}
Fornberg, B. 1988, Mathematics of computation, 51, 699

\bibitem[{{Gehan} {et~al.}(2018){Gehan}, {Mosser}, {Michel}, {Samadi}, \&
  {Kallinger}}]{Gehan2018}
{Gehan}, C., {Mosser}, B., {Michel}, E., {Samadi}, R., \& {Kallinger}, T. 2018,
  \aap, 616, A24

\bibitem[{Harris {et~al.}(2020)Harris, Millman, van~der Walt, Gommers,
  Virtanen, Cournapeau, Wieser, Taylor, Berg, Smith, Kern, Picus, Hoyer, van
  Kerkwijk, Brett, Haldane, Fernández~del Río, Wiebe, Peterson,
  Gérard-Marchant, Sheppard, Reddy, Weckesser, Abbasi, Gohlke, \&
  Oliphant}]{numpy}
Harris, C.~R., Millman, K.~J., van~der Walt, S.~J., {et~al.} 2020, Nature, 585,
  357–362

\bibitem[{{Hekker} \& {Christensen-Dalsgaard}(2017)}]{Hekker2017}
{Hekker}, S. \& {Christensen-Dalsgaard}, J. 2017, \aapr, 25, 1

\bibitem[{{Hekker} {et~al.}(2018){Hekker}, {Elsworth}, \&
  {Angelou}}]{Hekker2018}
{Hekker}, S., {Elsworth}, Y., \& {Angelou}, G.~C. 2018, \aap, 610, A80

\bibitem[{{Howell} {et~al.}(2014){Howell}, {Sobeck}, {Haas}, {Still},
  {Barclay}, {Mullally}, {Troeltzsch}, {Aigrain}, {Bryson}, {Caldwell},
  {Chaplin}, {Cochran}, {Huber}, {Marcy}, {Miglio}, {Najita}, {Smith},
  {Twicken}, \& {Fortney}}]{K2}
{Howell}, S.~B., {Sobeck}, C., {Haas}, M., {et~al.} 2014, \pasp, 126, 398

\bibitem[{Hunter(2007)}]{matplotlib}
Hunter, J.~D. 2007, Computing in Science \& Engineering, 9, 90

\bibitem[{{Jiang} {et~al.}(2020){Jiang}, {Cunha}, {Christensen-Dalsgaard}, \&
  {Zhang}}]{Jiang2020}
{Jiang}, C., {Cunha}, M., {Christensen-Dalsgaard}, J., \& {Zhang}, Q. 2020,
  \mnras, 495, 621

\bibitem[{{Jiang} {et~al.}(2022){Jiang}, {Cunha}, {Christensen-Dalsgaard},
  {Zhang}, \& {Gizon}}]{Jiang2022}
{Jiang}, C., {Cunha}, M., {Christensen-Dalsgaard}, J., {Zhang}, Q.~S., \&
  {Gizon}, L. 2022, \mnras, 515, 3853

\bibitem[{{Kjeldsen} \& {Bedding}(1995)}]{Kjeldsen1995}
{Kjeldsen}, H. \& {Bedding}, T.~R. 1995, \aap, 293, 87

\bibitem[{{Kuszlewicz} {et~al.}(2023){Kuszlewicz}, {Hon}, \&
  {Huber}}]{Kuszlewicz2023}
{Kuszlewicz}, J.~S., {Hon}, M., \& {Huber}, D. 2023, \apj, 954, 152

\bibitem[{{Matteuzzi} {et~al.}(2023){Matteuzzi}, {Montalb{\'a}n}, {Miglio},
  {Vrard}, {Casali}, {Stokholm}, {Tailo}, {Ball}, {van Rossem}, \&
  {Valentini}}]{Matteuzzi2023}
{Matteuzzi}, M., {Montalb{\'a}n}, J., {Miglio}, A., {et~al.} 2023, \aap, 671,
  A53

\bibitem[{{Miglio} {et~al.}(2021){Miglio}, {Chiappini}, {Mackereth}, {Davies},
  {Brogaard}, {Casagrande}, {Chaplin}, {Girardi}, {Kawata}, {Khan}, {Izzard},
  {Montalb{\'a}n}, {Mosser}, {Vincenzo}, {Bossini}, {Noels}, {Rodrigues},
  {Valentini}, \& {Mandel}}]{Miglio2021}
{Miglio}, A., {Chiappini}, C., {Mackereth}, J.~T., {et~al.} 2021, \aap, 645,
  A85

\bibitem[{{Montalb{\'a}n} {et~al.}(2013){Montalb{\'a}n}, {Miglio}, {Noels},
  {Dupret}, {Scuflaire}, \& {Ventura}}]{Montalban2013}
{Montalb{\'a}n}, J., {Miglio}, A., {Noels}, A., {et~al.} 2013, \apj, 766, 118

\bibitem[{{Montalb{\'a}n} {et~al.}(2010){Montalb{\'a}n}, {Miglio}, {Noels},
  {Scuflaire}, \& {Ventura}}]{Montalban2010}
{Montalb{\'a}n}, J., {Miglio}, A., {Noels}, A., {Scuflaire}, R., \& {Ventura},
  P. 2010, \apjl, 721, L182

\bibitem[{{Montalb{\'a}n} \& {Noels}(2013)}]{Montalban2013a}
{Montalb{\'a}n}, J. \& {Noels}, A. 2013, in European Physical Journal Web of
  Conferences, Vol.~43, European Physical Journal Web of Conferences, 03002

\bibitem[{{Mosser} {et~al.}(2011){Mosser}, {Barban}, {Montalb{\'a}n}, {Beck},
  {Miglio}, {Belkacem}, {Goupil}, {Hekker}, {De Ridder}, {Dupret}, {Elsworth},
  {Noels}, {Baudin}, {Michel}, {Samadi}, {Auvergne}, {Baglin}, \&
  {Catala}}]{Mosser2011}
{Mosser}, B., {Barban}, C., {Montalb{\'a}n}, J., {et~al.} 2011, \aap, 532, A86

\bibitem[{{Mosser} {et~al.}(2014){Mosser}, {Benomar}, {Belkacem}, {Goupil},
  {Lagarde}, {Michel}, {Lebreton}, {Stello}, {Vrard}, {Barban}, {Bedding},
  {Deheuvels}, {Chaplin}, {De Ridder}, {Elsworth}, {Montalban}, {Noels},
  {Ouazzani}, {Samadi}, {White}, \& {Kjeldsen}}]{Mosser2014}
{Mosser}, B., {Benomar}, O., {Belkacem}, K., {et~al.} 2014, \aap, 572, L5

\bibitem[{{Mosser} {et~al.}(2012){Mosser}, {Elsworth}, {Hekker}, {Huber},
  {Kallinger}, {Mathur}, {Belkacem}, {Goupil}, {Samadi}, {Barban}, {Bedding},
  {Chaplin}, {Garc{\'\i}a}, {Stello}, {De Ridder}, {Middour}, {Morris}, \&
  {Quintana}}]{Mosser2012a}
{Mosser}, B., {Elsworth}, Y., {Hekker}, S., {et~al.} 2012, \aap, 537, A30

\bibitem[{{Mosser} {et~al.}(2018){Mosser}, {Gehan}, {Belkacem}, {Samadi},
  {Michel}, \& {Goupil}}]{Mosser2018}
{Mosser}, B., {Gehan}, C., {Belkacem}, K., {et~al.} 2018, \aap, 618, A109

\bibitem[{{Mosser} {et~al.}(2017){Mosser}, {Pin{\c{c}}on}, {Belkacem},
  {Takata}, \& {Vrard}}]{Mosser2017}
{Mosser}, B., {Pin{\c{c}}on}, C., {Belkacem}, K., {Takata}, M., \& {Vrard}, M.
  2017, \aap, 607, C2

\bibitem[{{Mosser} {et~al.}(2015){Mosser}, {Vrard}, {Belkacem}, {Deheuvels}, \&
  {Goupil}}]{Mosser2015}
{Mosser}, B., {Vrard}, M., {Belkacem}, K., {Deheuvels}, S., \& {Goupil}, M.~J.
  2015, \aap, 584, A50

\bibitem[{{Noll} {et~al.}(2024){Noll}, {Basu}, \& {Hekker}}]{Noll2024}
{Noll}, A., {Basu}, S., \& {Hekker}, S. 2024, \aap, 683, A189

\bibitem[{{Ong} \& {Gehan}(2023)}]{Ong2023}
{Ong}, J.~M.~J. \& {Gehan}, C. 2023, \apj, 946, 92

\bibitem[{{Paxton} {et~al.}(2011){Paxton}, {Bildsten}, {Dotter}, {Herwig},
  {Lesaffre}, \& {Timmes}}]{Paxton2011}
{Paxton}, B., {Bildsten}, L., {Dotter}, A., {et~al.} 2011, \apjs, 192, 3

\bibitem[{{Paxton} {et~al.}(2013){Paxton}, {Cantiello}, {Arras}, {Bildsten},
  {Brown}, {Dotter}, {Mankovich}, {Montgomery}, {Stello}, {Timmes}, \&
  {Townsend}}]{Paxton2013}
{Paxton}, B., {Cantiello}, M., {Arras}, P., {et~al.} 2013, \apjs, 208, 4

\bibitem[{{Paxton} {et~al.}(2015){Paxton}, {Marchant}, {Schwab}, {Bauer},
  {Bildsten}, {Cantiello}, {Dessart}, {Farmer}, {Hu}, {Langer}, {Townsend},
  {Townsley}, \& {Timmes}}]{Paxton2015}
{Paxton}, B., {Marchant}, P., {Schwab}, J., {et~al.} 2015, \apjs, 220, 15

\bibitem[{{Paxton} {et~al.}(2018){Paxton}, {Schwab}, {Bauer}, {Bildsten},
  {Blinnikov}, {Duffell}, {Farmer}, {Goldberg}, {Marchant}, {Sorokina},
  {Thoul}, {Townsend}, \& {Timmes}}]{Paxton2018}
{Paxton}, B., {Schwab}, J., {Bauer}, E.~B., {et~al.} 2018, \apjs, 234, 34

\bibitem[{{Paxton} {et~al.}(2019){Paxton}, {Smolec}, {Schwab}, {Gautschy},
  {Bildsten}, {Cantiello}, {Dotter}, {Farmer}, {Goldberg}, {Jermyn}, {Kanbur},
  {Marchant}, {Thoul}, {Townsend}, {Wolf}, {Zhang}, \& {Timmes}}]{Paxton2019}
{Paxton}, B., {Smolec}, R., {Schwab}, J., {et~al.} 2019, \apjs, 243, 10

\bibitem[{{Pin{\c{c}}on} {et~al.}(2020){Pin{\c{c}}on}, {Goupil}, \&
  {Belkacem}}]{Pincon2020}
{Pin{\c{c}}on}, C., {Goupil}, M.~J., \& {Belkacem}, K. 2020, \aap, 634, A68

\bibitem[{{Pin{\c{c}}on} {et~al.}(2019){Pin{\c{c}}on}, {Takata}, \&
  {Mosser}}]{Pincon2019}
{Pin{\c{c}}on}, C., {Takata}, M., \& {Mosser}, B. 2019, \aap, 626, A125

\bibitem[{{Ricker} {et~al.}(2015){Ricker}, {Winn}, {Vanderspek}, {Latham},
  {Bakos}, {Bean}, {Berta-Thompson}, {Brown}, {Buchhave}, {Butler}, {Butler},
  {Chaplin}, {Charbonneau}, {Christensen-Dalsgaard}, {Clampin}, {Deming},
  {Doty}, {De Lee}, {Dressing}, {Dunham}, {Endl}, {Fressin}, {Ge}, {Henning},
  {Holman}, {Howard}, {Ida}, {Jenkins}, {Jernigan}, {Johnson}, {Kaltenegger},
  {Kawai}, {Kjeldsen}, {Laughlin}, {Levine}, {Lin}, {Lissauer}, {MacQueen},
  {Marcy}, {McCullough}, {Morton}, {Narita}, {Paegert}, {Palle}, {Pepe},
  {Pepper}, {Quirrenbach}, {Rinehart}, {Sasselov}, {Sato}, {Seager},
  {Sozzetti}, {Stassun}, {Sullivan}, {Szentgyorgyi}, {Torres}, {Udry}, \&
  {Villasenor}}]{TESS}
{Ricker}, G.~R., {Winn}, J.~N., {Vanderspek}, R., {et~al.} 2015, Journal of
  Astronomical Telescopes, Instruments, and Systems, 1, 014003

\bibitem[{{Serenelli} {et~al.}(2009){Serenelli}, {Basu}, {Ferguson}, \&
  {Asplund}}]{Serenelli2009}
{Serenelli}, A.~M., {Basu}, S., {Ferguson}, J.~W., \& {Asplund}, M. 2009,
  \apjl, 705, L123

\bibitem[{{Shibahashi}(1979)}]{Shibahashi1979}
{Shibahashi}, H. 1979, \pasj, 31, 87

\bibitem[{{Takata}(2005)}]{Takata2005}
{Takata}, M. 2005, \pasj, 57, 375

\bibitem[{{Takata}(2006)}]{Takata2006}
{Takata}, M. 2006, \pasj, 58, 893

\bibitem[{{Takata}(2016{\natexlab{a}})}]{Takata2016}
{Takata}, M. 2016{\natexlab{a}}, \pasj, 68, 109

\bibitem[{{Takata}(2016{\natexlab{b}})}]{Takata2016-2}
{Takata}, M. 2016{\natexlab{b}}, \pasj, 68, 91

\bibitem[{Virtanen {et~al.}(2020)Virtanen, Gommers, Oliphant, Haberland, Reddy,
  Cournapeau, Burovski, Peterson, Weckesser, Bright, {van der Walt}, Brett,
  Wilson, Millman, Mayorov, Nelson, Jones, Kern, Larson, Carey, Polat, Feng,
  Moore, {VanderPlas}, Laxalde, Perktold, Cimrman, Henriksen, Quintero, Harris,
  Archibald, Ribeiro, Pedregosa, {van Mulbregt}, \& {SciPy 1.0
  Contributors}}]{scipy}
Virtanen, P., Gommers, R., Oliphant, T.~E., {et~al.} 2020, Nature Methods, 17,
  261

\bibitem[{{Vrard} {et~al.}(2016){Vrard}, {Mosser}, \& {Samadi}}]{Vrard2016}
{Vrard}, M., {Mosser}, B., \& {Samadi}, R. 2016, \aap, 588, A87

\end{thebibliography}

\begin{appendix} \label{appendix}
\section{Calculation of $q$ in \MESA} \label{sec:calcq}

As we are interested in the behaviour of the coupling coefficient $q$ as a star evolves, the Takata prescription for strong coupling has been implemented in \MESA. The calculation in \MESA is done in several steps and is structured as follows:
\begin{enumerate}[(1)]
	\item Calculate $J$, $\mathcal{A}$ and $\mathcal{\Nu}$ in the star (Eqs.~\eqref{eq:J}, \eqref{eq:A}, and \eqref{eq:Nu}), where $J$ is the perturbation to the gravitational potential, and $\mathcal{A}$ and $\mathcal{\Nu}$ are variables used in the oscillation equations and depend on $N$ and $S$ respectively.
	\item Calculate the angular frequency, $\omega$, of the frequencies at which the coupling is to be calculated and then for each angular frequency $\omega$ do the following:
	\item Calculate $P$ and $Q$ (Eqs.~\eqref{eq:P} and \eqref{eq:Q}).
	\item Find the initial estimate of the evanescent zone boundaries, $r_1$ and $r_2$, using $P$ and $Q$ respectively.
	\item Smooth $P$ and $Q$ as a function of $s$ (Eq.~\ref{eq:smoothPQ}) fully inside and a few points outside the evanescent zone.
	\item Find the final evanescent zone boundaries, $r_1$ and $r_2$, using the smoothed $P$ and $Q$ respectively from step 5.
	\item Calculate $X$ (Eq.~\ref{eq:X}) and evaluate $q$ (Eq.~\ref{eq:q}).
\end{enumerate}
Steps 1--3 are straightforward calculations and steps 4--7 are explained in more detail below. In steps 4 and 6, the edges of the evanescent zones are found by searching for the zeros of $P$ and $Q$. As \MESA is cell based we search for where the signs of $P$ and $Q$ change, and then linearly interpolate to find $r_1$ and $r_2$. However, as \Nred becomes large at the surface (e.g.~Fig.~\ref{fig:propdiagramexplanation}), the outer 10\% by radius is not considered when searching for these zeros. As we are interested in the evanescent zone between the convective envelope and radiative core we only search for zeros above the area where the maximum hydrogen-burning rate takes place. This is above the centre of the star during the MS or above the hydrogen-burning shell in later evolutionary phases.

\begin{figure}
	\centering
	\includegraphics[width=\linewidth]{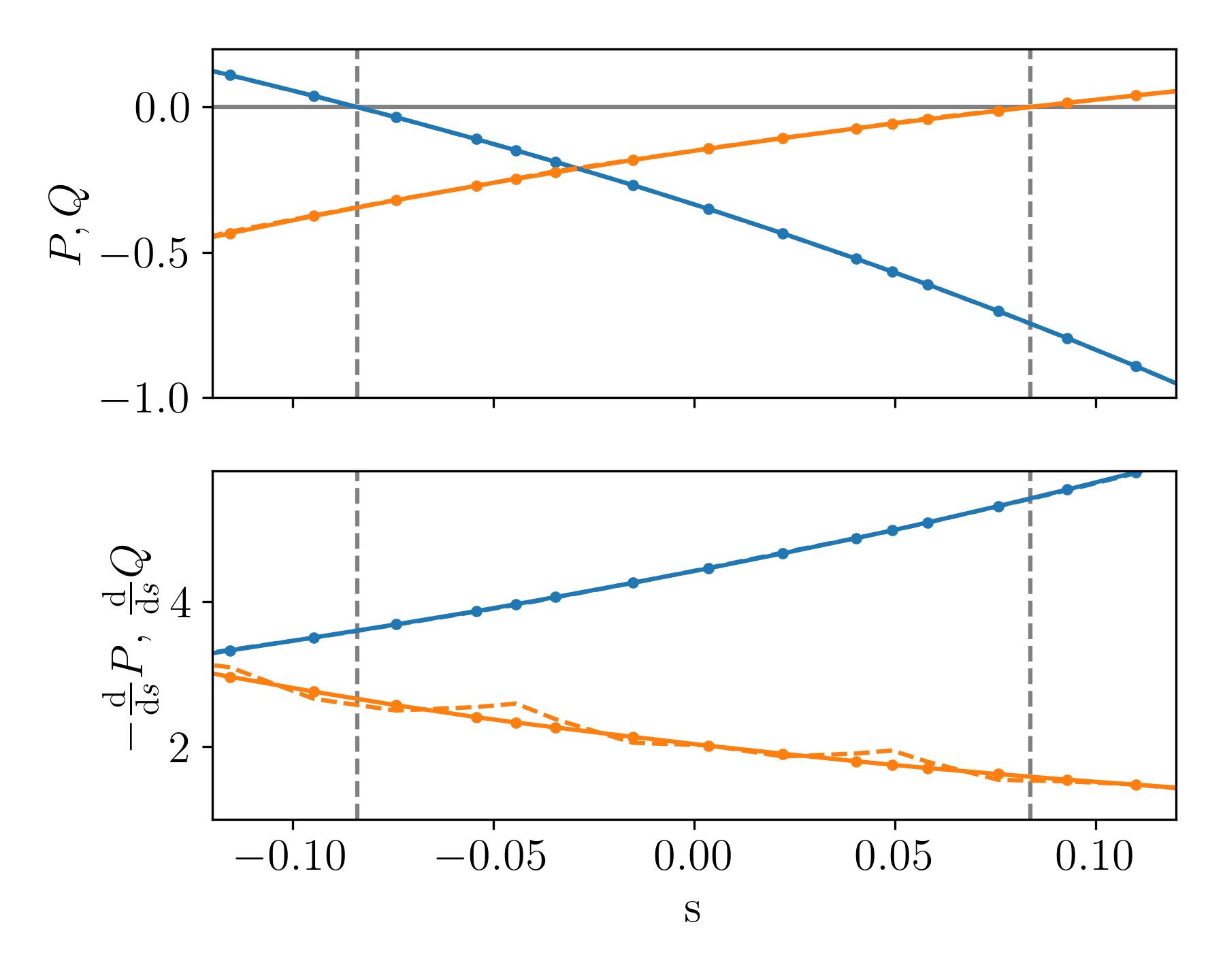}
	\caption{Before (dashed) and after (solid with points) smoothing of $P$ (blue) and $Q$ (orange) in the top panel and their gradients in the bottom panel. The vertical grey dashed lines show $\pm s_0$.}
	\label{fig:smoothing}
\end{figure}

In step 5, using Eqs.~\eqref{eq:P} and \eqref{eq:Q}, $P$ and $Q$ are evaluated and smoothed. The smoothing is done around and inside the evanescent zone, but not in the convective zone. The smoothing is done in two steps. First, a weighted moving average is taken around the central point as follows:
\begin{equation} \label{eq:smoothPQ}
	P_k = \frac{w_2 P_{k-1} + P_k + w_1 P_{k+1}}{1 + w_1 + w_2},
\end{equation}
where $k$ is the cell index, and $w_1$ and $w_2$ are the weights above and below the central point respectively. These weights are defined by
\begin{equation}
	w_1 = \frac{s_{k-1} - s_k}{s_{k-1} - s_{k+1}} \;\;\mathrm{and}\;\;w_2 = \frac{s_{k} - s_{k+1}}{s_{k-1} - s_{k+1}},
\end{equation}
and
\begin{equation}
	s_k = \ln \left(\frac{r_k}{\sqrt{r_1 r_2}}\right).
\end{equation}
The same is done for $Q$. Second, a quadratic is fit to the points $k-2$, $k-1$, and $k+1$ and evaluated at $k$, which becomes the new value $y_\mathrm{new}$ for all the points to be smoothed. The new value is only accepted if $\min (y_{k-2},\, y_{k-1},\, y_{k+1}) \leq y_\mathrm{new} \leq \max (y_{k-2},\, y_{k-1},\, y_{k+1})$. Additionally, when smoothing $Q$, $y_\mathrm{new} = \min(y_\mathrm{new}, 1)$ is included in the fit. This procedure is then repeated for points $k-1$, $k+1$, and $k+2$. This pair of interpolations is repeated 5 times. Figure~\ref{fig:smoothing} shows the result of this smoothing process, where the solid lines with points show the smoothed quantities, and the dashed lines show the raw quantities.
In the top panel the smoothed and unsmoothed $P$ and $Q$ overlap, and in the bottom panel the smoothed and unsmoothed $\dd P /\dd s$ also overlap. However, $-\dd Q /\dd s$ shows a good improvement in smoothness.

In step 7 the calculation of $X$ (Eq.~\ref{eq:X}) is split into two components, an integral part and a gradient part. The integral part is calculated first, which is done in two steps. First, a numerical integral is done using the composite Simpson's rule. This works well with relatively large evanescent zones. However, when the evanescent zone is narrow and contains only a few cells, the value returned is dominated by noise as the mesh does not resolve $\sqrt{PQ}$. As a work-around, we construct a polynomial from 3 points. From the way that $P$ and $Q$ are defined we know that $P(s_0) = 0$ and $Q(-s_0) = 0$. We use $P(0) = P_0$ and $Q(0) = Q_0$, and $P(-s_0) = P_1$ and $Q(s_0) = Q_1$. This allows us to construct the following equations for $P$ and $Q$:
\begin{alignat}{6}
	&P(s_0) &&= a_2 {s_0}^2 + a_1 s_0 + a_0 &&= 0, \\
	&P(0) &&= a_0 &&= P_0, \\
	&P(-s_0) &&= a_2 {s_0}^2 - a_1 s_0 + a_0 &&= P_1,
\end{alignat}
and
\begin{alignat}{6}
	&Q(s_0) &&= a_2 {s_0}^2 + a_1 s_0 + a_0 &&= Q_1, \\
	&Q(0) &&= a_0 &&= Q_0, \\
	&Q(-s_0) &&= a_2 {s_0}^2 - a_1 s_0 + a_0 &&= 0.
\end{alignat}
Solving for the polynomial coefficients for $P$ gives:
\begin{align}
	a_2 &= \frac{P_1 - 2P_0}{2{s_0}^2},\\
	a_1 &= \frac{-P_1}{2s_0},\\
	a_0 &= P_0
\end{align}
and for $Q$:
\begin{align}
	a_2 &= \frac{Q_1 - 2Q_0}{2{s_0}^2},\\
	a_1 &= \frac{Q_1}{2s_0},\\
	a_0 &= Q_0.
\end{align}
$P_0, P_1, Q_0$, and $Q_1$ are found by interpolating using three mesh points on either side of the point of interest using the algorithm described by \citet{fornberg_generation_1988}, and the implementation by \citet{bjodah2021}. Nominally this is an algorithm for the calculation of weights in finite difference formulas for arbitrarily spaced grids, however, in the special case of approximating the zeroth derivative, it provides a fast procedure for polynomial interpolation.

\begin{figure}
	\centering
	\includegraphics[width=\linewidth]{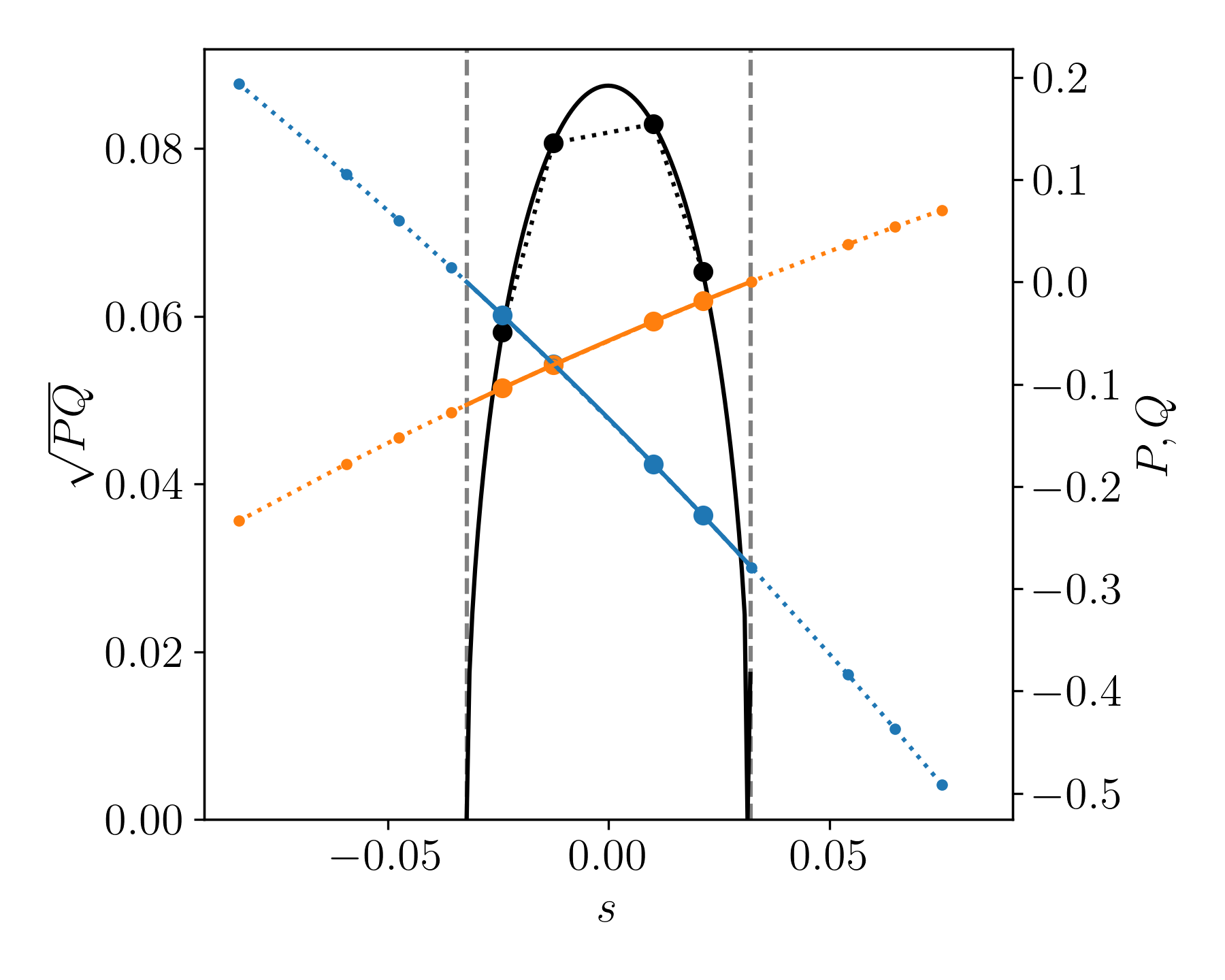}
	\caption{$P$, $Q$, and $\sqrt{PQ}$ are shown as blue, orange, and black lines respectively. Solid lines show values calculated using the quadratic fit, and dotted lines show values using linear interpolation. The coloured circles show values at mesh points, with small circles being outside the evanescent zone and large circles inside. The vertical grey dashed lines show $\pm s_0$.}
	\label{fig:pq}
\end{figure}

\begin{figure}
	\centering
	\includegraphics[width=\linewidth]{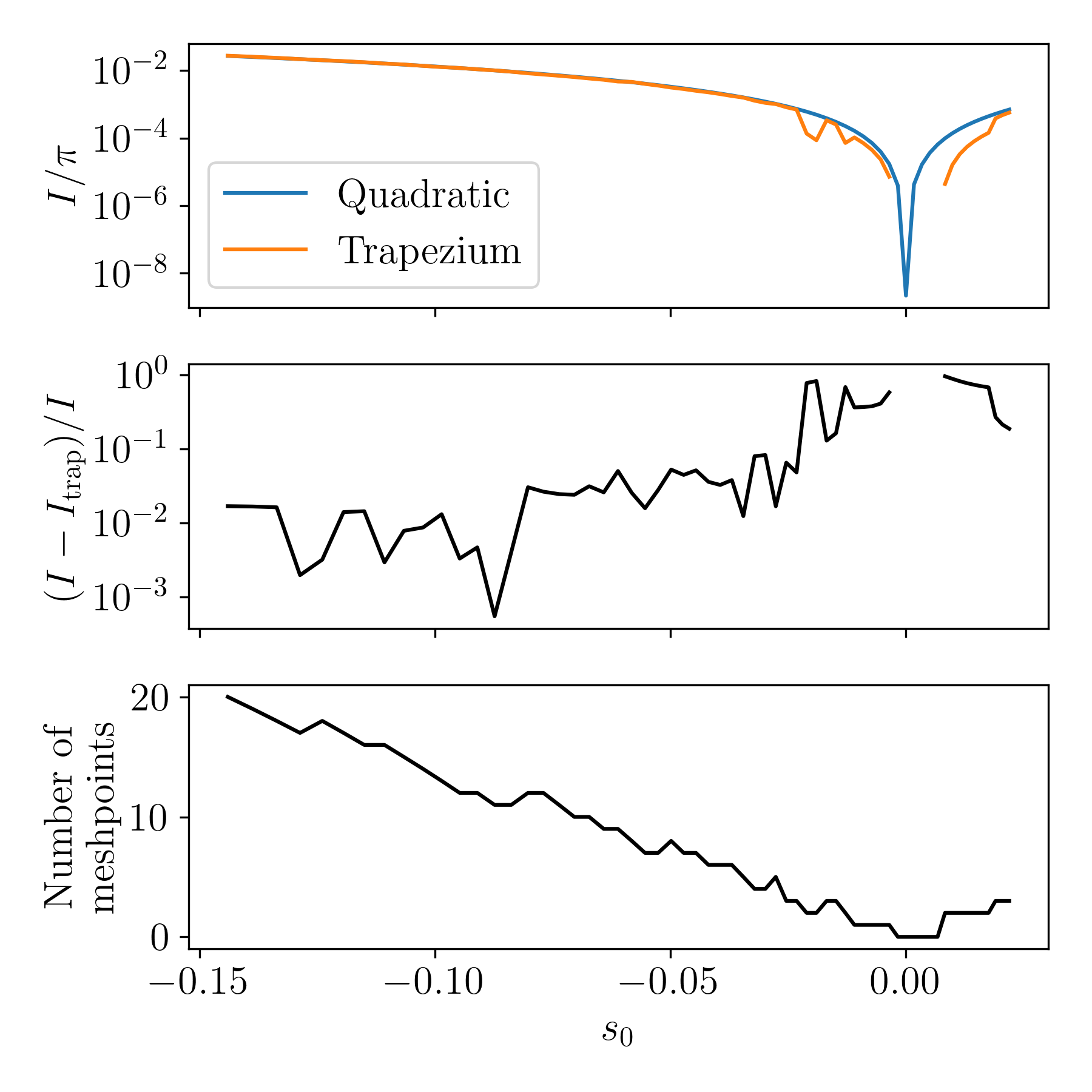}
	\caption{Integral component of $X$, $I/\pi$ (top), fractional difference between the quadratic approximation and the trapezium integration (middle), and number of meshpoints in the evanescent zone (bottom) as a function of $s_0$ of a 1~\msol solar metallicity star in the SGB. The quadratic approximation is shown in blue and the trapezium integration in orange.}
	\label{fig:compareintmethods}
\end{figure}

The integral in Eq.~\eqref{eq:X} is then performed using the composite Simpson's rule with 20 sub-intervals. The numerical integral is used when either of the following empirically determined conditions are true:
\begin{enumerate}[(1)]
	\item the evanescent zone is partially convective and there are at least six meshpoints in the evanescent zone, or
	\item the mean absolute difference between the quadratic fit of $P$ and $Q$ and the values of $P$ and $Q$ at each meshpoint in the evanescent zone is greater than 5\% and there are at least six meshpoints in the evanescent zone.
\end{enumerate}
If neither condition is true, then the quadratic approximation is used. The first condition ensures that the large or partially convective case is avoided when using the quadratic approximation as $Q$ is not approximated well by a quadratic. The second condition avoids using the quadratic approximation in case of a bad fit. Using these thresholds for switching between the quadratic approximation and the numerical integral does not result in noticeable jumps when transitioning between the two. However, using these thresholds does result in a reduction of noise due to the low number of meshpoints.
Figure~\ref{fig:pq} shows the resulting polynomials for $P$, $Q$, and the integrand $\sqrt{PQ}$ in Eq.~\eqref{eq:X}. The resulting equations for $P$ and $Q$ work well and can easily be integrated even though there are only four meshpoints inside the evanescent zone.
Comparing this method to a trapezium integration when there are between 6 and 20 meshpoints available, the integrals are consistent with each other to within 3\%. As fewer meshpoints become available the difference between the two methods grows significantly, as shown in Fig.~\ref{fig:compareintmethods}. The trapezium integration uses the meshpoints in the evanescent zone as well as two additional points at $s = \pm s_0$, where $\sqrt{PQ} = 0$. When the number of meshpoints in the evanescent zone is less than three the trapezium integration is dominated by noise, and when there are zero meshpoints in the evanescent zone the trapezium integral cannot be computed.
However, when any part of the evanescent zone is convective, $P$ and $Q$ are no longer well approximated by a quadratic.

The calculation of the strong-coupling coefficient using the Takata prescription is sensitive to the gradient term when the evanescent zone is small ($|s_0| \lesssim 0.1$) as $\frac{I}{\pi} \ll \frac{\mathcal{G}^2_{s=0}}{2\kappa_{s=0}}$ in this regime. The gradient part of the $\mathcal{G}$ term defined in Eq.~\eqref{eq:G} can be split in two parts:
\begin{equation}
	\frac{\dd}{\dd s}\left[\ln\left(\frac{P}{Q}\frac{s + s_0}{s_0 - s}\right)\right] = \frac{\dd}{\dd s}\ln\left(\frac{P}{s_0 - s}\right) - \frac{\dd}{\dd s} \ln\left(\frac{Q}{s + s_0}\right).
\end{equation}
To accurately evaluate these gradient components at $s=0$ a cubic polynomial is fit to the mesh around $s=0$. However, there are some difficulties when simply fitting this cubic directly to the points around $s=0$. The first issue is that the behaviour of the points around $s=s_0$ for $P$ and $s=-s_0$ for $Q$ introduces noise because both the numerator and the denominator approach 0 as $s$ goes to $s_0$ for $P$ and $s$ goes to $-s_0$ for $Q$. This issue is avoided by cutting out the problematic points and inserting a point at $\pm s_0$ using de l'H\^{o}pital's rule, which states that indeterminate forms of $f(x)/g(x)$ such as 0/0 or $\infty/\infty$ can be determined by evaluating the limit of the quotient of the derivatives $f'(x)/g'(x)$.
The second issue is that the point where $s=0$ moves from one cell to another during the evolution of the star. This, in turn, causes jumps in the polynomial coefficients, which in turn cause jumps in $q$. To avoid these issues, we remesh the points we fit to have an exponential spacing with endpoints between $s=\pm s_0/2$ or $s=\pm 0.05$, whichever is greater. This results in the following points in $s$:
\begin{equation}
	s_i = i \Delta s \exp \left(\frac{\left\lfloor\frac{n}{2}\right\rfloor - |i|}{\frac{1}{2}\left\lfloor\frac{n}{2}\right\rfloor}\right),
\end{equation}
where
\begin{equation}
	\Delta s = \frac{-\max (|s_0|, 0.05)}{n},
\end{equation}
and
\begin{equation}
	i \in -\left\lfloor\frac{n}{2}\right\rfloor \dots \left\lfloor\frac{n}{2}\right\rfloor,
\end{equation}
and $n$ is the number of points to remesh to and is set to $n=11$.

\begin{figure}
	\centering
	\includegraphics[width=\linewidth]{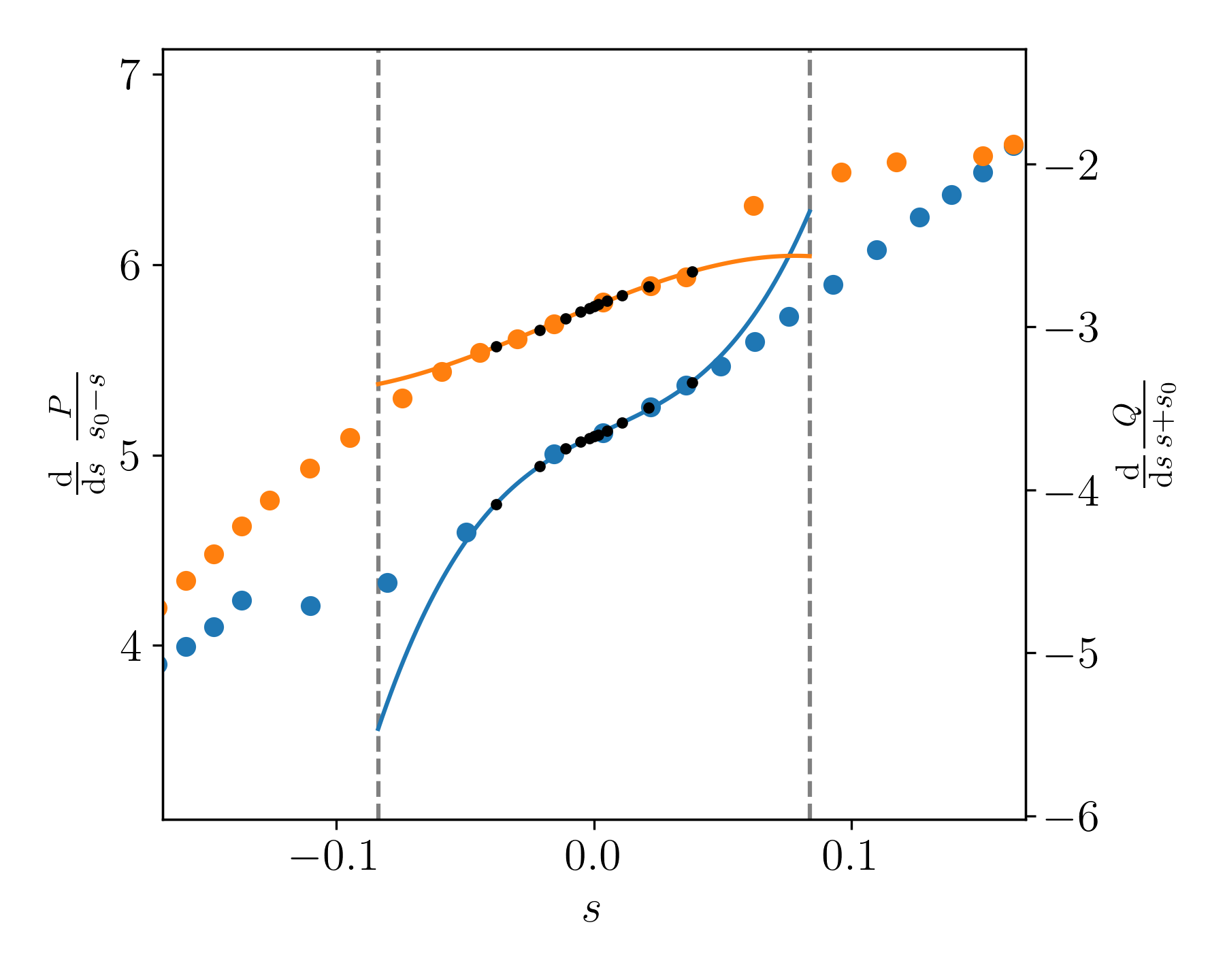}
	\caption{$\frac{\dd}{\dd s}\left(\frac{P}{s_0 - s}\right)$ is shown in blue, and $\frac{\dd}{\dd s}\left(\frac{Q}{s + s_0}\right)$ is shown in orange. Large coloured dots are the calculated gradient after inserting the point using de~l'H\^{o}pital's rule. The small black dots are the remeshed points, and the solid lines show the fitted cubic. The vertical grey dashed lines show $\pm s_0$.}
	\label{fig:meshdpq}
\end{figure}

A cubic polynomial is fitted to the resulting mesh using LAPACK's \texttt{dgels} method \citep{LAPACK}, which solves a real linear system of equations, giving us the best fitting polynomial coefficients $a_{k,P}$ and $a_{k,Q}$. Finally, to calculate the values for the gradient components we use the following equations:
\begin{equation}
	\frac{\dd}{\dd s}\ln\left(\frac{P}{s_0 - s}\right) = \frac{a_{0,P}}{P_0/s_0},
\end{equation}
\begin{equation}
	\frac{\dd}{\dd s}\ln\left(\frac{Q}{s + s_0}\right) = \frac{a_{0,Q}}{Q_0/s_0},
\end{equation}
where $P_0$ and $Q_0$ are the value of $P$ and $Q$ at $s=0$ calculated during the fit for the integral part. Figure~\ref{fig:meshdpq} shows how the fitted cubic performs. The cubic polynomial approximates the gradient around $s=0$ well whilst also being resilient against changes in which meshpoints are used to fit the cubic.

The methods for finding the polynomial coefficients for the integral and gradient terms of Eq.~\eqref{eq:X} are different as they require different properties. In the integral term, it is required that one of the zeros of $P$ and $Q$ coincides with one of the zeros of the quadratic used to fit them, as we are interested in the behaviour near $P=0$ and $Q=0$ in the evanescent zone. However, this is not the case when calculating the gradient. When evaluating the gradient we are interested in the behaviour at $s=0$. Therefore, the fact that the zeros do not coincide exactly has no effect when determining the gradient term. A second, more qualitative reason for not using the integral component's method for evaluating the gradient term, is that it would introduce significant oscillations in $\mathcal{G}$ around crossings of $r_1$ and $r_2$.

\section{Effect of helium abundance on coupling}
\label{sec:helium}
\begin{figure}
	\centering
	\includegraphics[width=\linewidth]{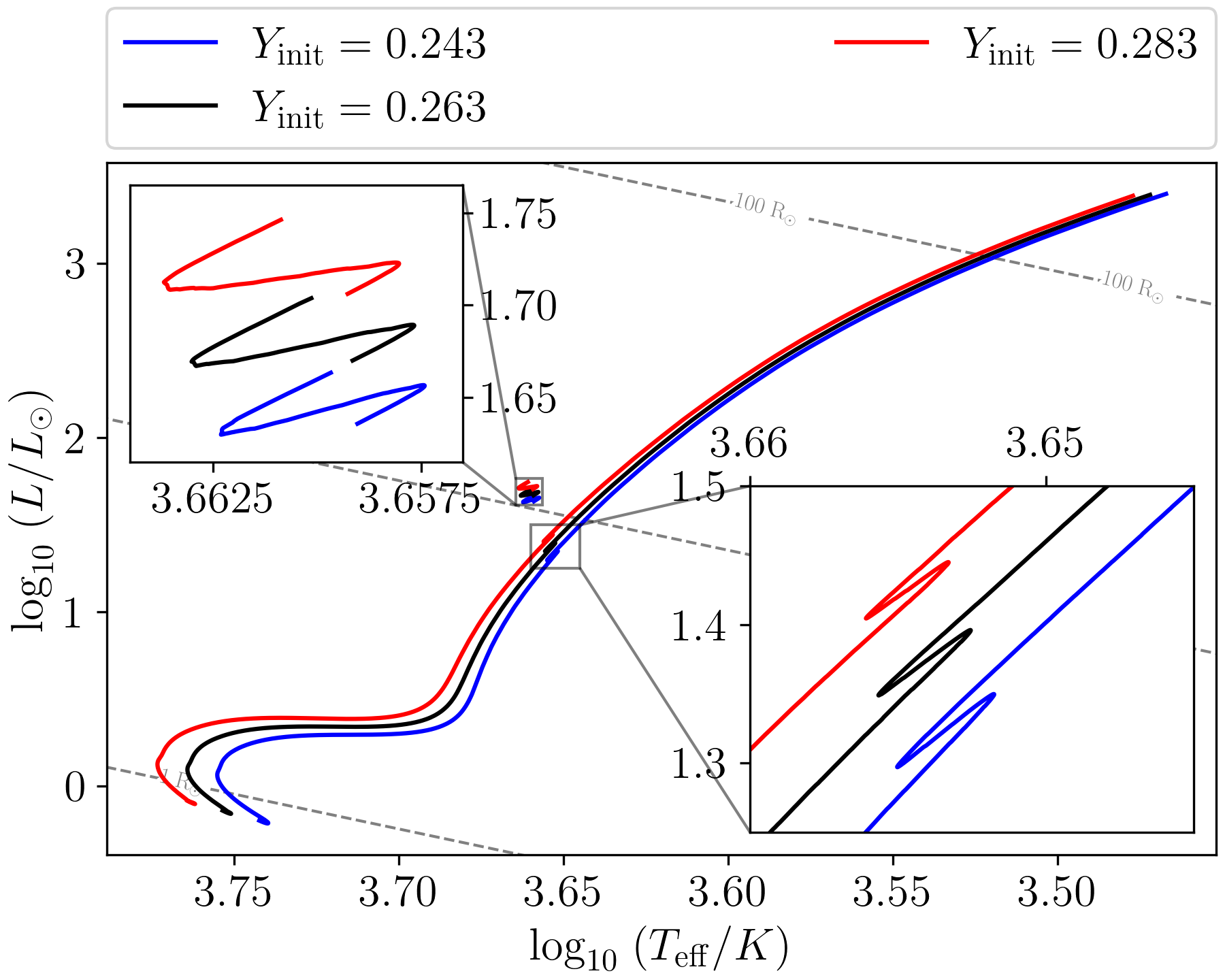}
	\caption{HRD tracks of modelled 1~\msol stars with initial helium abundances $Y_\mathrm{init}$ of 0.243 (blue), 0.263 (black), and 0.283 (red) . The helium flashes are removed for clarity. The upper left inset zooms in on the RC, and the lower right inset zooms in on the RGBb.}
	\label{fig:hehrd}
\end{figure}

An additional test was performed to determine how helium abundance affects $q$. We did this by varying the initial helium abundance of the solar analogue model by $\pm 0.02$. This results in $Y_\mathrm{init}$ of 0.243, 0.263, and 0.283, and keeping $Z_\mathrm{init}$ constant at 0.01448, for a 1~\msol mass with $\FeH=0$. Figure~\ref{fig:hehrd} shows the HRDs of these three tracks where the helium flashes are removed for clarity. During the MS, SGB, and RGB the helium-rich model (red) is the hottest and the helium-poor (blue) the coolest. This is due to helium-rich mixtures having lower opacities compared to helium-poor mixtures, which increases the efficiency of radiative energy transport. The helium-rich RGBb is slightly hotter by just 0.0006 dex and more luminous by 0.06 dex compared to the default $Y_\mathrm{init}$, whilst the helium-poor RGBb is 0.0007 dex cooler and 0.05 dex less luminous. Finally, the RC is at approximately the same effective temperature (differing by 0.0007 dex), whilst the helium-rich RC is 0.04 dex more luminous and the helium-poor RC is 0.04 dex less luminous.

\begin{table}
	\caption{Summary of the changes in $q$ due to varying the initial helium abundance. \label{tab:helium}}
	\centering
	\small
	\begin{tabular}{lccc}
		\hline \hline
		& & Coupling $q$& \\
		Evolutionary Phase & low $Y_\mathrm{init}$ & default $Y_\mathrm{init}$ & high $Y_\mathrm{init}$\\
		\hline
		SGB  & 0.385& 0.399& 0.459\\
		RGBb & 0.060& 0.055& 0.050\\
		RC   & 0.238& 0.252& 0.258\\
		\hline
	\end{tabular}
    \tablefoot{The values of the SGB and RGBb are the maximum $q$ reached during that phase, and for the RC it is the mean.}
\end{table}

Table \ref{tab:helium} shows the observed general effects on the coupling caused by the changes in initial helium abundance showing that during the SGB $q$ is correlated with the initial helium abundance, during the RGB it is anti-correlated, and during the RC it is correlated again with $Y_\mathrm{init}$. However, many of these changes are likely caused by the change in \numax due to the increased \Teff due to a higher helium abundance. For example, qualitatively, in the RC all three modelled stars have approximately the same $q$ of 0.22 when $\nu_\mathrm{q} \simeq 30~\uHz$ and a central helium mass fraction of around 0.5. However, the range $q$ takes when comparing similar frequencies during the RC is larger for helium-rich models.

During the ascent up the RGB and RGBb, the helium-rich core mass is correlated with the helium abundance at a given effective temperature with $\Delta m_\mathrm{He} \simeq 0.1 \Delta Y_\mathrm{init}$. The location of the RGBb in our models is separated mainly by luminosity, with the helium-poor model being the faintest at around $21~\Lsol$, the default model at around $23~\Lsol$, and the helium-rich model at around $27~\Lsol$. These different luminosities in the RGBb are due to the hydrogen-burning shell reaching the chemical discontinuity left by the first dredge-up later in the helium-rich models than in the helium-poor models. This is due to the convective envelope not plunging as deep as in the helium-poor cases as the opacity of the envelope decreases as the helium abundance increases. The luminosity at the tip of the RGB decreases only slightly with increasing $Y_\mathrm{init}$ (2510~\Lsol vs. 2450~\Lsol), and the helium-core mass at the tip is largest in the helium-poor model at 0.476~\msol and smallest in the helium-rich model at 0.469~\msol. During the RC, the helium-poor model has the lowest luminosity of around 44~\Lsol whilst the helium-rich model has a luminosity of around 52~\Lsol with the \Teff of the three cases differing by approximately 10~K during this phase.

\section{Dependence of $q$ with [Fe/H]: posterior distributions of the fitting parameters}
\label{sec:posteriors}
\begin{figure}
	\centering
	\includegraphics[width=\linewidth]{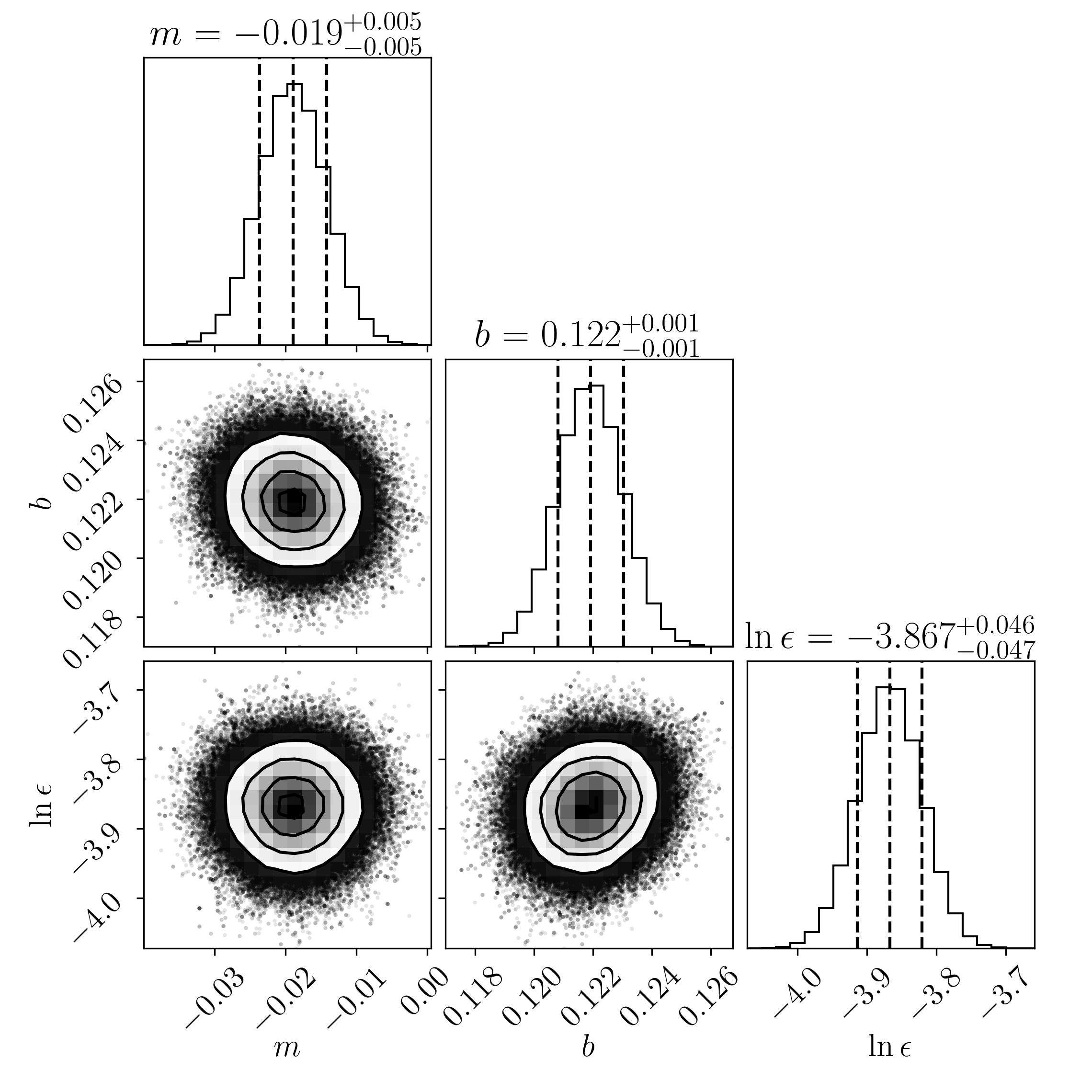}
	\caption{Corner plot of the slope $m$, y-intercept $b$, and intrinsic scatter $\epsilon$ posterior distributions from the RGB fit. The marginalized posterior distributions are shown in the panels in the diagonal with the 16$\mathrm{th}$, 50$\mathrm{th}$, and 84$\mathrm{th}$ percentiles shown as vertical dashed lines. The other panels show the 2-d histogram of the joint posterior distributions.}
	\label{fig:fehqcorner0}
\end{figure}

\begin{figure}
	\centering
	\includegraphics[width=\linewidth]{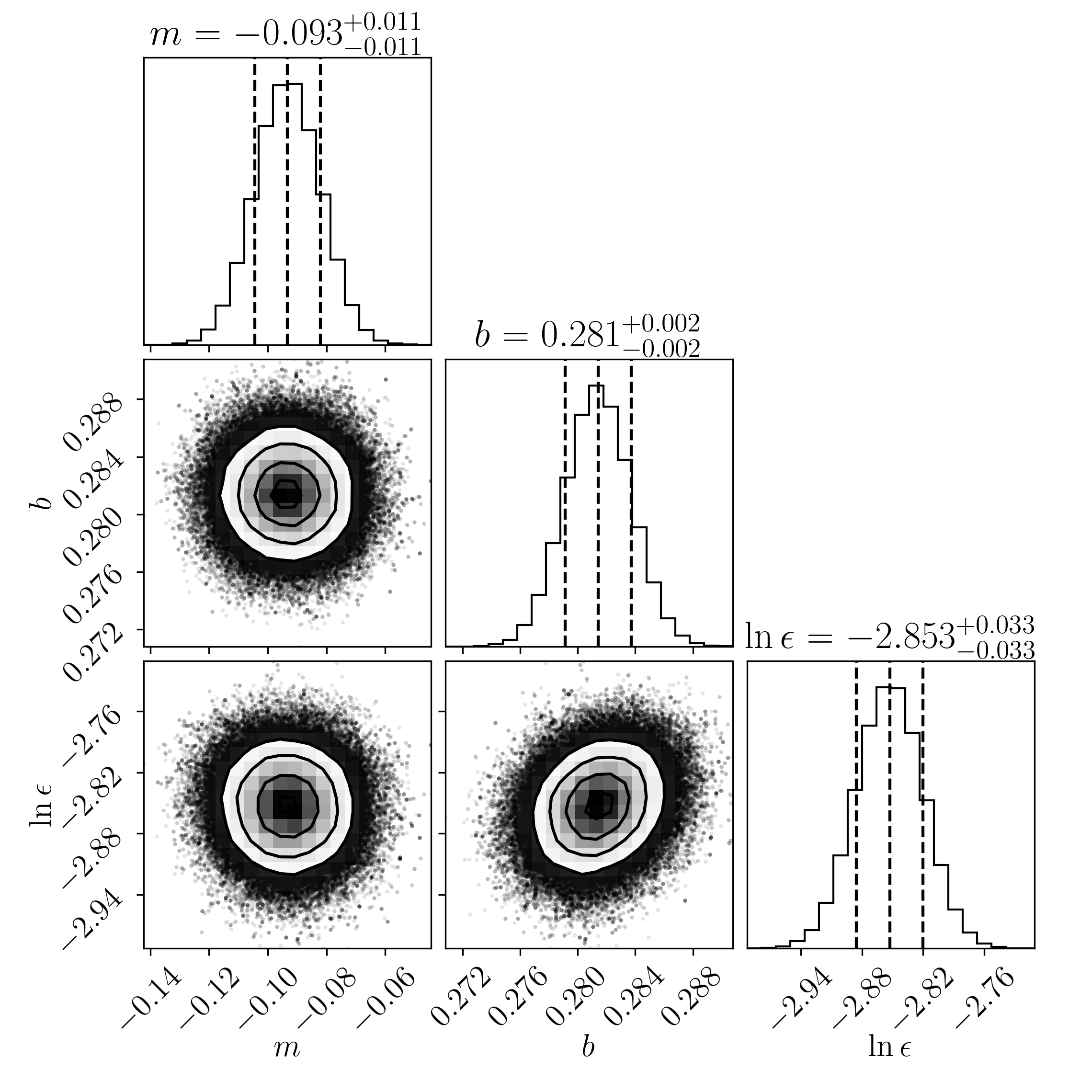}
	\caption{As Fig.~\ref{fig:fehqcorner0} but for RC with masses ${M < 1.8~\msol}$.}
	\label{fig:fehqcorner1}
\end{figure}

\begin{figure}
	\centering
	\includegraphics[width=\linewidth]{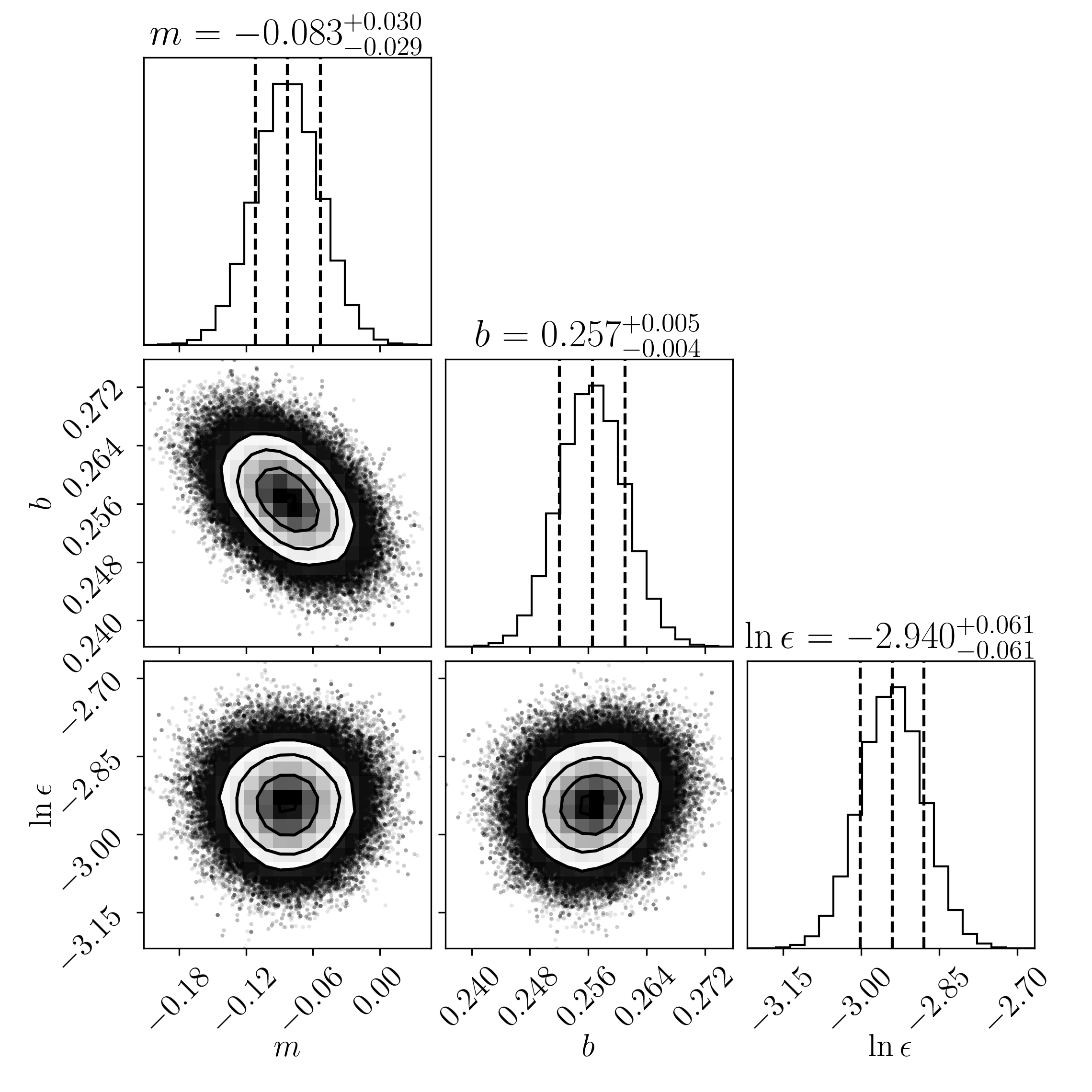}
	\caption{As Fig.~\ref{fig:fehqcorner0} but for RC with masses ${M \geq 1.8~\msol}$.}
	\label{fig:fehqcorner2}
\end{figure}

Figures~\ref{fig:fehqcorner0}, \ref{fig:fehqcorner1}, and \ref{fig:fehqcorner2} show corner plots \citep[e.g.][]{corner} with the posterior distributions of the fitted parameters defined in Eq. \ref{eq:linfit} and their marginalized posterior distributions.

\end{appendix}
\end{document}